%% file: pearceIspyDebrisPltCnstrntsArxiv.tex
\documentclass{aa}

\usepackage[varg]{txfonts}
\usepackage{graphics,graphicx,epsfig,ulem,longtable,pdflscape,lscape,
multirow,appendix,amsmath,amssymb,color,enumitem,natbib,url}
\usepackage{upgreek}
\usepackage[none]{hyphenat}	% Turn off hyphenation

\bibpunct{(}{)}{;}{a}{}{,} % to follow the A&A style

\mathchardef\shorthyphen="2D						% Short hyphen for system names
\newcommand{\mSun}{\; {\rm M_\odot}}				% Solar mass
\newcommand{\mJup}{\; {\rm M_{Jup}}}				% Jupiter mass
\newcommand{\mEarth}{\; {\rm M_\oplus}}				% Earth mass
\newcommand{\lSun}{\; {\rm L_\odot}}				% Solar luminosity
\newcommand{\magnitude}{\; {\rm mag}}				% Magnitude
\newcommand{\au}{\; {\rm au}}						% Astronomical unit
\newcommand{\pc}{\; {\rm pc}}						% Parsec
\newcommand{\m}{\; {\rm m}}							% Metre
\newcommand{\km}{\; {\rm km}}						% Kilometre
\newcommand{\um}{\; {\rm \mu m}}					% Micron
\newcommand{\myr}{\; {\rm Myr}}						% Megayear
\newcommand{\gyr}{\; {\rm Gyr}}						% Gegayear
\newcommand{\K}{\; {\rm K}}							% Kelvin
\newcommand{\mPerS}{\; {\rm m \; s^{-1}}}			% Metres per second
\newcommand{\gPerCmCubed}{\; {\rm g \; cm^{-3}}}	% Grams per cubic centimetre
\newcommand{\percent}{\; {\rm per \; cent}}			% Percent
\newcommand{\mSinI}{M_{\rm p}\sin(i_{\rm p})}		% Planet M sin(i)
\newcommand{\HD}{{\rm HD} \;}						% HD number
\newcommand{\HR}{{\rm HR} \;}						% HR number
\newcommand{\MML}{{\rm MML} \;}						% MML number

\input{journal_defs}

\title{Planet populations inferred from debris discs}

\subtitle{Insights from 178 debris systems in the ISPY, LEECH and LIStEN\\planet-hunting surveys}
  
\author{{\mbox{Tim D. Pearce\inst{\ref{jena}}\thanks{timothy.pearce@uni-jena.de}}
\and \mbox{Ralf Launhardt\inst{\ref{mpia}}}
\and \mbox{Robert Ostermann\inst{\ref{jena}}}
\and \mbox{Grant M. Kennedy\inst{\ref{warwickUni}, \ref{warwickExo}}}
\and \mbox{Mario Gennaro\inst{\ref{stsc}}}
\and \mbox{Mark Booth\inst{\ref{jena}}}
\and \mbox{Alexander V. Krivov\inst{\ref{jena}}}
\and \mbox{Gabriele Cugno\inst{\ref{zurich}}}
\and \mbox{Thomas K. Henning\inst{\ref{mpia}}}
\and \mbox{Andreas Quirrenbach\inst{\ref{landessternwarte}}}
\and \mbox{Arianna Musso Barcucci\inst{\ref{mpia}}}
\and \mbox{Elisabeth C. Matthews\inst{\ref{geneva}}}
\and \mbox{Henrik L. Ruh\inst{\ref{landessternwarte}, \ref{gottingen}}}
\and \mbox{Jordan M. Stone\inst{\ref{navalResearchLab}}}
}}

\institute{Astrophysikalisches Institut und Universit\"{a}tssternwarte, Friedrich-Schiller-Universit\"{a}t Jena, Schillerg\"{a}{\ss}chen 2-3, D-07745 Jena, Germany\label{jena}
\and
{Max Planck Institut f\"{u}r Astronomie, K\"{o}nigstuhl 17, 69117 Heidelberg, Germany}\label{mpia}
\and
{Department of Physics, University of Warwick, Gibbet Hill Road, Coventry CV4 7AL, UK}\label{warwickUni}
\and
{Centre for Exoplanets and Habitability, University of Warwick, Gibbet Hill Road, Coventry CV4 7AL, UK}\label{warwickExo}
\and
{Space Telescope Science Institute, 3700 San Martin Drive Baltimore, MD 21218, USA}\label{stsc}
\and
{ETH Zurich, Institute for Particle Physics and Astrophysics, Wolfgang-Pauli-Strasse 27, 8093 Zurich, Switzerland}\label{zurich}
\and
{Landessternwarte, Zentrum f\"{u}r Astronomie der Universit\"{a}t Heidelberg, K\"{o}nigstuhl 12, D-69117 Heidelberg, Germany}\label{landessternwarte}
\and
{Observatoire de l'Universit\'{e} de Gen\`{e}ve, Chemin Pegasi 51, 1290 Versoix, Switzerland}\label{geneva}
\and
{Institut f\"{u}r Astrophysik, Friedrich-Hund Platz 1, D-37077 G\"{o}ttingen, Germany}\label{gottingen}
\and
{Naval Research Laboratory, Remote Sensing Division, 4555 Overlook Ave.
SW, Washington, DS, USA}\label{navalResearchLab}
}

\titlerunning{Planets inferred from debris discs}
\authorrunning{T. D. Pearce et al.}

\date{Received DATE /
Accepted DATE} 

\abstract
{We know little about the prevalence and properties of the outermost exoplanets in planetary systems, because current detection methods are insensitive to moderate-mass planets on wide orbits. However, debris discs provide indirect probes of the outer-planet population, because dynamical modelling of observed discs can reveal properties of undetected perturbing planets. In this paper we use four sculpting and stirring arguments to infer planet properties in 178 debris-disc systems from the ISPY, LEECH and LIStEN planet-hunting surveys. Similar analyses are often conducted for individual discs, but we consider a large number of systems in a consistent manner. We aim to predict the population of wide-separation planets, gain insight into the formation and evolution histories of planetary systems, and determine the feasibility of detecting perturbing planets in the near future. We show that a `typical' cold debris disc likely requires a Neptune- to Saturn-mass planet at ${10-100\au}$, with some needing Jupiter-mass perturbers. Our predicted planets lie below current detection limits, but modest detection-limit improvements (expected from \textit{JWST}) should finally reveal many such perturbers. We find that planets thought to be perturbing debris discs at late times are very similar to those inferred to be forming in protoplanetary discs, so these could be the same population if newly formed planets do not migrate as far as currently thought. Alternatively, young planets could rapidly sculpt debris before migrating inwards, in which case the responsible planets are much more massive (and located much further inwards) than debris-disc studies assume. We also combine self-stirring and size-distribution modelling, showing that many debris discs cannot be self-stirred without having unreasonably high masses; planet- or companion-stirring may therefore be the dominant stirring mechanism in many (potentially all) debris discs instead. Finally, we provide catalogues of planet predictions for our 178 systems, and identify promising targets for future planet searches.
}


\begin{document}

\maketitle

%%%%%%%%%%%%%%%%%%%%%%%%%%%%%%%%%%%%%%%%%%%%%%%%%%%%%%%%%%%%%%%%%%%%%%%%%%%%%%%

\section{Introduction}
\label{sec: Introduction}

\noindent Whilst the number of detected exoplanets has increased sharply in recent years, relatively few have been found in the outer regions of planetary systems. This is likely due to the limitations of current detection techniques; radial-velocity and transit methods are insensitive to wide-separation planets, and direct imaging has yet to achieve the contrasts and sensitivities required to really probe outer-planet populations. However, debris discs provide an alternative, indirect method to explore system outskirts; discs in these regions can be imaged at high resolution, and their locations and morphologies can bear signatures of dynamical interactions with planets (e.g. \citealt{Wyatt1999, Wyatt2008, Krivov2010, Matthews2014}). These disc features can be used to infer the presence and properties of otherwise-undetectable planets, yielding unique insights into the outer regions of planetary systems.

A debris disc is therefore a planet probe, provided that planet-disc interactions are either ongoing or have previously occurred. Such interactions may be common, for a number of reasons. Firstly, the edges of both the Asteroid and Kuiper Belts are sculpted by dynamical interactions with Solar System planets (e.g. \citealt{Moons1996, Malhotra2019}), so it is reasonable to expect similar processes to occur in extrasolar systems. This is consistent with the emerging picture that many extrasolar debris discs may have edges consistent with planet sculpting (e.g. \citealt{Faramaz2021, Marino2021}). Secondly, many extrasolar systems contain several distinct debris discs at different radial locations, and the broad cavities between these are most-simply explained by multi-planet clearing \citep{Faber2007, Su2013}. Thirdly, some debris discs bear features that are directly attributable to interactions with known companions in the same system (e.g. \citealt{Mouillet1997, Lagrange2010}). Finally, the fact that we observe extrasolar debris discs at all could imply that planet-disc interactions occur; the observed dust is thought to be released through collisions between larger debris, and a popular explanation is that planetary perturbations excite debris random velocities sufficiently for such collisions to be destructive (known as `planet-stirring', \citealt{Mustill2009}, although alternative stirring mechanisms are also possible). If disc-planet interactions are therefore common, and possibly ubiquitous, then studying a large sample of debris discs should offer insights into the population of unseen planets.

\begin{figure}
  \centering
   \includegraphics[width=8cm]{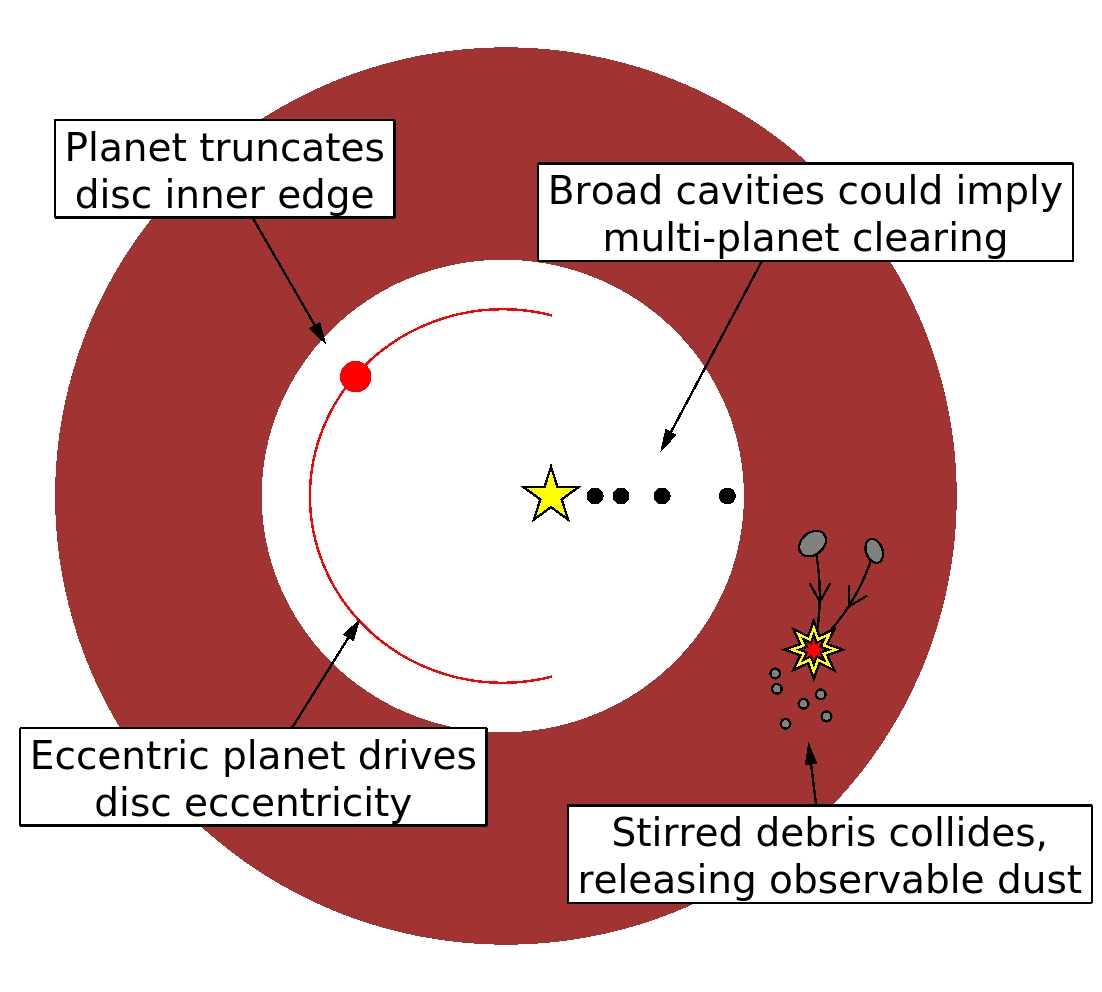}
   \caption{Cartoon showing the dynamical arguments used in this paper to constrain unseen planets from debris discs. At least one planet is assumed to have sculpted the inner edge of each disc (brown); this could be a single, large planet (red circle), or multiple, smaller planets (black circles). If the disc is eccentric, then we ascribe this to perturbations by a planet on an eccentric orbit (red line). The fact that we see dust implies that debris undergoes destructive collisions, which requires the disc to have been stirred; stirring may be performed by unseen planets (although other mechanisms are also possible). Planets can induce other debris morphologies in addition to those shown here, such as warps, resonant structures, and gaps; however, for this paper we only consider the above dynamical arguments.}
   \label{fig: pltConstraintsConcepts}
\end{figure}

In this paper, we use dynamical arguments to constrain planet properties from debris discs around 178 stars. We consider the planet(s) required to truncate and shape each of these discs, and also investigate whether planets are required to stir them; the dynamical arguments we use are outlined on Fig. \ref{fig: pltConstraintsConcepts}. We aim to explore the expected population of wide-separation exoplanets, in order to gain insight into the formation and evolution histories of planetary systems, and determine the feasibility of detecting perturbing planets in the near future. Attempts to constrain planets from debris disc features are common in the literature, but most target individual systems (e.g. \citealt{Mouillet1997, Chiang2009, Backman2009, Read2018, Booth2021, Pearce2021}); we consider a large sample of secure debris-disc stars, and infer properties of the planet population in a consistent manner. Some of the planets that we predict should soon be detectable with the \textit{James Webb Space Telescope} (\textit{JWST}), which will finally provide observational tests of much of the interaction theory discussed over recent decades.

The paper layout is as follows. Sect. \ref{sec: systemSample} describes the system sample, stellar parameters, planet-detection constraints and debris-disc parameters that we use. Sect. \ref{sec: discTruncation} considers the planets required to sculpt our discs, and Sect. \ref{sec: stirring} determines whether planets are also needed to stir them. We discuss our results in Sect. \ref{sec: discussion}, including comparisons of our predicted planet populations to both known planets and those inferred to be forming in protoplanetary discs, as well as identifying systems most favourable for future planet searches. We conclude in Sect. \ref{sec: conclusions}. The Appendix provides a catalogue of our planet predictions, and notes for specific systems.

%%%%%%%%%%%%%%%%%%%%%%%%%%%%%%%%%%%%%%%%%%%%%%%%%%%%%%%%%%%%%%%%%%%%%%%%%%%%%%%%%%%%%%%%%%%%%%%%%%%

\section{System sample and debris disc data}
\label{sec: systemSample}

To infer planet properties from debris discs, we first require information about the discs and their host stars. We also require planet-detection limits for later comparison with our predicted planets. Sect. \ref{subsec: sampleSystems} describes the system sample, stellar properties and planet-detection limits, and Sect. \ref{subsec: debrisDiscData} the debris-disc data. We also quote ${1\sigma}$ uncertainties on these data; we will later use Gaussian uncertainty propagation to derive ${1\sigma}$ uncertainties on any values calculated from these data.

%---------------------------------------------------------
\subsection{System sample, stellar parameters and planet-detection limits}
\label{subsec: sampleSystems}

%---------------------------------------------------------
\subsubsection{System sample}

Our sample of debris disc stars is drawn from three large, high-contrast, L$'$-band imaging surveys: the NaCo Imaging Survey for Planets around Young stars (NaCo-ISPY: \citealt{Launhardt2020}), the Large Binocular Telescope Interferometer Exozodi Exoplanet Common Hunt (LEECH: \citealt{Stone2018}), and the L$'$-band Imaging Survey for Exoplanets in the North (LIStEN: \citealt{MussoBarcucci2021}). LEECH and LIStEN were conducted using LMIRCam on the Large Binocular Telescope Interferometer (LBTI), while ISPY used NaCo on the Very Large Telescope (VLT). The target selection criteria for these surveys are described in their respective papers, but can roughly be summarised as follows: targets were restricted to nearby stars including low-mass star-forming regions (distances $\lesssim 160\pc$), exclude extreme spectral types (those later than M5 or earlier than B8), were observable by our telescopes (declinations between $-73$ and $+70^\circ$), had good adaptive-optics performance (K-band magnitudes below 10), and were not too young (ages ${\gtrsim 10\myr}$), focussing on pre-main-sequence stars up to several ${100\myr}$ but also including main-sequence stars with Gyr ages. For each star in these samples, we performed an SED fitting (Sect. \ref{subsec: sedSystems}) and selected only those stars with excess infrared emission that confirmed with high confidence the existence of at least one debris disc; stars with no significant excesses, or possible blending with background objects, were discarded. In total we selected 178 debris disc targets, of which 136 were in the ISPY survey, 33 in LEECH (including eight stars overlapping with ISPY), and 22 in LIStEN (including five overlapping with ISPY). These 178 stars are listed in Table \ref{tab: systemObsData}.

%---------------------------------------------------------
\subsubsection{Stellar parameters}
\label{subsec: stellarParameters}

For each star, its distance $d$ was derived from its \textit{Gaia}-DR2 parallax according to the formalism described by \cite{BailerJones2018}. The effective temperature $T_{\rm eff}$ and luminosity ${L_*}$ of the star, as well as the fractional luminosity and blackbody temperature ${T_{\rm BB}}$ of its associated debris disc(s), were derived by fitting simultaneous stellar ({\sc phoenix}; \citealt{husser2013}) and blackbody models to the observed photometry and spectra (see Sect. \ref{subsec: sedSystems}). The photometry was obtained from multiple catalogues and publications, including 2MASS, APASS, \textit{Hipparcos}/Tycho-2, \textit{Gaia}, \textit{AKARI}, \textit{WISE}, \textit{IRAS}, \textit{Spitzer}, \textit{Herschel}, JCMT, and ALMA. In some cases photometry was excluded, for example due to saturation or confusion with other objects. The fitting method used synthetic photometry of grids of models to find best fits with the {\sc MultiNest} code \citep{Feroz2009,Buchner2014, Yelverton2019}. For uncertainties on effective temperature $T_{\rm eff}$, we adopted a lower limit of ${70\K}$ (even if the formal fitting uncertainty was smaller); this lower limit was derived from the rms scatter of the difference between photometric and spectroscopic effective temperatures. 

Stellar metallicities were taken from \cite{Gaspar2016} where available (133 stars); for the remaining stars we adopted ${\rm [Fe/H] = -0.03 \pm 0.19}$, which is the median and standard deviation of the 133 stars with metallicities. L$'$ magnitudes were derived by weighted interpolation of the 2MASS \citep{Cutri2003} JHK and WISE \citep{Cutri2013} W1-3 fluxes to the central wavelength of the L$'$ filter (${3.80\um}$), taking into account saturated flux measurements and upper limits.

To derive stellar ages, we first checked each target for membership of known associations using the {\sc banyan $\Sigma$} tool\footnote{http://www.exoplanetes.umontreal.ca/banyan/banyansigma.php} \citep{Gagne2018}. If the membership probability was ${> 80 \percent}$, then we assigned the stellar age $t_*$ to be the mean age of the association\footnote{For seven stars with membership probabilities between 55 and ${76 \percent}$, we also assigned the mean age of the association because these ages are widely used in the literature and our isochronal ages did not contradict the association ages. These systems are marked in Table \ref{tab: systemObsData}.}. In total we assigned association ages to 78 of our 178 stars. For our remaining 100 field stars, ages were assigned by compiling various literature estimates (particularly from large surveys with diverse age-determination methods). If these various estimates yielded contradictory ages, then we considered possible reasons before assigning an age (for example, the luminosity could be corrupted by unresolved binarity, or some methods may be inappropriate for the spectral type and age, or the literature uses an outdated association membership that appears unlikely given newer photometry, parallaxes, radial velocities and astrometry). If multiple valid age estimates were available, then we adopted an age (and conservative uncertainty range) that accounted for the spread in estimates.

To validate our adopted ages, and to derive stellar masses, we performed Hertzsprung-Russell diagram isochrone fits. We adopted evolutionary models from the MIST project\footnote{http://waps.cfa.harvard.edu/MIST/} \citep{Dotter2016, Choi2016, Paxton2011, Paxton2013, Paxton2015, Paxton2018}. Our method consisted of finding combinations of star mass $M_*$, age and metallicity that reproduced the measured effective temperature and luminosity, using the MIST models (despite having metallicity estimates, we still needed to sample metallicity to predict $T_{\rm eff}$ and ${L_*}$ properly). The likelihood of the predicted $T_{\rm eff}$ and $L_*$ given the data was modelled as a two-dimensional normal distribution, with standard deviations set equal to the ${1\sigma}$ uncertainties on $T_{\rm eff}$ and $L_*$ (the two were assumed to be uncorrelated, based on inspection of the $T_{\rm eff}$ and $L_*$ samples produced by the {\sc MultiNest} fits). We assumed a \cite{Chabrier2003} single-star prior on mass. For field stars, we assumed a $1/t_*$ prior on age (uniform in ${\log t_*}$); for stars deemed to belong to associations, we substituted the $1/t_*$ prior with a normal one based on the association mean age and its standard deviation. For metallicity, we assumed a normal prior based on the \cite{Gaspar2016} measurements. The posterior distribution of mass, age and metallicity was derived using the {\sc emcee} sampler\footnote{https://emcee.readthedocs.io/en/stable/}. The resulting stellar masses (the medians of the posterior distributions), along with their ${1\sigma}$ uncertainties, were adopted for each star. The corresponding ages and uncertainties (which are naturally large for stars close to the main sequence) were not adopted; however, they were used to verify the consistency of our adopted ages with the isochrone fits\footnote{Our adopted and isochronal ages agree to within ${2\sigma}$ for all but two stars, ${\MML8}$ and ${\MML36}$. For these two stars, the isochronal ages and the adopted (association) ages differ only by factors of 2-3, and the isochrone fits confirm their youth (${\sim10\myr}$). \cite{Gaspar2016} do not provide metallicities for these stars, so we adopted the median metallicity of our sample; this could affect the isochronal age estimates for such young stars, and potentially explain the discrepancy. Also note that the formal uncertainties do not reflect the model uncertainties, which may be large at young ages.}. Only for one star (${\HD218340}$), for which we could not find any literature age, did we adopt the isochrone age directly. The corresponding metallicities were also not used; instead we adopted the \cite{Gaspar2016} measurements, which were typically very similar.

Close binary stars were not excluded \textit{a priori}, and were only avoided in certain circumstances (for example, where they would hamper coronagraphic observations). However, for each system we checked \textit{a posteriori} for close binarity (${\rm separations < 1 \; arcsec}$) using our L$'$-band images and the SB9 and WDS catalogues \citep{Pourbaix2004, Mason2020}. While these resources may be incomplete (in particular for very close companions with ${\rm separations < 0.2 \; arcsec}$), they are the most complete repository to verify whether our calculated luminosity might be affected by previously unresolved sources, which would also affect our mass and age determination (see above). Moreover, physical binarity would affect the system dynamics and consequently our derived planet parameters, although additional data are often required to show physical association (e.g. \citealt{Pearce2015Orbits}). Targets where such close sources could affect our stellar parameters are flagged by asterisks in Table \ref{tab: systemObsData}, and some possible effects of binarity on our results are discussed in Sect. \ref{subsec: binaryDiscussion}.

Our adopted distances, luminosities, effective temperatures, masses and ages are listed in Table \ref{tab: systemObsData}, along with their ${1\sigma}$ uncertainties and the references and associations considered in our age determination. For our 178 targets, the distances range from 3.2 to ${160 \pc}$ (median ${42 \pc}$), luminosities from ${4.7 \times 10^{-3}}$ to ${390 \lSun}$ (median ${3.6 \lSun}$), stellar temperatures from 3000 to ${12000 \K}$ (median ${6700\K}$), masses from 0.22 to ${3.7\mSun}$ (median ${1.4\mSun}$), and ages from ${10\myr}$ to ${6\gyr}$ (median ${200\myr}$). 

%---------------------------------------------------------
\subsubsection{Planet-detection limits}
\label{subsec: planetDetectionLimits}

To get upper limits on the masses of undetected planets, we first extract contrast-detection-threshold curves from the L$'$-band images. For comparability with other surveys, we employ a classical ${5\sigma}$ detection threshold; as a function of angular separation, we determine the mean contrast required for a PSF-shaped planet to exceed this ${5\sigma}$ threshold by injecting ‘fake planets’ into the images as described in \cite{Launhardt2020}. We then determine the physical relation between this contrast-detection threshold and planet mass, given the star's distance, age and L$'$ magnitude, together with the AMES ‘hot start’ COND (no photospheric dust opacity; \citealt{Baraffe2003}) and DUSTY (maximal dust opacity; \citealt{Chabrier2000}) evolutionary model isochrones for the L$'$ passband. The mass-detection limits from the two models were then stitched together according to their validity range (COND: ${<1400\K}$, DUSTY: ${>1700\K}$) based on the best-fit COND effective temperatures, with a linear interpolation between the two validity ranges. Our resulting L$'$ contrast limits have median and standard deviation of ${6.8\pm1.3\magnitude}$ at ${0.15\; \rm arcsec}$ and ${9.6\pm1.2\magnitude}$ at ${0.5\; \rm arcsec}$, with median and standard deviation mass-detection limits of ${40 ^{+130}_{-10} \mJup}$ at ${10\au}$ and ${8^{+15}_{-4} \mJup}$ at ${100\au}$. These detection limits are not always as constraining as those from other instruments (e.g. SPHERE; see Fig. \ref{fig: hd9672PltCnstrnts}), but the strength of our data arises from the size and homogeneity of our sample of secure debris disc stars.

We also consider Gaussian uncertainty ranges on our mass-detection curves. For each image, the uncertainty on the azimuthally averaged ${5\sigma}$ contrast curve was taken to be the azimuthal variance of the contrasts. To derive ${1\sigma}$ uncertainties on the mass-detection limits, this contrast variance was combined with uncertainties on other parameters (propagated via Monte-Carlo sampling); these uncertainty sources are the L$'$ magnitude of the star, its distance, and its age (the latter in most cases being the dominating uncertainty factor). Throughout the paper we will use the mass-detection limits calculated above, together with these associated ${1\sigma}$ uncertainties.

Several of our systems already have low-mass stellar/brown dwarf companions published, and those that could affect our debris disc dynamics are described in Appendix \ref{app: specificSystems}. In addition, some potentially planet-mass companions are present in our images, but these are undergoing verification and will not be discussed in this paper.
 
%---------------------------------------------------------
\subsection{Debris disc data}
\label{subsec: debrisDiscData}

Constraining planet parameters from a debris disc requires knowledge of the debris morphology. Our models will use the spatial distribution of large parent bodies (e.g. asteroids), which we infer from observations of dust thermal emission; we therefore derive our disc properties from either ALMA, \textit{Herschel} or Spectral Energy Distribution (SED) data. Thermal emission is used because, unlike scattered light, the former is expected to represent the underlying parent-body morphology reasonably well; thermal emission traces larger grains that are less affected by non-gravitational forces, and its shallower radial dependence (${\propto r^{-1/2}}$ rather than ${\propto r^{-2}}$ for scattered light, where $r$ is the stellocentric distance) makes it better at constraining the outer edges of broad discs. Hence we do not use scattered light data in our modelling. Whilst this means some additional features are neglected, such as the sweeping morphology of `The Moth' (${\HD61005}$; \citealt{Hines2007}) in scattered light, considering only thermal emission means that all of our disc morphologies (and resulting planet constraints) will be derived in a consistent manner.

For each system, debris disc parameters are taken from the best-available data source; ALMA data are preferred, then \textit{Herschel}, and finally unresolved SED data. Only one debris model, derived from one data source, is used per disc. If multiple discs are present, then we only consider the outermost disc. For each disc we aim to predict planets located interior to the disc inner edge; to do this, the most important disc parameters are the locations and shapes of the disc inner and outer edges. We describe these edges using four parameters: the pericentre $q_{\rm i}$ and apocentre $Q_{\rm i}$ of an ellipse tracing the disc inner edge, and similarly the pericentre $q_{\rm o}$ and apocentre $Q_{\rm o}$ of an ellipse tracing the outer edge (see Fig. \ref{fig: hd202628Image}). These parameters are derived using different models according to whether the system has ALMA, \textit{Herschel} or SED data available, as described in Sects. \ref{subsec: almaSystems}-\ref{subsec: sedSystems}. 

%---------------------------------------------------------
\subsubsection{Systems with ALMA data}
\label{subsec: almaSystems}

\begin{figure}
  \centering
   \includegraphics[width=8cm]{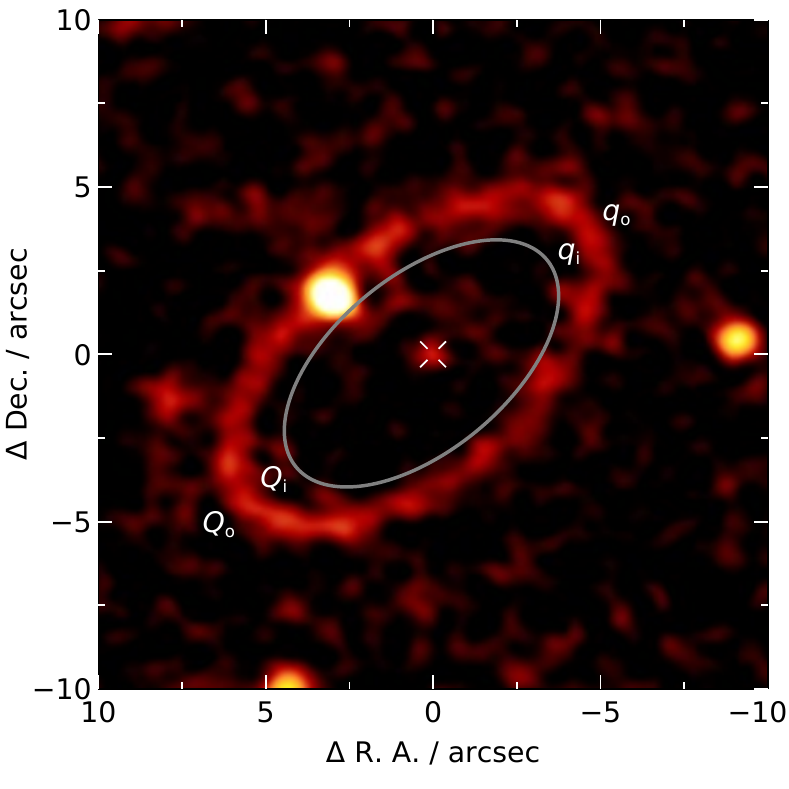}
   \caption{ALMA Band 6 ($1.3 \; \rm mm$) image of ${\HD202628}$ \citep{Faramaz2019}, as an example of the ALMA data used in this paper (Sect. \ref{subsec: almaSystems}). North is up, east is left. The debris disc is well-resolved and clearly eccentric, with the host star (crosshair) offset from the geometric centre of the disc. Considering the pericentre ${q_{\rm i}}$ and apocentre ${Q_{\rm i}}$ of the disc inner edge, plus equivalent values ${q_{\rm o}}$ and ${Q_{\rm o}}$ for the outer edge, yields the minimum mass of a single planet required to sculpt and stir the disc within the system lifetime (Sects. \ref{subsec: singlePlanet} and \ref{subsec: planetStirring}). This minimum-mass planet would have the orbit shown by the grey line; alternatively, a higher-mass planet located further inwards could also sculpt and stir the disc. It is unclear whether the bright point in the north-east part of the ring is a background object or associated with the system \citep{Faramaz2019}.}
   \label{fig: hd202628Image}
\end{figure}

32 of our 178 systems have reliable ALMA disc models in the literature. For these systems we take the disc parameters from literature ALMA fits, which are expected to be the most-reliable disc parameters currently available. These are also the only systems for which we consider asymmetric disc models, and these systems are expected to yield the tightest planet constraints of all systems in our sample. An example ALMA image that we use is shown on Fig. \ref{fig: hd202628Image}.

If several literature ALMA models are available for one disc, then we pick one of them. In choosing, we consider whether the model allows the disc to be asymmetric; some discs have well-constrained eccentricities, and we favour these models over axisymmetric fits because the former potentially allow planet eccentricities to be constrained. We also consider the data and fit quality, and whether system-specific considerations have been included. Systems with disc parameters derived from ALMA data are identified in Table \ref{tab: systemObsData}, and the literature sources for individual disc models are given in Appendix \ref{app: specificSystems}. Some of our ALMA fits have not yet been published but come from the upcoming REASONS analysis (\citealt{Sepulveda2019}; Matr\`{a} et al. in prep.); their spatial dust model follows equation 1 in \cite{Matra2020}, and we refer the reader to that paper for details of the fitting process for these systems.

Since we use ALMA models from a range of literature sources, the fitted disc edges are not defined in the same way across all ALMA systems. Some works, such as \cite{LiemanSifry2016}, use a broad disc model with sharp edges (similar to our \textit{Herschel} models in Sect. \ref{subsec: herschelSystems}); in these cases we use their fitted inner and outer edges. Others, such as REASONS, model the dust spatial morphology as Gaussian; in this case we take the disc width as the fitted full width at half maximum, and calculate the edges accordingly.

%---------------------------------------------------------
\subsubsection{Systems with \textit{Herschel} data, but no ALMA data}
\label{subsec: herschelSystems}

\noindent Of the 146 systems without ALMA data, many still have resolved \textit{Herschel} PACS data to which disc models can be fitted. The lower \textit{Herschel} resolution makes these disc parameters less reliable than those in ALMA systems, but they are still favoured over those derived from unresolved SED data alone (Sect. \ref{subsec: sedSystems}). Whilst our PACS data have shorter wavelengths than ALMA (${\sim100\um}$ for PACS vs. ${\sim1 \; \rm mm}$ for ALMA), both instruments are expected to trace the cold populations of parent planetesimals (e.g. \citealt{Krivov2010, Pawellek2019}). The main differences between disc parameters inferred from \textit{Herschel} and ALMA data are expected to be dominated by differences in resolution, rather than different disc morphologies at different wavelengths; for this reason, we can compare planets inferred from ALMA discs to those from \textit{Herschel} discs.

Using data from the \textit{Herschel} Science Archive\footnote{http://archives.esac.esa.int/hsa/whsa/}, we identify every system in our sample with PACS 70 or ${100\um}$ data. If level 2.0 and 2.5 data were available, we chose 2.5 (which combines scans to create a single image). For these systems, a parametric debris disc model was fitted to each individual 70 and ${100\um}$ image. The model is similar to that in \cite{Yelverton2019}, and consists of a star plus a single axisymmetric disc with sharp inner and outer edges. The locations of these edges were treated as free parameters, along with the disc orientation, disc flux and star position (to account for non-perfect \textit{Herschel} pointing). The surface brightness profile was fixed as ${r^{-1.5}}$, and the discs were assumed to be optically and geometrically thin. Note that, unlike \cite{Yelverton2019}, no unresolved inner component was included (discussed later in this section).  

The fitting procedure is described in \cite{Yelverton2019}, and was conducted using publicly available code\footnote{http://github.com/drgmk/pacs-model}. We refer the reader to that paper for a detailed description of the fitting process, but briefly it involved using a Markov Chain Monte Carlo method to fit the disc-plus-star model, which had been convolved with a point-spread function (PSF) estimated using a calibration star. The disc fitting was only performed if PSF subtraction yielded significant residuals around the target star, which could be indicative of a resolved debris disc. Since each available 70 and ${100\um}$ dataset was considered separately, the analysis yields multiple \textit{Herschel} models for some systems: one for each ${70\um}$ image, and one for each ${100\um}$ image.

\begin{figure*}
  \centering
   \includegraphics[width=17cm]{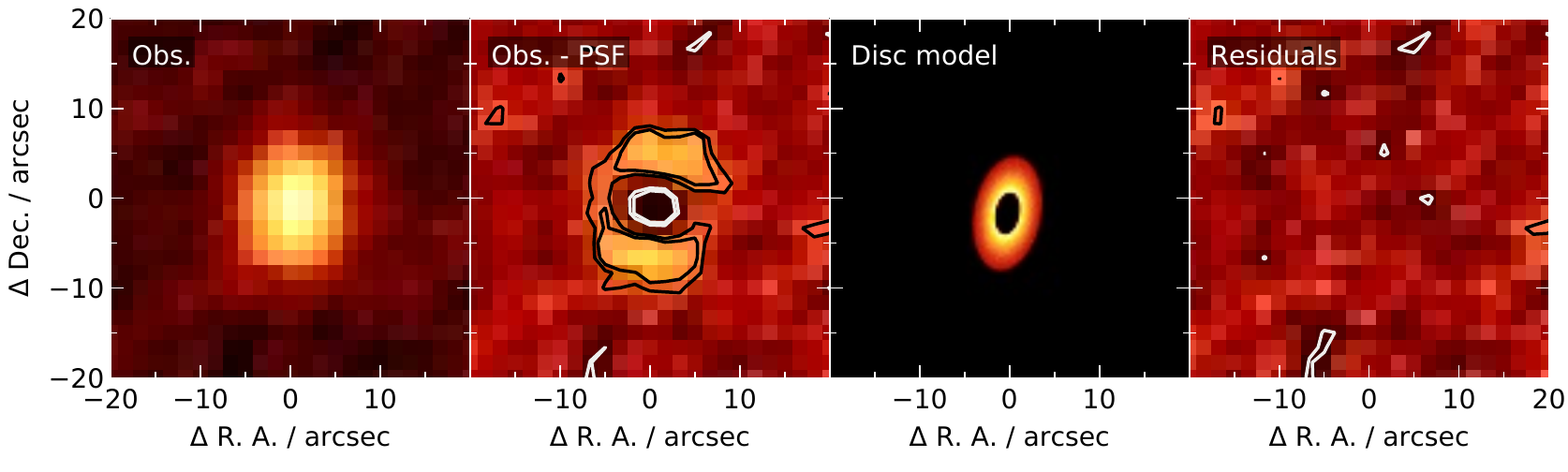}
   \caption{\textit{Herschel} PACS ${100\um}$ image of ${\HD1461}$, as an example of the \textit{Herschel} data used in this paper (Sect. \ref{subsec: herschelSystems}). North is up, east is left. From left to right, the plots show the observation, observation after PSF subtraction, best-fitting disc model, and residuals after model subtraction. Contours show ${\pm2}$ and ${\pm3}$ times the root-mean-square intensity. Whilst the disc is resolved, it clearly has lower resolution than the example ALMA image on Fig. \ref{fig: hd202628Image}. Our \textit{Herschel} disc models are axisymmetric; the offset between the disc model and the origin reflects the imperfect \textit{Herschel} pointing, rather than a physical offset between the disc and the star.}
   \label{fig: hd1461Image}
\end{figure*}

Given the fitting results, we conducted a visual inspection to remove models where the fitting was likely confused by background objects. We also omitted models where the image-minus-model flux within ${15 \; \rm arcsec}$ of the image centre had a greater than ${5\percent}$ probability of having a non-Gaussian distribution in an Anderson-Darling test, which could be indicative of a poor fit. The surviving models were deemed reliable. For systems with multiple reliable \textit{Herschel} models, we selected one fit; we chose ${70\um}$ fits over ${100\um}$ owing to better resolution, and if multiple models still remained (i.e. there were multiple observations at the same wavelength), then we chose the fit to the observation with the longest duration. This yielded 35 systems with reliable \textit{Herschel} disc models (in addition to those resolved with ALMA), and the edges of these fitted discs, as well as the wavelengths of the PACS images used, are listed in Table \ref{tab: systemObsData}. An example \textit{Herschel} image and disc model that we use is shown on Fig. \ref{fig: hd1461Image}.

\begin{figure}
  \centering
   \includegraphics[width=8cm]{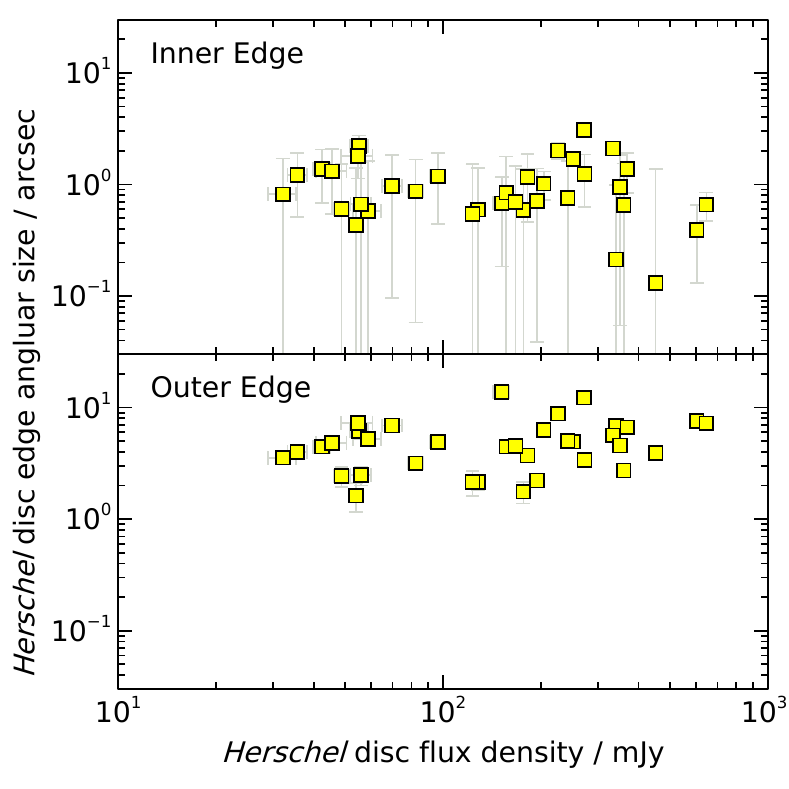}
   \caption{Fitted debris disc parameters for systems where \textit{Herschel} PACS data are used. The discs are modelled as axisymmetric with sharp inner and outer edges, as described in Sect. \ref{subsec: herschelSystems}. The outer edges of these discs appear resolved (bottom plot), but the majority of the inner edges are unresolved (top plot).}
   \label{fig: herschelDiscSizeAndFlux}
\end{figure}

The outer edges of the \textit{Herschel} discs appear resolved and well-constrained (bottom plot of Fig. \ref{fig: herschelDiscSizeAndFlux}), but the majority of the inner edges are unresolved and less reliable (top plot of Fig. \ref{fig: herschelDiscSizeAndFlux}). This will later lead to large uncertainties on planet parameters derived from \textit{Herschel} data. The inner-edge determination is further compounded by the modelling not including an unresolved inner component; were separate inner discs also present, then we may underestimate the locations of the inner edges of the outer discs in the \textit{Herschel} fits. This underestimation would not invalidate our later calculations of the minimum planet masses required to sculpt the inner edges, because the required masses increase with inner edge location (Sect. \ref{sec: discTruncation}); underestimating the inner edges would mean the planets must be larger than the minimum masses we calculate, so these minimum masses still hold. However, underestimating the inner edges could mean that we overestimate the planet masses required to stir \textit{Herschel} discs, as discussed in Sect. \ref{subsec: herschelDiscussion}.

%---------------------------------------------------------
\subsubsection{Systems with only SED data}
\label{subsec: sedSystems}

\noindent The majority of our systems (111 out of 178) are neither resolved by ALMA nor \textit{Herschel}, and for these systems we infer disc properties from SED data alone. These data provide only rough disc locations, with no information on disc widths or degrees of asymmetry. The planet parameters later predicted for individual SED systems may therefore be unreliable; however, we will show that the \textit{distributions} of predicted sculpting planets across all SED systems are probably reasonable, since they are similar to those derived from more accurate ALMA and \textit{Herschel} data. For systems with ALMA or \textit{Herschel} data, we also use SED fits in the self-stirring calculations in Sect. \ref{subsec: selfStirring}.

\begin{figure}
  \centering
   \includegraphics[width=8cm]{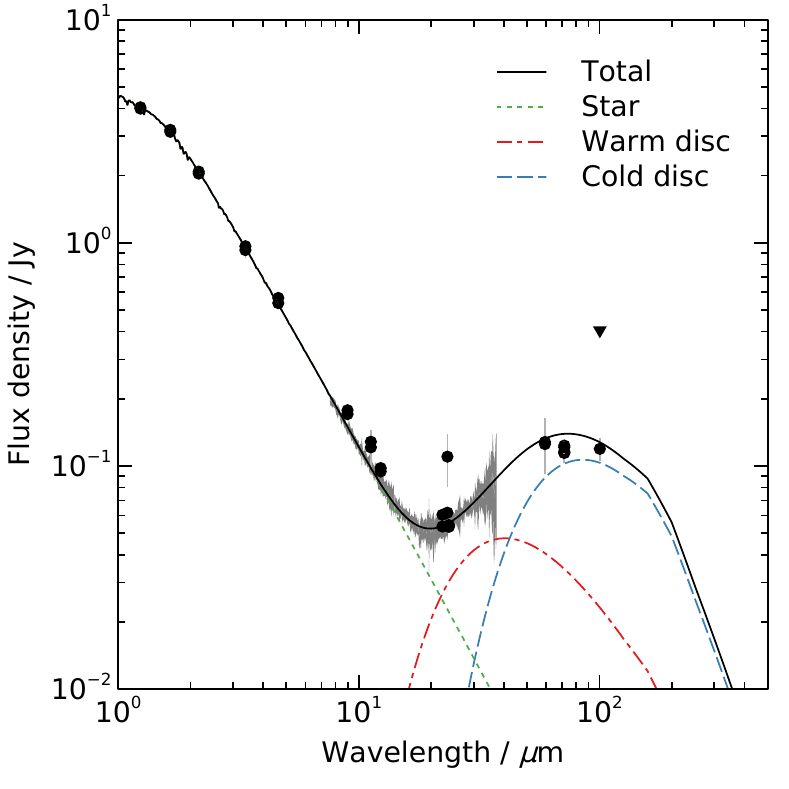}
   \caption{SED data for one example system, ${\HD37484}$ (Sect. \ref{subsec: sedSystems}). Black circles show photometric data, triangles are upper limits, and the grey shaded region the \textit{Spitzer} IRS spectrum. The best-fit model (black line) comprises the stellar flux (green dotted line) plus additional emission from two debris discs: a cold disc with blackbody temperature ${60 \pm 7 \K}$ (blue dashed line), and a warm disc with blackbody temperature ${130 \pm 10 \K}$ (red dash-dot line). This paper considers only the outermost (cold) discs.}
   \label{fig: hd37484SED}
\end{figure}

The SED fitting process is described in \cite{Launhardt2020}, and we refer the reader to that paper for details and data sources (see also \citealt{Yelverton2019}). For each system, the SED is fitted with a stellar model, plus one or two blackbody components representing one or two debris discs. An example is shown on Fig. \ref{fig: hd37484SED}. To estimate the radial location of the disc(s), the temperature of each blackbody component $T_{\rm BB}$ is first converted to a blackbody radius

\begin{equation}
r_{\rm BB} = 1 \au \left(\frac{T_{\rm BB}}{278 \K}\right)^{-2} \left(\frac{L_*}{\rm L_\odot}\right)^{1/2}.
\label{eq: uncorrectedBlackbodyRadius}
\end{equation}

\noindent However, this blackbody radius is often a poor estimate of the `true' disc location, since it fails to account for grain optical properties and blowout size; in line with \cite{Booth2013} and \cite{Pawellek2015}, we therefore use a correction factor $\Gamma$ to better-estimate the disc location as ${r_{\rm SED} \equiv \Gamma r_{\rm BB}}$. The correction factor depends on stellar luminosity as

\begin{equation}
\Gamma = A \left(\frac{L_*}{L_\odot} \right)^B,
\label{eq: gammaFactorGeneral}
\end{equation}

\noindent where \cite{Launhardt2020} used values for $A$ and $B$ derived from resolved \textit{Herschel} data ($A=7.0$, $B=-0.39$; \citealt{Pawellek2017}), with $\Gamma$ limited to 4 for stars with ${L_* < 4 \lSun}$. However, this trend does not appear to hold for disc radii derived from better-resolved ALMA data \citep{Matra2018}, and \cite{Pawellek2021} have since argued that a shallower function is better for discs around ${1-20\lSun}$ stars:

\begin{equation}
\Gamma = \left(2.92 \pm 0.50 \right) \times \left(\frac{L_*}{L_\odot} \right)^{-0.13 \pm 0.07}.
\label{eq: gammaFactorPaw2021}
\end{equation}

\begin{figure}
  \centering
   \includegraphics[width=8cm]{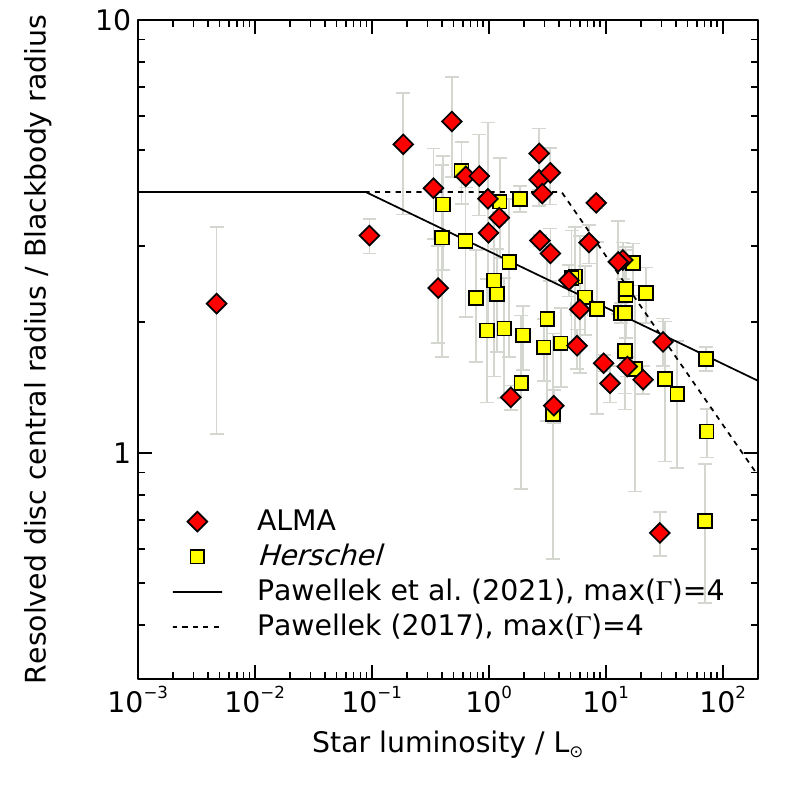}
   \caption{Ratios of resolved debris disc radii to blackbody radii for the systems in our sample, as described in Sect. \ref{subsec: sedSystems}. Red diamonds and yellow squares show systems where resolved disc radii were derived from ALMA and \textit{Herschel} data, respectively. Each system is plotted once, using only the best-available resolved data as described in Sect. \ref{subsec: debrisDiscData}. The lines show $\Gamma$ correction factors; the solid line is derived from ALMA data (where we have limited ${\Gamma}$ to a maximum of 4), and is that used in this paper \citep{Pawellek2021}, and the dashed line is from \textit{Herschel} data \citep{Pawellek2017} as used by \citet{Launhardt2020}. Note that the vertical axis uses the central radius of the resolved disc, defined as halfway between pericentre of the disc inner edge and apocentre of the outer edge. The outlier on the left is the M-type Fomalhaut C (${\rm NLTT \; 54872}$). The two systems in the bottom right of the plot may be complicated by either an underestimated cold disc location due to an additional warm component (${\beta \; \rm UMa}$: ${\HD95418}$, \citealt{Moerchen2010}), or background contamination (${\gamma \rm \; Oph}$: ${\HD161868}$, \citealt{Holland2017}).}
   \label{fig: blackbodyVsResolvedDiscRadii}
\end{figure}

\noindent On Fig. \ref{fig: blackbodyVsResolvedDiscRadii} we plot these two $\Gamma$ relations, and also show the ratios of ALMA- or \textit{Herschel}-derived radii to blackbody radii for the systems in our sample.

The figure shows that discs around less luminous stars tend to be located exterior to where their blackbody temperatures imply, and the discrepancy is smaller for more luminous stars, in agreement with other studies (e.g. \citealt{Booth2013, Pawellek2015, Marshall2021}). Neither $\Gamma$ correction factor provides a strong fit across our entire sample, but since the form in Eq. \ref{eq: gammaFactorPaw2021} \citep{Pawellek2021} is derived from well-resolved ALMA data, we choose to use this $\Gamma$ correction over the \textit{Herschel}-derived factor from \citet{Pawellek2017} (although we again limit $\Gamma$ to a maximum of 4). It should be noted that, for our sample at least, the ALMA and \textit{Herschel} data could actually be in agreement, suggesting a $\Gamma$ steepness in line with \citet{Pawellek2017} but with a smaller offset (e.g. ${\Gamma \sim 4 (L_*/\lSun)^{-0.39}}$); however, we stress that we do not account for the various nuances involved in the estimation of $\Gamma$, and that this is an active area of research, so we take $\Gamma$ from Eq. \ref{eq: gammaFactorPaw2021} rather than that roughly calculated from our sample. We also note that $\Gamma$ is simply a trend used to estimate disc size; 
factors omitted here, such as disc width, may also be important, so we expect a reasonable scatter on Fig. \ref{fig: blackbodyVsResolvedDiscRadii} regardless of the $\Gamma$ prescription used.

For each system without resolved ALMA or \textit{Herschel} data, we estimate the location of the (outermost) disc by taking the blackbody radius and scaling it by $\Gamma$ from Eq. \ref{eq: gammaFactorPaw2021} (where we assert ${\Gamma \leq 4}$ as in \citealt{Launhardt2020}). This gives a rough location for the discs in our 111 SED systems. Since we have no width information, we are forced to set the inner and outer edges to be equal, i.e. ${Q_{\rm i} = q_{\rm i} = Q_{\rm o} = q_{\rm o} = r_{\rm SED}}$, and we discuss the effect of this throughout the paper. Disc inclination is also unconstrained for SED systems.

%%%%%%%%%%%%%%%%%%%%%%%%%%%%%%%%%%%%%%%%%%%%%%%%%%%%%%%%%%%%%%%%%%%%%%%%%%%%%%%%%%%%%%%%%%%%%%%%%%%
\section{Planet constraints from disc sculpting}
\label{sec: discTruncation}

Using the star and debris disc parameters from Sect. \ref{sec: systemSample}, we now use dynamical arguments to infer the properties of unseen planets that could be interacting with our discs. In this section we consider planets that could be sculpting these discs; later, in Sect. \ref{sec: stirring}, we will examine planets required to stir each disc. The discs in our sample are typically located at tens or hundreds of au, and it is generally assumed that planets exist interior to such discs and sculpt their inner edges. In this section we consider the planets required to sculpt the disc inner edges, assuming that the shapes and locations of these edges result from interactions with unseen planets. 

For each system we assume that the sculpting planet(s) orbits just interior to the disc inner edge, in the same plane as the disc, and that the planet orbits have not evolved significantly over the stellar lifetime; this simple model should provide reasonable planet constraints for the great majority of systems, although note that more complicated interactions are also possible, such as those proposed for Fomalhaut \citep{Faramaz2015, Pearce2021}, ${\HD107146}$ \citep{Pearce2015HD107146, Yelverton2018, Sefilian2021}, or the Solar System \citep{Gomes2005}. However, these more complicated interactions often produce system-specific features, so to ensure generality across our sample we only consider basic interactions in this paper. We also again emphasise that this section concerns planets sculpting the \textit{inner edges} of debris discs; for broad discs with radial structure (e.g. ${\HD15115}$, ${\HD92945}$, ${\HD107146}$; \citealt{Macgregor2019, Marino2019, Marino2018}) we are \textit{not} constraining planets that potentially produce gaps in those broad discs, and care must therefore be taken when comparing our results to others to ensure the same planets are being considered.

If the disc inner edges are truncated by planets, then these planets must be sufficiently massive to sculpt the edges within the lifetime of the star. However, the specific planet masses required depend on the system architecture, history, and debris properties and morphology; many different mass estimation methods exist in the literature, each making different assumptions about the nature and degree of debris sculpting, and each predicting different masses for the planet(s) responsible (e.g. \citealt{Wisdom1980, Gladman1990, Faber2007, Mustill2012, Pearce2014, Morrison2015, Nesvold2015, Shannon2016, Lazzoni2018, Regaly2018}). In this analysis we use two different sets of assumptions to calculate the minimum mass of the sculpting planet(s), which should provide reasonable upper and lower bounds on the minimum planet masses required in each system. We first use a modified version of the \cite{Pearce2014} method, which assumes that a single planet has ejected ${95 \percent}$ of unstable debris from near the disc inner edge (Sect. \ref{subsec: singlePlanet}). This method is chosen because it is valid for planets on either circular or eccentric orbits, with the latter favoured in some systems with asymmetric discs. We then employ an alternative method based on \cite{Faber2007} and \cite{Shannon2016}, which assumes that multiple, equal-mass planets on circular orbits have removed ${50 \percent}$ of unstable debris (Sect. \ref{subsec: multiplePlanets}). This second method is used because it yields conservative lower limits on planet masses, which are generally an order-of-magnitude below those predicted by the \cite{Pearce2014} method (since a single, high-mass planet is less efficient at clearing than multiple, lower-mass planets, and it takes longer to remove ${95 \percent}$ of unstable debris than ${50 \percent}$). Taken together, these two methods likely bound the minimum planet masses required to sculpt the disc inner edges, with the actual required masses depending on the specific system architecture and degree of clearing.

\begin{figure}
  \centering
   \includegraphics[width=8cm]{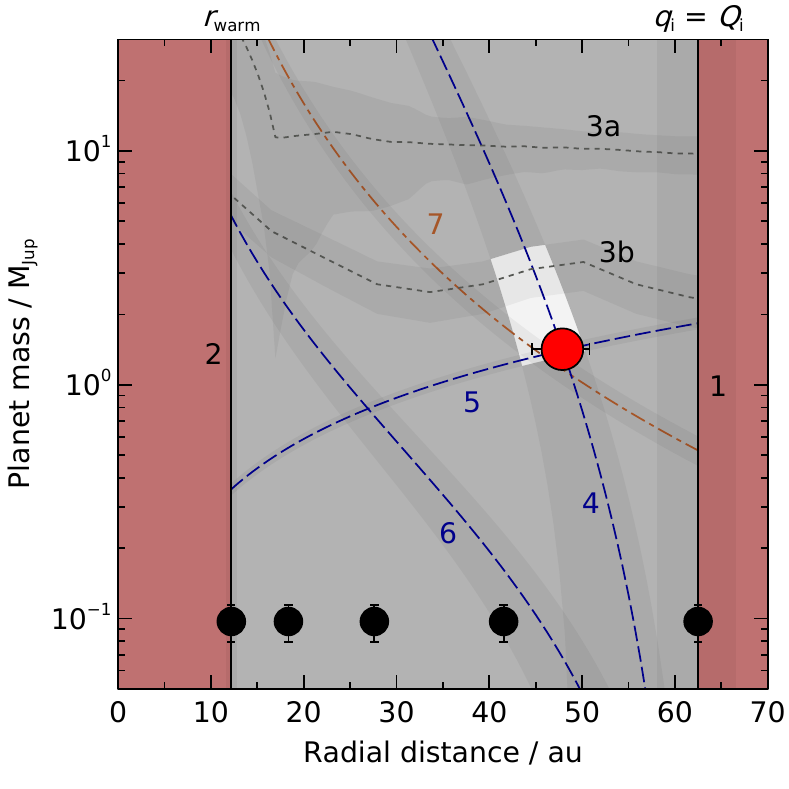}
   \caption{Masses and locations of the smallest planets needed to sculpt or stir the outer debris disc of ${49 \; \rm Cet}$ (${\HD9672}$), according to different models. The system hosts two debris discs; the outer disc has its resolved inner edge at ${62\pm4 \au}$ (line 1), and the unresolved inner disc is located around ${12.2 \pm 0.6 \au}$ (line 2). Lines 3a and 3b show ${5\sigma}$ detection limits; 3a is our NaCo-ISPY limit, and 3b is the SPHERE H2-band limit from \citet{Choquet2017} (with uncertainties derived as in Sect. \ref{subsec: planetDetectionLimits}). The red circle is the minimum-mass planet predicted by the \citet{Pearce2014} model, which assumes that the outer disc is sculpted by a single planet (Sect. \ref{subsec: singlePlanet}). Lines 4, 5 and 6 show constraints from the Hill radius (Eq. \ref{eq: massConstraintFromHillAndInnerEdgeEccentricity}), diffusion time (Eq. \ref{eq: massConstraintFromDiffusion}) and secular time (Eq. \ref{eq: massConstraintFromSecularTime}) respectively, used to predict the properties of this single planet; if the outer disc is sculpted by one planet, then that planet is predicted to lie in the white region. The black circles denote the minimum-mass planets predicted by the alternative \citet{Shannon2016} model, which assumes that the outer disc is sculpted by multiple, equal-mass planets spanning the gap between the inner and outer discs (Eqs. \ref{eq: shannonNumberOfPlanets} and \ref{eq: shannonPltMassK16} in Sect. \ref{subsec: multiplePlanets}). Line 7 is the minimum planet mass required to stir the outer disc (Eq. \ref{eq: pltMassToStir} in Sect. \ref{subsec: planetStirring}); note that this is a separate analysis to the sculpting analysis, because we do not necessarily assume that the disc is stirred by the sculpting planet(s). The ${1\sigma}$ uncertainties are shown as error bars around points and dark grey regions around lines. Similar analyses to those shown here are applied to all systems in our sample. Note that grain dynamics could be complicated by the presence of gas in this system (much less than in a protoplanetary disc, but still potentially enough to couple to dust; \citealt{Moor2019}).}
   \label{fig: hd9672PltCnstrnts}
\end{figure}

Throughout this section we will refer to an example system, ${49 \; \rm Cet}$ (${\HD9672}$), to demonstrate the concepts used, before applying those methods to all systems in our sample. Fig. \ref{fig: hd9672PltCnstrnts} shows the various arguments used to constrain planets around ${49 \; \rm Cet}$, as described below. Note that the grain dynamics in this particular system may be complicated by the presence of gas (much less than in a protoplanetary disc, but still potentially enough to couple to dust; \citealt{Moor2019}). Nonetheless, this system should serve as an adequate example for our purposes.

%---------------------------------------------------------
\subsection{Disc sculpting by a single planet}
\label{subsec: singlePlanet}

We first consider a scenario where one planet sculpts the inner edge of each disc. We calculate the minimum mass of the planet required to do this, as well as its maximum semimajor axis and minimum eccentricity. The {\sc python} program used for this analysis is publicly available for download\footnote{tdpearce.uk/public-code}.

%---------------------------------------------------------
\subsubsection{Disc sculpting by a single planet: method}
\label{subsec: singlePlanetMethod}

A single planet would destabilise non-resonant debris either side of its orbit, and the larger the planet mass, semimajor axis and eccentricity, the wider the unstable region. \cite{Pearce2014} show that non-resonant material orbiting within five eccentric Hill radii of planet apocentre would eventually be ejected. Rearranging their equations 9 and 10 yields the mass a single planet would need to sculpt the inner edge of an external disc, as a function of planet semimajor axis $a_{\rm p}$ and eccentricity $e_{\rm p}$: 

\begin{equation}
M_{\rm p} \approx 8.38 \mJup \left( \frac{M_*}{\mSun}\right) \left[\frac{Q_{\rm i}}{a_{\rm p} (1+e_{\rm p})} -1 \right]^3 (3-e_{\rm p}),
\label{eq: massConstraintFromHill}
\end{equation}

\noindent where $Q_{\rm i}$ is the apocentre of an ellipse tracing the disc inner edge. The apocentre is used because it equals the semimajor axis of the innermost disc particle if the disc eccentricity arises through a secular interaction with an internal planet \citep{Pearce2014}; if the disc is axisymmetric, then $Q_{\rm i}$ is just the disc inner edge radius.

Eq. \ref{eq: massConstraintFromHill} shows that the larger the planet eccentricity, the wider the unstable region. However, in addition to removing unstable debris, a coplanar eccentric planet would also drive surviving debris into an eccentric structure, apsidally aligned with the planet orbit \citep{Wyatt1999, Faramaz2014, Pearce2014}; this means that ${a_{\rm p}}$ and ${e_{\rm p}}$ cannot be varied independently in Eq. \ref{eq: massConstraintFromHill} because an eccentric planet with apocentre close to the disc inner edge would also make the disc eccentric, which may not be compatible with observations. We can use this idea to eliminate ${e_{\rm p}}$ in Eq. \ref{eq: massConstraintFromHill}, whilst retaining the dependence on planet eccentricity. If an internal eccentric planet sculpts the disc then, provided the planet eccentricity is not too high, the disc inner and outer edge dimensions are related to the perturbing planet orbit through

\begin{equation}
a_{\rm p} e_{\rm p} \approx \frac{2}{5} \left(Q_{\rm i} - q_{\rm i} \right) \approx \frac{2}{5} \left(Q_{\rm o} - q_{\rm o} \right),
\label{eq: planetSemimajorAxisEccentricityFromDiscEdges}
\end{equation}

\noindent recalling that ${q_{\rm i}}$ is the pericentre of an ellipse tracing the disc inner edge, and ${q_{\rm o}}$ and ${Q_{\rm o}}$ are the pericentre and apocentre of an ellipse tracing the disc outer edge, respectively (Eq. \ref{eq: planetSemimajorAxisEccentricityFromDiscEdges} is derived from equations 1 and 5-8 in \citealt{Pearce2014}). Note that the disc inner edge would be more elliptical than the outer edge, and for a broad, moderately eccentric disc, Eq. \ref{eq: planetSemimajorAxisEccentricityFromDiscEdges} can be used to assess whether that eccentricity could be driven by a single eccentric planet. Combining Eqs. \ref{eq: massConstraintFromHill} and \ref{eq: planetSemimajorAxisEccentricityFromDiscEdges} yields the general planet mass (as a function of semimajor axis) required to sculpt an external disc, where that disc may or may not be eccentric:

\begin{multline}
M_{\rm p} \approx 8.38 \mJup \left( \frac{M_*}{\mSun}\right) \left[\frac{Q_{\rm i}}{a_{\rm p} + 0.4 (Q_{\rm i} - q_{\rm i})} -1 \right]^3 
\\
\times \left(3-0.4 \frac{Q_{\rm i} - q_{\rm i}}{a_{\rm p}}\right).
\label{eq: massConstraintFromHillAndInnerEdgeEccentricity}
\end{multline}

\noindent This is line 4 on Fig. \ref{fig: hd9672PltCnstrnts}. The equation holds for both circular planets, and those with eccentricities up to moderate values\footnote{\cite{Pearce2014} show that the assumptions behind Eq. \ref{eq: planetSemimajorAxisEccentricityFromDiscEdges} begin to break down for forcing eccentricities above ${\sim 0.4}$, so Eq. \ref{eq: massConstraintFromHillAndInnerEdgeEccentricity} will be less accurate for planet eccentricities greater than ${e_{\rm p} \gtrsim 0.3 \; Q_{\rm i} / a_{\rm p}}$; none of the planet eccentricities we derive will violate this, so Eqs. \ref{eq: planetSemimajorAxisEccentricityFromDiscEdges} and \ref{eq:  massConstraintFromHillAndInnerEdgeEccentricity} are accurate for our systems.}. 

For the resolved discs in our sample, the vast majority do not show significant asymmetries. The deprojected edges of these discs do not deviate significantly from circles, and they have usually been fitted with axisymmetric models. This does not necessarily mean that these discs are axisymmetric; many could be asymmetric, but we lack the resolution and sensitivity necessary to detect this. Nonetheless, for this analysis any discs without significant asymmetries are treated as axisymmetric. In this case a circular planet could sculpt the disc inner edge; terms relating to planet eccentricity in Eqs. \ref{eq: planetSemimajorAxisEccentricityFromDiscEdges} and \ref{eq: massConstraintFromHillAndInnerEdgeEccentricity} disappear (i.e. ${Q_{\rm i} = q_{\rm i}}$ and ${Q_{\rm o} = q_{\rm o}}$). For almost all resolved discs in this analysis, the eccentricity of the sculpting planet is therefore set to zero. Likewise, for unresolved discs we have no information on widths or asymmetries, so for these discs we also assume axisymmetry and non-eccentric sculpting planets (using ${Q_{\rm i} = q_{\rm i} = Q_{\rm o} = q_{\rm o} = r_{\rm SED}}$ in the above equations). However, there are several resolved discs that do display significant eccentricities, for which we can also constrain the eccentricity of the minimum-mass sculpting planet.

Eqs. \ref{eq: planetSemimajorAxisEccentricityFromDiscEdges} and \ref{eq: massConstraintFromHillAndInnerEdgeEccentricity} yield the sculpting planet mass and eccentricity as functions of its semimajor axis. However, this equation alone does not yield a lower bound on the planet mass required to sculpt a disc, which tends to zero as the planet apocentre approaches the disc inner edge. We can, however, combine this equation with a timescale argument to calculate the minimum planet mass required; if a planet sculpts a disc, then it must be massive enough to clear debris within the stellar lifetime. For an eccentric planet there are two timescales involved; the diffusion timescale sets how quickly debris close to the planet is removed via scattering, and the secular timescale determines how quickly more distant debris particles are driven onto eccentric, planet-crossing orbits. \citet{Pearce2014} show that the time taken for a planet to clear ${95 \; \rm per \; cent}$ of unstable debris is ten times the longer of the secular and diffusion timescales at the outer edge of the unstable region. Rearranging their equation 18 yields the planet mass required to clear the unstable region within the stellar lifetime $t_*$ if the process is set by scattering:

\begin{equation}
M_{\rm p} \geq 0.331 \mJup \left( \frac{a_{\rm p}}{\au} \right) \left( \frac{Q_{\rm i}}{\au} \right)^{-1/4} \left( \frac{t_*}{\myr} \right)^{-1/2} \left( \frac{M_*}{\mSun} \right)^{3/4}.
\label{eq: massConstraintFromDiffusion}
\end{equation}

\noindent This is line 5 on Fig. \ref{fig: hd9672PltCnstrnts}. Similarly, if the clearing is limited by secular evolution, then rearranging equation 17 in \citet{Pearce2014} yields the minimum planet mass required to clear the unstable region as 

\begin{multline}
M_{\rm p} \geq 0.0419 \mJup \left( \frac{a_{\rm p}}{\au} \right)^{-1} \left( \frac{Q_{\rm i}}{\au} \right)^{5/2} \left( \frac{t_*}{\myr} \right)^{-1}
\\ \times \left( \frac{M_*}{\mSun} \right)^{1/2} \left[ b^{(1)}_{3/2} \left( \frac{a_{\rm p}}{Q_{\rm i}} \right) \right]^{-1},
\label{eq: massConstraintFromSecularTime}
\end{multline}

\noindent where

\begin{equation}
b^{(j)}_{s}(\alpha) \equiv \frac{1}{\pi} \int^{2\pi}_{0} \frac{\cos(j \psi)}{(1 - 2 \alpha \cos \psi + \alpha^2)^s} {\rm d}\psi
\label{eq: leplaceCoeff}
\end{equation}

\noindent is a Laplace coefficient \citep{Murray1999}. Equation \ref{eq: massConstraintFromSecularTime} is line 6 on Fig. \ref{fig: hd9672PltCnstrnts}. This secular timescale can be important for highly eccentric planets, but is less important for low-eccentricity planets because these do not significantly increase debris eccentricities through secular interactions. Furthermore, for planetary-mass objects the secular timescale is shorter than the diffusion timescale unless the planet semimajor axis is much smaller than the disc inner edge location (equations 17 and 18 in \citealt{Pearce2014}), and in this case the planet is unlikely to sculpt the disc unless the former has very high eccentricity. For the systems considered in this analysis we will find that the required eccentricities of the sculpting planets are low even for asymmetric discs (Sect. \ref{subsec: singlePlanetResults}), and so the clearing times are set by the diffusion timescale rather than the secular timescale for all of our systems.

Eqs. \ref{eq: massConstraintFromDiffusion} and \ref{eq: massConstraintFromSecularTime} put lower bounds on the sculpting planet mass as functions of planet semimajor axis, and Eq. \ref{eq: massConstraintFromHillAndInnerEdgeEccentricity} relates planet mass and semimajor axis. We can therefore find the minimum planet mass required to sculpt a disc by equating Eq. \ref{eq: massConstraintFromHillAndInnerEdgeEccentricity} and the larger of Eqs. \ref{eq: massConstraintFromDiffusion} and \ref{eq: massConstraintFromSecularTime} (for our sample, the larger is always Eq. \ref{eq: massConstraintFromDiffusion}). This is done numerically. This also yields the semimajor axis of the minimum-mass perturbing planet, which is the \textit{maximum} semimajor axis that a perturbing planet could have. These minimum mass and maximum semimajor axis predictions are shown for ${49 \; \rm Cet}$ (${\HD9672}$) as the red circle on Fig. \ref{fig: hd9672PltCnstrnts}. Finally, Eq. \ref{eq: planetSemimajorAxisEccentricityFromDiscEdges} gives the eccentricity of the minimum-mass planet (in cases where significant disc asymmetry is observed), which is the \textit{minimum} possible eccentricity of a perturbing planet; in general, since we model the planet pericentre as being internal to the pericentre of the disc inner edge, Eq. \ref{eq: planetSemimajorAxisEccentricityFromDiscEdges} also shows that the eccentricity of a sculpting planet must be at least

\begin{equation}
e_{\rm p} \gtrsim \frac{4 e_{\rm i}}{5 - e_{\rm i}},
\label{eq: lowerLimitOnPlanetEccentricity}
\end{equation}

\noindent where ${e_{\rm i}}$ is the eccentricity of an ellipse tracing the disc inner edge.

%---------------------------------------------------------
\subsubsection{Disc sculpting by a single planet: results}
\label{subsec: singlePlanetResults}

\begin{figure}
  \centering
   \includegraphics[width=8cm]{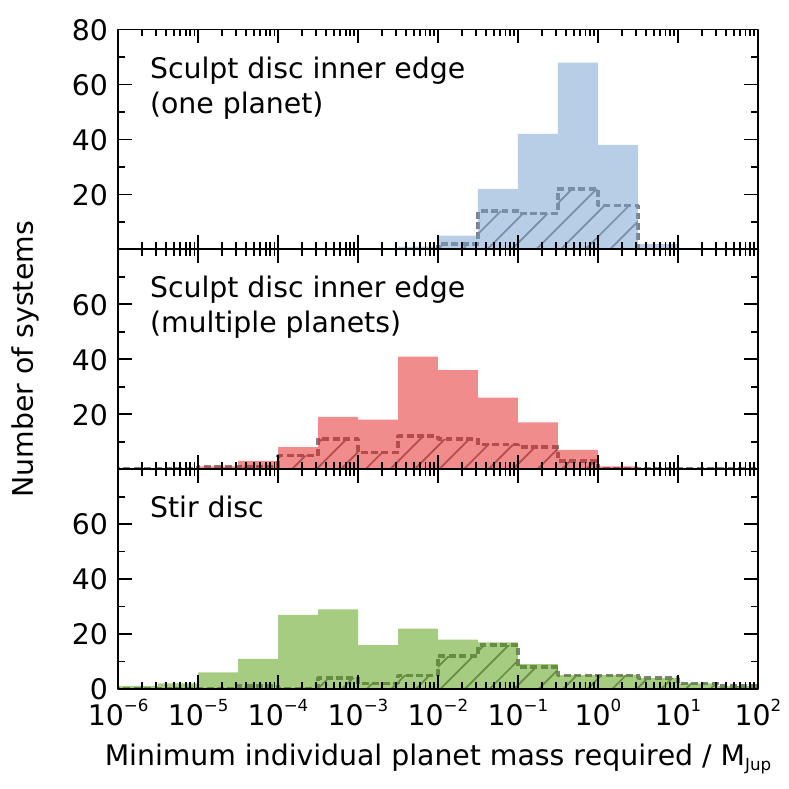}
   \caption{Minimum masses of the planet(s) predicted to exist in our debris disc systems, for three different scenarios: a single planet sculpts the inner edge of the disc (top plot), multiple planets sculpt the inner edge of the disc (middle), or a single planet stirs the disc (bottom). The solid bars show predicted planet parameters in all of our systems, and the hatched bars show only systems where the disc parameters come from resolved images (ALMA or \textit{Herschel}); the results for resolved and unresolved discs are similar for sculpting planets (top two plots), but differ for stirring planets because the results for unresolved discs are conservative lower bounds (Sect. \ref{subsec: planetStirringMethod}). For the multi-planet scenario (middle plot), note that the plot shows the mass of each \textit{individual} planet.}
   \label{fig: planetMassesHistogram}
\end{figure}

We apply the above method to every system in our sample, to constrain the minimum masses of planets required to sculpt the inner edges of our discs (assuming that one planet is responsible in each case), as well as the maximum semimajor axes and minimum eccentricities of those planets. The results for each system are listed in Table \ref{tab: systemAnalysisData}. The predicted minimum masses are shown on the top plot of Fig. \ref{fig: planetMassesHistogram}, with a median of ${0.4 \mJup}$ and first and third quartiles of 0.2 and ${0.8 \mJup}$, respectively. These values were calculated from our whole sample, but the distribution is similar if we instead only consider more reliable constraints from systems where the discs are resolved; for systems with ALMA- or \textit{Herschel}-resolved discs, the minimum sculpting planet masses have a median of ${0.3 \mJup}$ and first and third quartiles of 0.1 and ${0.9 \mJup}$, respectively. This suggests that the `typical' individual planet required to sculpt a debris disc is around half a Jovian mass; such bodies are unlikely to be detectable in the outer regions of systems with current instruments, but could be in the near future (as discussed below).

\begin{figure*}
  \centering
   \includegraphics[width=17cm]{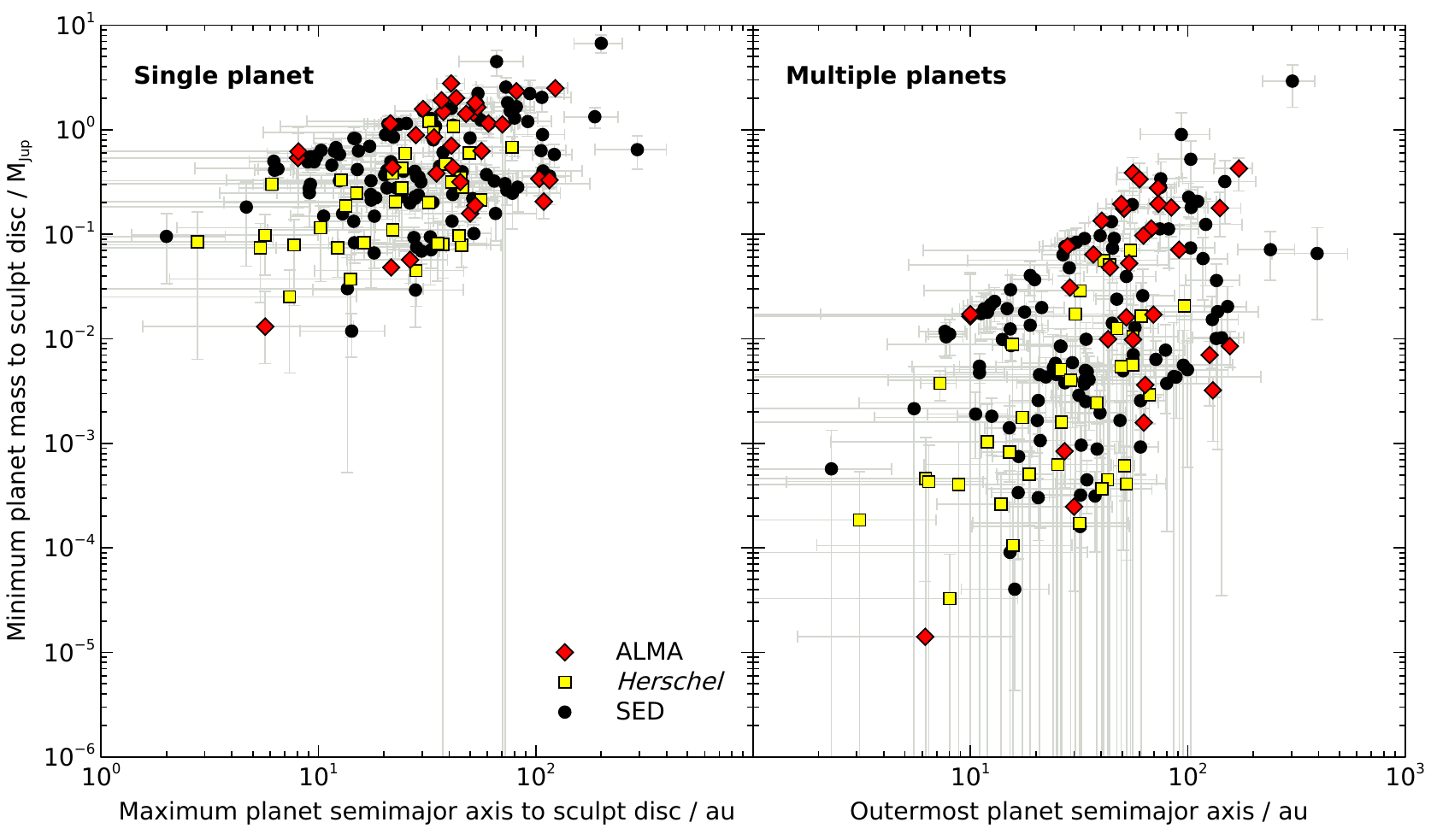}
   \caption{Masses and locations of the smallest planets required to sculpt our debris disc inner edges, assuming the planets lie interior to those discs (Sect. \ref{sec: discTruncation}). Left plot: minimum planet masses required if each disc is sculpted by a single planet \citep{Pearce2014}. More massive planets located further inwards could also sculpt the discs to the same degree. Right plot: minimum individual planet masses required if each disc is instead sculpted by multiple, equal-mass planets \citep{Shannon2016}, where the outermost planet location is fixed at the disc inner edge. The single-planet model (left plot) requires larger planet masses to sculpt discs than the multi-planet model does (right plot), mainly because multiple planets are more efficient at removing debris. Symbols denote the disc data sources, where the best-available source has been used for each system; systems with ALMA data (red diamonds) are expected to yield the most reliable planet constraints, followed by those with \textit{Herschel} data (yellow squares), then finally those with only unresolved SED data (black circles). The larger uncertainties on the multi-planet masses are driven by uncertainties on the disc inner edge locations; the multi-planet model has a stronger dependence on this than the single-planet model does.}
   \label{fig: truncatingPlanetsMassVsSemimajorAxis}
\end{figure*}

The left plot of Fig. \ref{fig: truncatingPlanetsMassVsSemimajorAxis} shows the minimum masses and maximum semimajor axes of the sculpting planets. The two are correlated because the more distant discs require more distant planets to sculpt them, and these sculpting planets would need higher masses because the clearance timescale increases with stellocentric distance (line 5 on Fig. \ref{fig: hd9672PltCnstrnts}). For each system, comparing the predicted semimajor axis of the minimum-mass sculpting planet and the location of the disc inner edge yields a median ratio of ${81 \percent}$ (first and third quartiles: 78 and ${84\percent}$, respectively), which is comparable to Neptune and the Kuiper Belt (although care is required with this comparison due to the complicated history of our Solar System; \citealt{Levison2008}). Note that we have only considered \textit{minimum-mass} planets here; higher-mass planets located further inwards could also sculpt the discs to the same degree (line 4 on Fig. \ref{fig: hd9672PltCnstrnts}). For the few discs with significant asymmetries in thermal emission, the sculpting planets typically require minimum eccentricities of around 0.1. These eccentricities are listed in Table \ref{tab: systemAnalysisData}.

\begin{figure*}
  \centering
   \includegraphics[width=17cm]{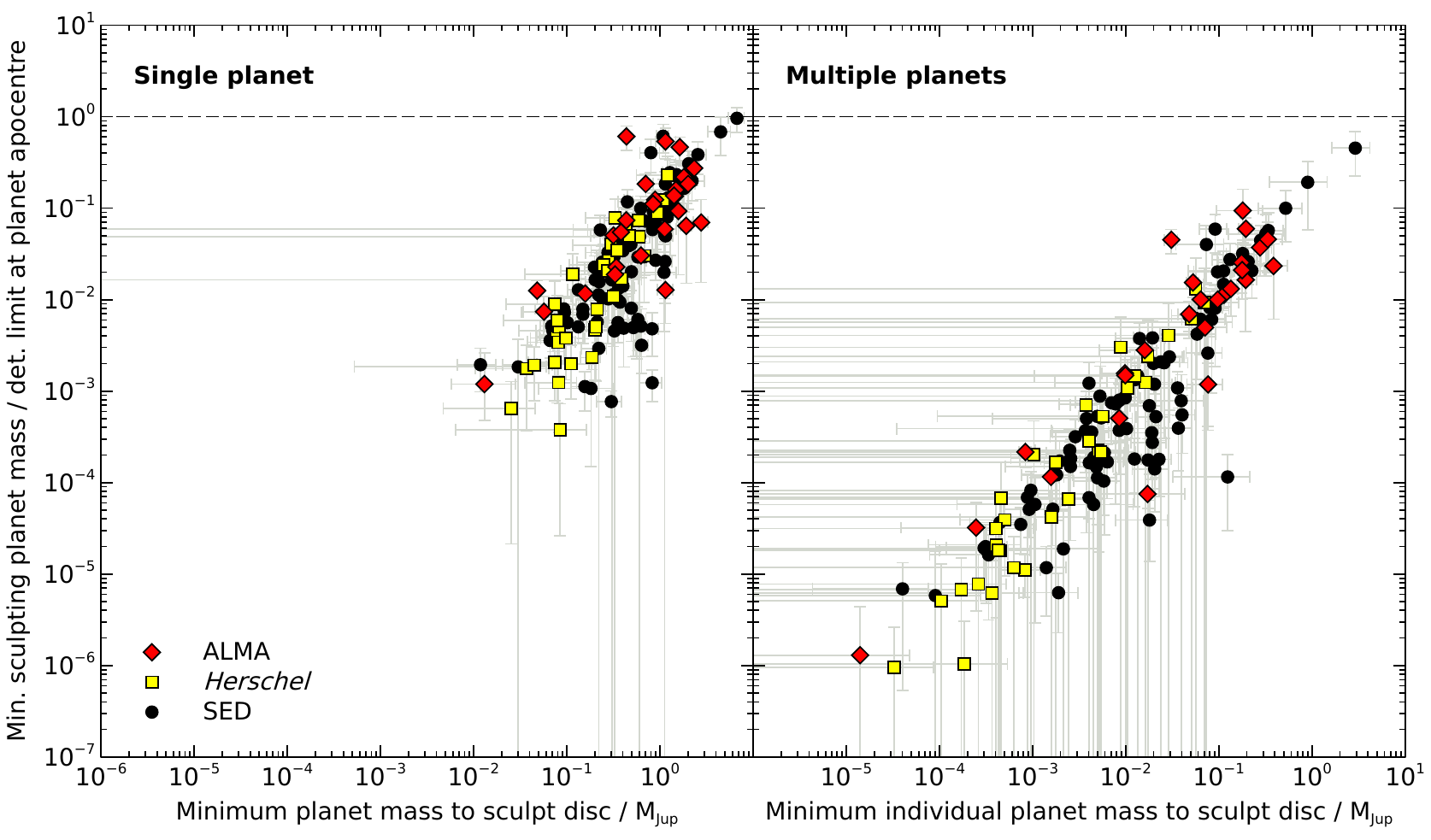}
   \caption{Detectability of the predicted disc-sculpting planets (Sect. \ref{sec: discTruncation}). Horizontal axes: minimum planet masses required to sculpt the discs, if debris clearing is performed by one planet (left plot) or by multiple, equal-mass planets (right plot). Vertical axis: sculpting planet masses divided by our L$'$-band detection limits at the apocentres of those planets (assuming the best-case scenario where the systems are face-on to the observer). The dashed lines show unity. Symbols were defined on previous figures. If the discs are each sculpted by one planet (left plot), then some of these planets would lie at or close to the current detection limits, and a factor of 10 improvement in mass detection limits could yield many more predicted planets. Conversely, if instead each disc is sculpted by multiple planets (right plot), then these planets could be significantly smaller and many might not be observable even with significant improvements in detection limits.}
   \label{fig: truncatingPlanetsObsLimits}
\end{figure*}

Fig. \ref{fig: truncatingPlanetsObsLimits} shows the detectability of our predicted sculpting planets, given our survey sensitivities. The figure shows the minimum sculpting-planet masses, divided by our L$'$-band AMES-COND/DUSTY mass-detection limits at the apocentres of those planets (if we have multiple contrast curves, for example if the target is in more than one of our surveys, then we use the most constraining detection limit at that location). Any planet lying above the horizontal line marking unity could therefore be detectable. The left plot shows our predictions if each disc is sculpted by a single planet (the right plot, for multi-planet sculpting, will be discussed in Sect. \ref{subsec: multiplePlanetsResults}). We see that, to $1\sigma$, \textit{all} predicted planets on the left plot are consistent with being below the detection limits; this means that our discs could each be sculpted by a single, unseen planet. However, some of these planets would be at the limits of detectability; for 23 systems, the predicted planet masses are greater than ${10\percent}$ of our detection limits (to $1\sigma$), including eight systems with reliable ALMA or \textit{Herschel} disc data. These planets would be detectable following a factor 10 improvement in mass-detection limits. The planets in a further 23 systems could be above ${10\percent}$ of our detection limits too, but their $1\sigma$ uncertainties mean they could also be below ${10\percent}$ of these limits. The conclusion is that, if a single planet sculpts each disc, then many such planets could be observable in the near future; we will discuss planet-detection possibilities further in Sect. \ref{subsec: futureObs}. However, also note that the sculpting planets could be less massive than those predicted here if sculpting is not performed by a single planet, but instead by multiple planets; that scenario is considered in the following section.
 
%---------------------------------------------------------
\subsection{Disc sculpting by multiple planets}
\label{subsec: multiplePlanets}

\noindent The previous section considered a scenario where one planet sculpts each debris disc, and identified the minimum-mass planet required in each case. However, multiple planets can clear debris more efficiently than a single planet could; if disc sculpting is instead performed by multiple planets, then these planets could be less massive than those predicted in Sect. \ref{subsec: singlePlanet}.

In this section we aim to identify the smallest planets that could possibly sculpt each disc, by considering a somewhat-idealised scenario where multiple, equal-mass planets remove unstable debris. To this end we also relax our definition of a `sculpted' disc; here we consider the planets required to remove just $50\percent$ of unstable material, rather than ${95\percent}$ as in Sect. \ref{subsec: singlePlanet}. This lowers the required planet masses still further, although the mass reduction is predominantly caused by the use of multiple planets as argued later in this section. The planet masses predicted in this Section are the very minimum possible\footnote{Our two models for the minimum planet masses required for disc truncation could alternatively be classified as one `optimistic' scenario (multiple planets on closely spaced orbits eject only half of the unstable debris; this section), and one `pessimistic' scenario (a single planet ejects ${95\percent}$ of the unstable debris; Sect. \ref{subsec: singlePlanet}).}, and it is likely that the actual sculpting planets have masses more similar to those calculated in Sect. \ref{subsec: singlePlanet}. Nonetheless, calculating such extreme lower bounds is useful; it demonstrates that, even if a single planet predicted to sculpt a disc can be ruled out by observations, the disc could still be sculpted by multiple, lower-mass planets that evade detection.

%---------------------------------------------------------
\subsubsection{Disc sculpting by multiple planets: method}
\label{subsec: multiplePlanetsMethod}

We explore the scenario of \citet{Faber2007} and \citet{Shannon2016}, where multiple, equal-mass planets on circular orbits span a region interior to the disc, and clear that region of debris (the black circles on Fig. \ref{fig: hd9672PltCnstrnts} show such planets for ${49 \; \rm Cet}$ as an example). The planet orbits are assumed to be spaced such that each adjacent pair of orbits are separated by $K$ mutual Hill radii, where $K$ is a constant and the mutual Hill radius is

\begin{equation}
R_{\rm H} =  \frac{a_1 + a_2}{2} \left(\frac{M_1 + M_2}{3 M_*} \right)^{1/3};
\label{eq: mutualHillRadius}
\end{equation}

\noindent here the two planets have masses $M_1$ and $M_2$, and semimajor axes $a_1$ and $a_2$. If each planet has the same mass $M_{\rm p,n}$, and the innermost and outermost planets have orbital radii $r_1$ and $r_2$ respectively, then Eq. \ref{eq: mutualHillRadius} can be used to show that the number of planets between $r_1$ and $r_2$ is

\begin{equation}
n_{\rm p} = 1 + \frac{\log\left(r_2 / r_1\right)}{\log\left[\frac{1 + \eta(M_{\rm p,n})}{1 - \eta(M_{\rm p,n})}\right]},
\label{eq: shannonNumberOfPlanets}
\end{equation}

\noindent where

\begin{equation}
\eta(M_{\rm p,n}) \equiv 0.0430 K \left(\frac{M_{\rm p,n}}{\mJup} \right)^{1/3} \left(\frac{M_*}{\mSun} \right)^{-1/3}.
\label{eq: shannonNumberOfPlanetsEta}
\end{equation}

\noindent Eqs. \ref{eq: shannonNumberOfPlanets} and \ref{eq: shannonNumberOfPlanetsEta} are equivalent to equation 5 in \citet{Shannon2016}, only ours are written in terms of Jupiter masses, and with $K$ (the number of mutual Hill radii between each pair of adjacent planet orbits) explicitly stated. Note that the logarithms in Eq. \ref{eq: shannonNumberOfPlanets} can have any matching base.

Eqs. \ref{eq: shannonNumberOfPlanets} and \ref{eq: shannonNumberOfPlanetsEta} yield the number of planets of a given mass that can fit between radii $r_1$ and $r_2$, but we still need to establish how massive those planets would have to be to clear that region of debris. As in Sect. \ref{subsec: singlePlanet}, the planets must be massive enough to clear unstable debris within the stellar lifetime. \citet{Shannon2016} ran numerical simulations of this setup, and found that the time taken to clear ${50 \rm \; per \; cent}$ of debris depends on planet mass, star mass, outer radius $r_2$ and orbit spacing constant $K$:

\begin{equation}
t_{\rm clear} \approx \tau(K) \left( \frac{r_2}{\au} \right)^{3/2} \left( \frac{M_{\rm p,n}}{\mEarth} \right)^{-1} \left(\frac{M_*}{\mSun} \right)^{1/2},
\label{eq: shannonClearingTime}
\end{equation}

\noindent where ${\tau(K) \approx 4 \myr}$ if $K = 20$, or ${2 \myr}$ if $K = 16$. For this analysis we assume that each adjacent planet pair is separated by $K = 16$ mutual Hill radii, for several reasons; first, Eq. \ref{eq: shannonNumberOfPlanets} fails if ${K^3 \geq 12 M_* / M_{\rm p,n}}$, and we found that some of our systems encounter this problem if ${K=20}$, but not if ${K=16}$. Second, we are interested in finding the smallest planets that could possibly sculpt each disc, and taking ${K=16}$ rather than 20 reduces the planet masses required by a factor of two\footnote{ Using ${K=16}$ rather than 20 can also be physically justified. First, \cite{Weiss2018} show that Kepler systems with four or more planets have median ${K\sim 16}$, and we typically expect ${\sim 4+}$ planets in our multi-planet model (for our systems with both warm and cold discs, if equal-mass planets span the gaps then \mbox{Eqs. \ref{eq: shannonNumberOfPlanets}-\ref{eq: shannonClearingTime}} suggest a median of five planets occupy each gap if ${K=16}$, or four planets if ${K=20}$). Second, whilst we know of many compact (${\lesssim 1\au}$) multi-planet systems from Kepler, our paper considers systems on ${10-100\au}$ scales for which interplanetary spacings may differ; prominent examples include the Solar System and ${\HR8799}$ (both with ${K<16}$), so ${K=16}$ may be more appropriate for our analyses than ${K=20}$.}. We therefore estimate the minimum mass of a planet required to clear debris in a multi-planet system as

\begin{equation}
M_{\rm p,n} \geq 6.29 \times 10^{-3} \mJup \left(\frac{t_*}{\myr} \right)^{-1} \left(\frac{r_2}{\au} \right)^{3/2} \left(\frac{M_*}{\mSun} \right)^{1/2}.
\label{eq: shannonPltMassK16}
\end{equation}

\noindent This is equivalent to equation 4 in \cite{Shannon2016}, only ours is written in terms of Jupiter masses, and we use ${K=16}$ rather than 20.

Eq. \ref{eq: shannonPltMassK16} shows that the minimum planet mass required to sculpt a disc depends on the orbital radius of the outermost planet, which \cite{Shannon2016} argue is roughly the inner edge of the disc. Whilst such a planet would also clear debris beyond its orbit, and would actually be located interior to the disc edge as in Sect. \ref{subsec: singlePlanet}, for this analysis we follow \cite{Shannon2016} and assume that the outermost planet resides at the inner edge of our disc. The effect of this is that clearing may occur slightly faster in reality than assumed here, but the discrepancy is expected to be small. In the few cases where the disc is asymmetric, we approximate the outermost planet orbit to be circular with radius equal to the pericentre of the disc inner edge, because this provides a lower-limit for the required planet mass; whilst the assumption of circular planet orbits may not hold for asymmetric discs, the low degrees of asymmetry in our discs make this a reasonable approximation. We therefore use ${r_2 = q_{\rm i}}$.

\cite{Shannon2016} designed this model for two-disc systems, where planets clear a gap between a warm and a cold disc. They therefore argue that $r_1$ in Eq. \ref{eq: shannonNumberOfPlanets} can be set to the location of the inner disc (${r_1 = r_{\rm warm}}$), allowing the calculation of the number of planets in the gap. However, this number should be treated with caution, because it depends very strongly on model assumptions (equal-mass planets extending all the way in to the warm disc). Owing to this, plus the fact that the majority of our systems host only one detected disc, we choose not to calculate the total \textit{number} of planets expected from the \cite{Shannon2016} multi-planet model. Despite this, the minimum \textit{masses} of these planets are expected to be robust; the clearing timescale is set only by the outermost planet, so we can still calculate the masses required if multiple, equal-mass planets separated by ${K=16}$ mutual Hill radii sculpt the disc, without knowing the actual number of such planets (provided it is greater than one). We calculate these masses for each system in the following section (note that, for illustrative purposes only, we did calculate the number of planets between the inner and outer disc of ${49 \; \rm Cet}$ on Fig. \ref{fig: hd9672PltCnstrnts}, using ${r_{\rm warm}=12.2 \pm 0.6\au}$).

%---------------------------------------------------------
\subsubsection{Disc sculpting by multiple planets: results}
\label{subsec: multiplePlanetsResults}

We now apply the above method to every system in our sample, to constrain the minimum masses of planets required to sculpt the inner edges of discs in the most conservative scenario (sculpting by multiple, equal-mass planets on circular orbits). The results for each system are listed in Table \ref{tab: systemAnalysisData}. The minimum masses of the individual planets in the multi-planet scenario are shown on the middle plot of Fig. \ref{fig: planetMassesHistogram}, with a median of ${0.01 \mJup}$ (${3 \mEarth}$) and first and third quartiles of ${2 \times 10^{-3} \mJup}$ (${0.8 \mEarth}$) and ${0.05 \mJup}$ (${16 \mEarth}$), respectively. As in the single-planet case, the distribution is similar if instead we only consider the more reliable constraints from systems where the discs are resolved; for those systems the minimum masses have a median of ${9 \times 10^{-3} \mJup}$ (${3 \mEarth}$) with first and third quartiles of ${8 \times 10^{-4} \mJup}$ (${0.3 \mEarth}$) and ${0.05 \mJup}$ (${17 \mEarth}$), respectively. 

The planets required to sculpt each disc in the multi-planet scenario are on average 40 times less massive than a single planet would have to be to sculpt the same disc (first and third quartiles: 20 and 80 times less massive). This difference is predominantly caused by multiple planets being more efficient at clearing debris than a single planet is; the requirement that only ${50\percent}$ of unstable debris is removed in the multi-planet model (rather than ${95\percent}$ in the single-planet model) cannot account for the difference, since \cite{Yabushita1980} show that the number of particles surviving repeated close encounters with a single planet decreases over time as $t^{-1}$ or $t^{-2}$, and propagating this through Eq. \ref{eq: massConstraintFromDiffusion} shows that the required mass of a single planet would only decrease by a factor of ${2-3}$ due to this effect. Similarly, Eq. \ref{eq: shannonClearingTime} shows that the multi-planet masses required for sculpting depend strongly on the inter-planetary spacing; a modest reduction in the number of mutual Hill radii between planets from ${K=20}$ to 16 lowers the required planet masses by a factor of 2, implying that the presence of multiple planets strongly influences the interaction timescales. Nonetheless, the multi-planet model represents a somewhat-idealised scenario, so the minimum planet masses found in this section are conservative lower bounds. In reality, it is likely that the smallest planets capable of sculpting a `typical' cold debris disc lie between the super-Earths predicted in this section and the Jupiters predicted by the single-planet model (Sect. \ref{subsec: singlePlanet}); the former may remain undetectable in the outer regions of systems for the near future, but the latter could be observable soon.

The multi-planet model assumes that the outermost planet is located at the inner edge of the disc, and we show the predicted semimajor axes of these planets on the right plot of Fig. \ref{fig: truncatingPlanetsMassVsSemimajorAxis}; as in the single-planet case, the planet masses required to sculpt a disc increase with the disc inner edge radius. Given our L$'$-band detection limits at the inner edges, the right plot of Fig. \ref{fig: truncatingPlanetsObsLimits} shows the detectability of these planets; the minimum planet masses required in the multi-planet-sculpting scenario all lie below our detection limits, so all of our discs could be sculpted by multiple unseen planets. In fact, for the majority of systems such planets could lie very far below current detection limits; if multiple planets sculpt these discs, then only one system (${\HD126062}$) requires the outermost planet mass to be greater than ${10 \percent}$ of our detection limit (to ${1\sigma}$), and since no ALMA or \textit{Herschel} data are available for this system these specific constraints are not particularly reliable. A further three systems also have minimum outer-planet masses compatible with ${10 \percent}$ of the detection limits (to ${1\sigma}$), although their uncertainties mean they could also be below ${10 \percent}$ of these limits. Of these, just one (${\rm CPD \shorthyphen 72 \; 2713}$) has resolved disc data. Therefore, in a worst-case scenario, even a tenfold improvement in mass detection limits would not necessarily allow the observation of significantly more sculpting planets. However, it must again be recalled that the multi-planet model produces a conservative lower bound on planet mass, and the (less idealised) single planet model suggests that a tenfold improvement in mass detection limits could yield a substantial number of planets.

To summarise, in Sect. \ref{sec: discTruncation} we considered the planets required to sculpt the inner edges of our debris discs. We showed that, if single planets sculpt each disc, then many of those planets could be detectable with upcoming instrumentation. Conversely, if each disc is instead sculpted by multiple planets, then it is possible for those planets to evade detection even following a tenfold improvement in mass detection limits. Future planet searches should bear this in mind; even if observations can rule out an individual planet required to sculpt a disc, that does not mean planet sculpting is not occurring, since that disc could still be sculpted by multiple, lower-mass planets lying below the detection limits. However, this latter scenario is a somewhat-idealised case, where all planets have equal, minimal masses and are optimally spaced to facilitate clearing. Even if multiple planets are present, then (by analogy with Jupiter and Neptune in the Solar System) we suspect that clearing would often still be dominated by a single planet, and those planets would be more akin to those predicted in Sect. \ref{subsec: singlePlanet} (Neptune- to Jupiter-mass planets, as opposed to the super-Earths predicted in this section).

%%%%%%%%%%%%%%%%%%%%%%%%%%%%%%%%%%%%%%%%%%%%%%%%%%%%%%%%%%%%%%%%%%%%%%%%%%%%%%%%%%%%%%%%%%%%%%%%%%%
\section{Planet constraints from debris stirring}
\label{sec: stirring}

The previous section discussed the planets required to sculpt the inner edges of debris discs. We now use a different physical argument to constrain unseen planets: the requirement that the discs have been somehow stirred, such that collisions between large bodies are sufficiently common and violent to produce the dust that we see. We first show that disc self-gravity alone is insufficient to bring about these collisions in a number of systems, unless the disc masses were infeasibly high (Sect. \ref{subsec: selfStirring}). This could imply that planets perform the stirring instead, and in Sect. \ref{subsec: planetStirring} we show that unseen planets would indeed be capable of stirring our discs.

%---------------------------------------------------------
\subsection{Self-stirring by debris discs}
\label{subsec: selfStirring}

For a debris disc to be self-stirred, its mass must be sufficient to drive debris onto crossing orbits and induce destructive collisions at its outer edge within the system lifetime \citep{Kenyon2001, Kenyon2008, Kenyon2010, Kennedy2010, Krivov2018}. In this model the largest bodies in the disc, those with radii $s_{\rm max}$, perturb and induce destructive collisions between smaller bodies. However, the size of these largest bodies is unclear. We conduct a new analysis, combining the results of  \cite{Krivov2018} and \cite{Krivov2021}, to find the minimum-possible $s_{\rm max}$ required to stir a disc; in turn, this provides a self-consistent lower bound on the minimum disc mass required for self-stirring. The {\sc python} program used for this analysis is publicly available for download\footnote{tdpearce.uk/public-code}.

%---------------------------------------------------------
\subsubsection{Self-stirring by debris discs: method}
\label{subsec: selfStirringMethod}

\noindent Rearranging equation 28 of \cite{Krivov2018} yields a lower limit on the disc mass required to initiate self-stirring within the system lifetime:

\begin{multline}
M_{\rm disc, stir} \gtrsim 1.62 \times 10^{-3} \; \mEarth \left(\frac{t_*}{\myr}\right)^{-1} \left(\frac{M_*}{\mSun}\right)^{-1/2}
\\
\times \left(\frac{a_2-a_1}{\au}\right) \left(\frac{a_1+a_2}{\au}\right)^{5/2}
\\
\times \gamma^{-1} \left(\frac{\rho_{\rm s}}{1 \gPerCmCubed}\right)^{-1} \left[\frac{v_{\rm frag}(s_{\rm weak})}{\mPerS}\right]^{4} \left(\frac{s_{\rm max}}{\km}\right)^{-3}.
\label{eq: selfStirringMinimumMass}
\end{multline}

\noindent Here $a_1$ and $a_2$ are the innermost and outermost semimajor axes of the debris disc particles respectively, $\gamma$ parametrises the eccentricity of the stirring bodies (${1 \leq \gamma \leq 2}$), $\rho_{\rm s}$ is the planetesimal material density, and $v_{\rm frag}$ the collision speed required to fragment the \textit{weakest} colliding body (which has radius $s_{\rm weak}$). Eq. \ref{eq: selfStirringMinimumMass} strongly depends on $v_{\rm frag}$ and $s_{\rm max}$; \cite{Krivov2018} used ${30 \mPerS}$ and ${200 \km}$ respectively as examples, but the actual values are unclear. We now reconsider these values, in order to place self-consistent lower bounds on the minimum disc masses required for self-stirring.

To proceed, we must first consider the mass of a debris disc. Defining $n(s)$ as the number of particles of radius $s$ in the range ${s\rightarrow s+{\rm d}}s$, the debris disc size distribution is often modelled as a single powerlaw  ${n(s) \propto s^{-q}}$, with ${q\approx3.5}$ expected from destructive collisions \citep{Dohnanyi1969}. However, \cite{Krivov2021} argue that debris is unlikely to follow a single power law all the way from dust up to the largest planetesimals, and that a 3-slope model is more realistic; in this case there exist transitional sizes $s_{\rm mm}$ and $s_{\rm km}$, where the size distribution slope changes due to different physics affecting debris. For grains smaller than ${s_{\rm mm} \sim 1 {\; \rm mm}}$, radiation effects are important and the size distribution is defined to be ${n(s) \propto s^{-q}}$. Larger bodies are unaffected by radiation forces, so debris larger than $s_{\rm mm}$ but smaller than the largest colliding body $s_{\rm km}$ have their size distribution set by destructive collisions, with ${n(s) \propto s^{-q_{\rm med}}}$. Finally, bodies larger than $s_{\rm km}$ are not yet involved in this collisional cascade; it takes time to stir large bodies such that they undergo destructive collisions, so bodies larger than (time-dependent) $s_{\rm km}$ are primordial and not yet colliding. Hence the size distribution between the largest fragmenting body $s_{\rm km}$ and the largest body $s_{\rm max}$ goes as ${n(s) \propto s^{-q_{\rm big}}}$. As in \cite{Krivov2021}, we fix ${q=3.5}$, ${q_{\rm med} = 3.7}$, and ${q_{\rm big} = 2.8 \pm 0.1}$, noting that these values describe indices and should not be confused with the disc pericentre locations ${q_{\rm i}}$ and ${q_{\rm o}}$. Integrating over the size distribution yields the total disc mass in the 3-slope model as

\begin{equation}
M_{\rm disc} = M_{\rm dust} \frac{4-q}{4-q_{\rm big}} \left( \frac{s_{\rm km}}{s_{\rm mm}} \right)^{4-q_{\rm med}} \left( \frac{s_{\rm max}}{s_{\rm km}}\right)^{4-q_{\rm big}},
\label{eq: krivovWyattDiscMass}
\end{equation}

\noindent where ${M_{\rm dust}}$ is the mass in grains smaller than $s_{\rm mm}$. This can be derived from observations:

\begin{equation}
M_{\rm dust} \approx \frac{F_\nu d^2}{\kappa(\lambda) B_\nu(T_{\rm BB})},
\label{eq: dustMass}
\end{equation}

\noindent where $F_\nu$ is the disc specific flux per unit frequency range at observation wavelength $\lambda$, ${\kappa(\lambda)\equiv1.7 \; \rm cm^2\;g^{-1}} (850\um / \lambda)$ is the assumed dust opacity (noting that this involves various factors and is subject to uncertainty, e.g. \citealt{Krivov2021}), and $B_\nu(T_{\rm BB})$ is the blackbody emission intensity at $\nu$ for dust with temperature $T_{\rm BB}$, where $T_{\rm BB}$ is found from the SED disc fits (Sect. \ref{subsec: sedSystems}).

The total disc mass (Eq. \ref{eq: krivovWyattDiscMass}) therefore depends on the unknown sizes of both the largest body ($s_{\rm max}$) and the largest colliding body ($s_{\rm km}$). \cite{Krivov2021} show that $s_{\rm km}$ is set by $s_{\rm max}$; combining their equations 17 and 19 yields

\begin{multline}
\left(\frac{s_{\rm km}}{\rm km}\right)^{q_{\rm big} - q_{\rm med}} = 2.66 \times 10^{-3} \left(\frac{t_*}{\myr}\right)^{-0.59}
\\
\times \left[\frac{M_{\rm disc}'(s_{\rm max})}{\mEarth}\right]^{-0.59} \left(\frac{M_*}{\mSun}\right)^{-0.79} \left(\frac{a_1 + a_2}{\au}\right)^{2.56},
\label{eq: krivovWyattSKm}
\end{multline}

\noindent where $M_{\rm disc}'(s_{\rm max})$ is the mass that would arise from Eq. \ref{eq: krivovWyattDiscMass} if $s_{\rm km}$ were set to ${1\km}$. So if we knew the size of the largest body $s_{\rm max}$, we could uniquely calculate the corresponding disc mass from Equation \ref{eq: krivovWyattDiscMass}.

Let us now reconsider the disc mass required for self-stirring (Eq. \ref{eq: selfStirringMinimumMass}), which depends strongly on $s_{\rm max}$ and the fragmentation speed of the weakest colliding body, $v_{\rm frag}(s_{\rm weak})$. These are unknown, but by assuming the aforementioned size distribution from \cite{Krivov2021}, we can calculate $v_{\rm frag}(s_{\rm weak})$ as a function of $s_{\rm max}$ and therefore express Eq. \ref{eq: selfStirringMinimumMass} as a function of a single unknown, $s_{\rm max}$. Equation 1 in \cite{Krivov2018QDStar} gives the material strength $Q_{\rm D}^*$ of a colliding particle as

\begin{multline}
Q_{\rm D}^*(s, v_{\rm col}) = 9.13 \; {\rm J \; kg^{-1}} \left(\frac{v_{\rm col}}{\rm m \; s^{-1}}\right)^{1/2} 
\\
\times \left[ \left(\frac{s}{\rm m}\right)^{-0.36} + \left(\frac{s}{\km}\right)^{1.4} \right],
\label{eq: QDStar}
\end{multline}

\noindent which depends on particle size $s$ and collision speed ${v_{\rm col}}$, approximating the material as basalt. This is a v-shaped function of particle size, where the weakest-possible colliding particle would have radius ${110\m}$. Since the largest colliding particle has radius ${s_{\rm km}}$, the radius of the weakest colliding particle $s_{\rm weak}$ will therefore be the smaller of $s_{\rm km}$ and ${110\m}$. We can calculate the fragmentation speed of this particle; since destructive collisions would occur between equal-sized colliders at the onset of the collisional cascade, we can take ${v_{\rm frag}(s) = \sqrt{8 Q_{\rm D}^*[s, v_{\rm frag}(s)]}}$ from \cite{Krivov2005}, which yields

\begin{multline}
v_{\rm frag}(s_{\rm weak}) = 17.5 \mPerS \left[ \left(\frac{s_{\rm weak}}{\rm m}\right)^{-0.36} + \left(\frac{s_{\rm weak}}{\km}\right)^{1.4} \right]^{2/3}.
\label{eq: vFrag}
\end{multline}

\noindent So for a given $s_{\rm max}$, we can determine the size of the largest colliding particle ${s_{\rm km}}$, which then yields the size of the weakest colliding particle $s_{\rm weak}$, which in turn yields the fragmentation speed ${v_{\rm frag}(s_{\rm weak})}$ of that particle\footnote{Note that, for our $Q_{\rm D}^*$ prescription, we may actually expect a 4-slope size distribution with another transition at ${s=110\m}$ (e.g. \citealt{OBrien2003, Lohne2008}); however, we will find that our derived ${s_{\rm km}}$ values are typically not far below ${110\m}$, so omitting this additional transition is unlikely to significantly affect our results.}. 

Combining the above arguments, for a given $s_{\rm max}$ we can uniquely solve the disc mass (Eq. \ref{eq: krivovWyattDiscMass}) \textit{and} determine whether that mass is sufficient to stir the disc (Eq. \ref{eq: selfStirringMinimumMass}). Furthermore, inspection of Eqs. \ref{eq: selfStirringMinimumMass}-\ref{eq: vFrag} show that the disc mass $M_{\rm disc}$ monotonically increases with  $s_{\rm max}$, while the mass required for stirring, $M_{\rm disc, stir}$, monotonically decreases with  $s_{\rm max}$. This means that there will be a single critical value of $s_{\rm max}$ where the disc mass is \textit{just} sufficient for stirring; for any higher $s_{\rm max}$ the disc can be self-stirred, and for lower $s_{\rm max}$ it cannot. In turn, this critical $s_{\rm max}$ corresponds to the minimum mass that a disc requires to self-stir.

Using Eqs. \ref{eq: selfStirringMinimumMass}-\ref{eq: vFrag}, we can therefore compute the minimum masses that our discs would require to be self-stirred. We do this by equating Eqs. \ref{eq: selfStirringMinimumMass} and \ref{eq: krivovWyattDiscMass}, and numerically solving the result for the critical $s_{\rm max}$ required for self stirring (recalling that ${s_{\rm km}}$ and ${v_{\rm frag}(s_{\rm weak})}$ are also functions of $s_{\rm max}$). Note that we are considering the \textit{most-optimistic} case here, where the largest planetesimals are only just massive enough to stir the disc; even so, we will show that some of these required masses could still be unfeasibly large. Such discs would need to be stirred by other mechanisms, such as by unseen planets (Sect. \ref{subsec: planetStirring}).

For axisymmetric discs we set $a_1$ and $a_2$ in Eq. \ref{eq: selfStirringMinimumMass} to the inner and outer disc edges, respectively. If the disc is eccentric then we assume that the eccentricity is caused by a secular interaction with an eccentric planet; in this case the innermost and outermost debris semimajor axes will equal the apocentre of the disc inner edge and the pericentre of the disc outer edge, respectively (equations 5 and 7 in \citealt{Pearce2014}). We therefore generally set ${a_1 = Q_{\rm i}}$ and ${a_2 = q_{\rm o}}$ in Eq. \ref{eq: selfStirringMinimumMass}. Also, as in \cite{MussoBarcucci2021}, we set $\gamma \equiv 1.5$ and use ${\rho_{\rm s} = 1 \gPerCmCubed}$ \citep{Krivov2018}. Finally, since the required disc mass is proportional to the disc width (${a_2-a_1}$ in Eq. \ref{eq: selfStirringMinimumMass}), and no width information is available from disc SEDs, we omit systems with only SED data from this analysis.

%---------------------------------------------------------
\subsubsection{Self-stirring by debris discs: results}
\label{subsec: selfStirringResults}

Across our resolved discs, we find that the median $s_{\rm max}$ required for stirring is ${14 \km}$ (first and third quartiles: 12 and ${18\km}$). This means that, for a typical disc to be self-stirred, it must contain bodies larger than ${14 \km}$; for these values of $s_{\rm max}$, the largest colliding bodies would have median size ${s_{\rm km} = 55 \m}$ (first and third quartiles: 31 and ${94\m}$), and the median fragmentation speeds of the weakest colliding bodies ${v_{\rm frag}(s_{\rm weak}) = 7.0 \mPerS}$ (first and third quartiles: 6.6 and ${7.8\mPerS}$). Note that these minimum $s_{\rm max}$ and ${v_{\rm frag}(s_{\rm weak})}$ values are much smaller than the values of ${200 \km}$ and ${30 \mPerS}$ respectively used by \cite{Krivov2018}, and the latter values are not necessarily consistent with one another; for example, if the largest planetesimal in the disc of ${49 \; \rm Cet}$ (${\HD9672}$) really had ${s_{\rm max} = 200\km}$, then the fragmentation velocity of the weakest colliding body ${v_{\rm frag}(s_{\rm weak})}$ would be just ${7\mPerS}$. This difference in ${v_{\rm frag}(s_{\rm weak})}$ is very important due to the ${v_{\rm frag}^4}$ term in Eq. \ref{eq: selfStirringMinimumMass}. In general, we therefore argue that equation 28 in \cite{Krivov2018} (Eq. \ref{eq: selfStirringMinimumMass} here) should not be used with `standard' values of ${s_{\rm max} = 200\km}$ and ${v_{\rm frag}(s_{\rm weak}) = 30\mPerS}$ to calculate the minimum disc mass required to self-stir, but that these values should instead be calculated on a system-specific basis using a method akin to that in this paper.

\begin{figure*}
  \centering
   \includegraphics[width=17cm]{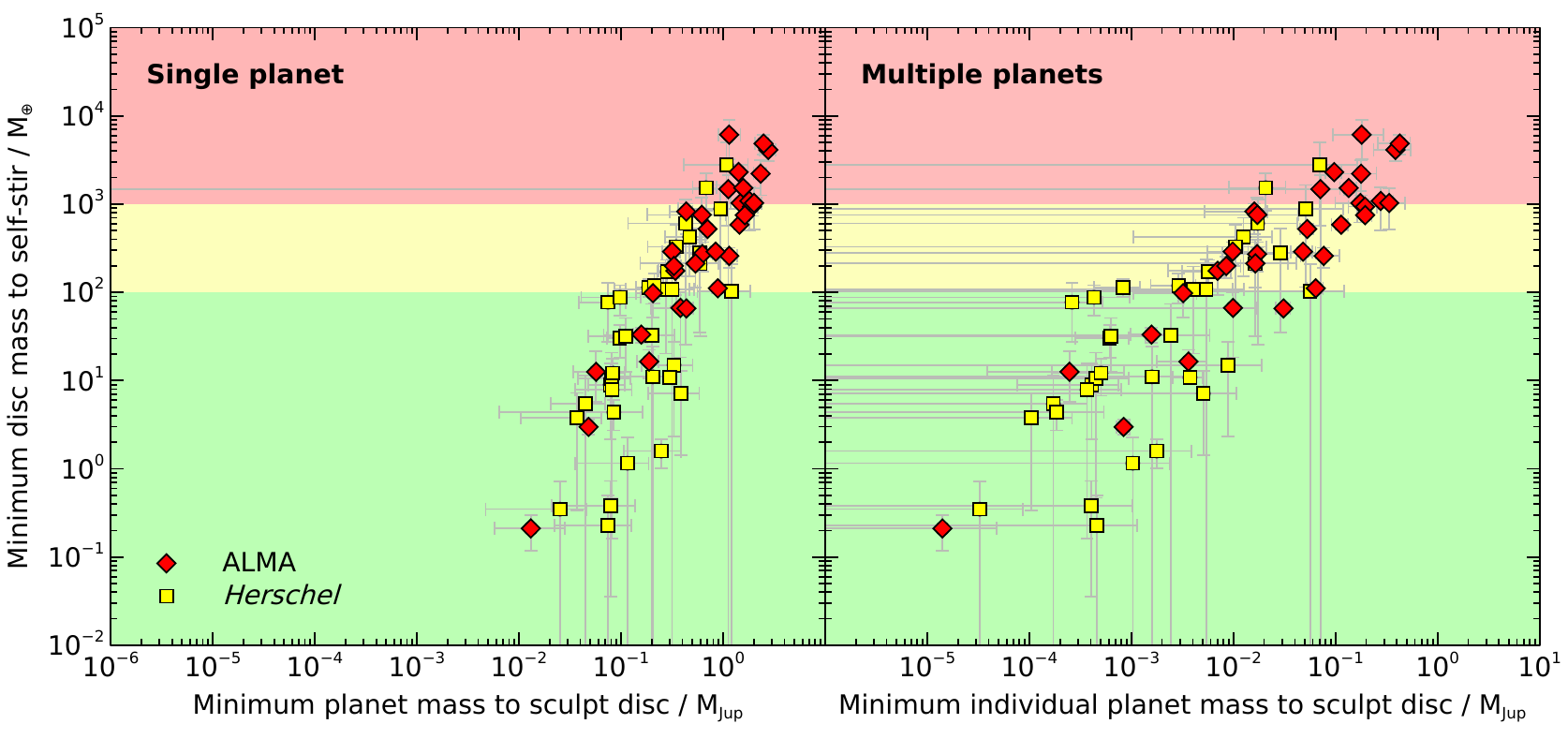}
   \caption{Self-stirring constraints in systems with resolved debris discs. Vertical axis: minimum masses of discs required for those discs to be self-stirred (Sect. \ref{subsec: selfStirring}). Horizontal axes: minimum planet masses required to sculpt the inner edges of those discs (Sect. \ref{sec: discTruncation}), if debris clearing is performed by one planet (left plot) or by multiple, equal-mass planets (right plot). Symbols were defined on previous figures. Background colours denote debris disc masses that may be unfeasibly large (red), borderline (yellow) or allowed (green), according to the quantity of solids that may have been available at the protoplanetary disc stage \citep{Krivov2021}. It may not be possible for discs in the red and yellow regions to be self-stirred, in which case other stirring mechanisms are required. Since these discs also typically require large planets to sculpt their inner edges, it is likely that planet-stirring operates in these systems, as shown in Sect. \ref{subsec: planetStirring}. The requirement for large planets to both stir and sculpt these discs makes systems in the yellow and red regions prime targets for future planet searches, as discussed in Sect. \ref{subsec: futureObs}.}
   \label{fig: selfStirringAndTruncatingPlanets}
\end{figure*}

If our resolved discs are self-stirred, then their minimum possible masses have a median of ${110 \mEarth}$ (first and third quartiles: 12 and ${680\mEarth}$). These minimum masses are shown on Fig. \ref{fig: selfStirringAndTruncatingPlanets} and listed in Table \ref{tab: systemAnalysisData}. Of our 67 resolved discs, 26 would need masses greater than ${100\mEarth}$ to self-stir (to ${1\sigma}$), including six requiring masses greater than ${1000\mEarth}$. Table \ref{tab: largeSelfStirringDiscMasses} lists the systems where the disc masses required for self-stirring are compatible with being ${1000\mEarth}$ or greater (to ${1\sigma}$). However, \cite{Krivov2021} argue that debris disc masses greater than ${100-1000\mEarth}$ may be unfeasible, since these could exceed the total mass in solids available in their parent protoplanetary discs. This implies that some, and potentially many, of our debris discs cannot be self-stirred, because this would require unrealistically high disc masses. In general, \cite{Krivov2021} argue that the resolution to the `debris disc mass problem' is that the largest planetesimals in some debris discs are smaller than previously assumed (${s_{\rm max} \sim 1 \km}$, rather than ${\sim 100 \km}$), but in this case these discs would be incapable of self-stirring. To summarise, even if we assume the optimal case where the largest planetesimals are \textit{just} large enough to stir the discs, it is likely that a reasonable fraction of our discs cannot be self-stirred without exceeding their allowed mass budget. If correct, these discs would need to be stirred by other mechanisms.

\begin{table*}
\caption{Systems where the debris-disc mass required for self-stirring, ${M_{\rm disc, stir}}$, could be ${1000 \mEarth}$ or more (to ${1\sigma}$), as described in Sect. \ref{subsec: selfStirringResults}. Such masses are potentially unfeasible, given the quantity of solids that would have been available at the protoplanetary disc stage \citep{Krivov2021}; this could imply that other stirring mechanisms operate in the systems listed. Also shown are the minimum planet masses required to stir the discs, ${M_{\rm p, stir}}$, from Sect. \ref{subsec: planetStirring}; all stirring planets are consistent with being below our L$'$-band detection limits at their predicted locations, implying that all of these discs can be planet-stirred, but not all can be self-stirred. Only systems with resolved discs are shown, where the `Disc data' column lists the observational data source for the disc edge locations. For axisymmetric discs, the `Disc location and extent' column lists the disc inner edge $a_{\rm i}$ and outer edge $a_{\rm o}$ as `${a_{\rm i} \rightarrow a_{\rm o}}$'. For asymmetric discs, this column shows the inner edge pericentre $q_{\rm i}$, inner edge apocentre $Q_{\rm i}$, outer edge pericentre $q_{\rm o}$ and outer edge apocentre $Q_{\rm o}$ as `${q_{\rm i}, Q_{\rm i} \ \rightarrow q_{\rm o}, Q_{\rm o}}$'. Uncertainties are ${1\sigma}$.}
\begin{tabular}{r r r r r r r}
\hline
\multicolumn{1}{c}{System} & \multicolumn{1}{c}{$M_*$ / $\rm M_\odot$} & \multicolumn{1}{c}{$t_*$ / Myr} & \multicolumn{1}{c}{Disc location and extent / au} & \multicolumn{1}{c}{Disc data} & \multicolumn{1}{c}{$M_{\rm disc, stir}$ / $\rm M_\oplus$} & \multicolumn{1}{c}{$M_{\rm p, stir}$ / $\rm M_{Jup}$} \\
\hline
CPD-72 2713 & $0.80\pm 0.02$ & $24\pm 3$ & $80^{+30}_{-20}$ $\rightarrow$ $180\pm 30$ & \multicolumn{1}{l}{\phantom{$^*$}ALMA} & $6000\pm 3000$ & $0.3^{+0.4}_{-0.5}$ \\
HD 9672 & $1.98\pm 0.01$ & $45\pm 5$ & $62\pm 4$ $\rightarrow$ $210\pm 4$ & \multicolumn{1}{l}{\phantom{$^*$}ALMA} & $2300\pm 200$ & $1.2^{+0.5}_{-0.4}$ \\
HD 16743 & $1.56\pm 0.01$ & $45\pm 5$ & $50\pm 50$ $\rightarrow$ $260\pm 70$ & \multicolumn{1}{l}{\phantom{$^*$}\textit{Herschel}} & $3000\pm 2000$ & $4\pm 10$ \\
HD 32297 & $1.68^{+0.05}_{-0.1}$ & $100^{+300}_{-70}$ & $91\pm 2$ $\rightarrow$ $153\pm 2$ & \multicolumn{1}{l}{\phantom{$^*$}ALMA} & $1500^{+600}_{-3000}$ & $0.04^{+0.03}_{-0.1}$ \\
HD 38206 & $2.36\pm 0.02$ & $42^{+6}_{-4}$ & $80\pm 20$, $140^{+30}_{-40}$ $\rightarrow$ $190\pm 30$, $320^{+50}_{-40}$ & \multicolumn{1}{l}{\phantom{$^*$}ALMA} & $2200^{+900}_{-1000}$ & $0.08\pm 0.09$ \\
HD 48370 & $0.96^{+0.05}_{-0.06}$ & $60^{+40}_{-20}$ & $30\pm 30$ $\rightarrow$ $160\pm 50$ & \multicolumn{1}{l}{\phantom{$^*$}\textit{Herschel}} & $600^{+500}_{-600}$ & $2\pm 5$ \\
HD 107146 & $1.03^{+0.02}_{-0.04}$ & $150^{+100}_{-50}$ & $52.2^{+0.6}_{-0.7}$ $\rightarrow$ $162.2^{+0.6}_{-0.7}$ & \multicolumn{1}{l}{\phantom{$^*$}ALMA} & $800^{+200}_{-400}$ & $0.14^{+0.06}_{-0.1}$ \\
HD 111520 & $1.32^{+0.1}_{-0.07}$ & $15\pm 3$ & $50\pm 10$ $\rightarrow$ $100\pm 10$ & \multicolumn{1}{l}{\phantom{$^*$}ALMA} & $1000\pm 400$ & $0.2\pm 0.2$ \\
HD 121617 & $1.90^{+0.06}_{-0.05}$ & $16\pm 3$ & $50\pm 7$ $\rightarrow$ $106\pm 7$ & \multicolumn{1}{l}{\phantom{$^*$}ALMA} & $900\pm 200$ & $0.3\pm 0.2$ \\
HD 131835 & $1.81^{+0.05}_{-0.04}$ & $16\pm 2$ & $40\pm 2$ $\rightarrow$ $127\pm 2$ & \multicolumn{1}{l}{\phantom{$^*$}ALMA} & $1500\pm 100$ & $1.2\pm 0.5$ \\
HD 138813 & $2.15^{+0.06}_{-0.07}$ & $10\pm 3$ & $56^{+9}_{-8}$ $\rightarrow$ $174^{+9}_{-8}$ & \multicolumn{1}{l}{\phantom{$^*$}ALMA} & $4000\pm 1000$ & $3\pm 2$ \\
HD 146181 & $1.28^{+0.09}_{-0.08}$ & $16\pm 2$ & $70^{+10}_{-20}$ $\rightarrow$ $90\pm 20$ & \multicolumn{1}{l}{\phantom{$^*$}ALMA} & $1100^{+500}_{-600}$ & $0.04\pm 0.05$ \\
HD 146897 & $1.32\pm 0.03$ & $10\pm 3$ & $60\pm 10$ $\rightarrow$ $80\pm 20$ & \multicolumn{1}{l}{\phantom{$^*$}ALMA} & $1000\pm 500$ & $0.06\pm 0.07$ \\
HD 156623 & $1.91^{+0.02}_{-0.03}$ & $16\pm 2$ & $10\pm 10$ $\rightarrow$ $150\pm 30$ & \multicolumn{1}{l}{\phantom{$^*$}ALMA} & $800\pm 400$ & $200\pm 500$ \\
HD 192425 & $1.94\pm 0.04$ & $400\pm 100$ & $100\pm 30$ $\rightarrow$ $420\pm 60$ & \multicolumn{1}{l}{\phantom{$^*$}\textit{Herschel}} & $1500\pm 700$ & $0.8\pm 1.0$ \\
HD 192758 & $1.62\pm 0.02$ & $45\pm 5$ & $40\pm 40$ $\rightarrow$ $180\pm 50$ & \multicolumn{1}{l}{\phantom{$^*$}\textit{Herschel}} & $900\pm 800$ & $2\pm 5$ \\
HD 218396 & $1.59\pm 0.02$ & $42^{+6}_{-4}$ & $170^{+30}_{-40}$ $\rightarrow$ $210^{+20}_{-10}$ & \multicolumn{1}{l}{\phantom{$^*$}ALMA} & $5000\pm 1000$ & $0.05^{+0.04}_{-0.03}$ \\
\hline
\end{tabular}
\label{tab: largeSelfStirringDiscMasses}
\end{table*}

Fig. \ref{fig: selfStirringAndTruncatingPlanets} also shows that discs requiring larger masses to self stir also need larger planets to sculpt their inner edges (Sect. \ref{sec: discTruncation}). This is expected, because both the sculpting planet mass and self-stirring disc mass increase with increasing disc size, and decrease with increasing system age. For the discs requiring potentially unfeasible masses to self-stir, this means that large planets are typically expected in those systems anyway, whatever the stirring mechanism. An obvious hypothesis is therefore that planets stir the discs that cannot be self-stirred, and these could be the same planets that sculpt the disc inner edges (although not necessarily). In the following section we calculate the planets required to stir the debris discs, and show that known or observationally allowed companions could stir all discs in our sample.

%---------------------------------------------------------
\subsection{Stirring the debris discs with planets}
\label{subsec: planetStirring}

The previous section examined the disc masses required for self-stirring, and showed that some would be unfeasibly high. There are two main alternatives to explain these discs; either they were `pre-stirred', i.e. debris orbits were already crossing at the end of the protoplanetary disc phase \citep{Wyatt2008}, or unseen planets are responsible for stirring \citep{Mustill2009}. Here we explore the latter scenario, where each disc is stirred by a single planet. Secular perturbations by this planet would cause debris eccentricities and orientations to evolve, eventually causing debris orbits to cross and initiating the collisional cascade. Since the secular interaction does not change the debris and planet semimajor axes, if the primordial debris has very low eccentricities then the perturbing planet would have to be eccentric to stir a disc. For this analysis we assume that primordial debris is unexcited, and therefore that any stirring is driven by a planet with non-zero eccentricity. Whilst a single planet could be responsible for both stirring and sculpting the disc, we treat these two analyses separately; the stirring constraints derived in this section are independent from the sculpting constraints from Sect. \ref{sec: discTruncation}.

%---------------------------------------------------------
\subsubsection{Stirring the debris discs with planets: method}
\label{subsec: planetStirringMethod}

We assume that the stirring planet is located interior to the inner edge of the disc. The stirring timescale decreases with planet mass $M_{\rm p, stir}$, and increases with the outer radius of the disc; a lower-bound on stirring planet mass is therefore the planet mass that is \textit{just} sufficient to stir the disc outer edge within the stellar age. \cite{Mustill2009} give timescales for an eccentric planet to stir debris at a given radius; rewriting their equation 15 in our units yields the time for a planet to stir the outer edge of an external disc as

\begin{multline}
t_{\rm p, stir} = 5.07 \times 10^{-5} \myr \left(\frac{a_2}{\au} \right)^{9/2} \left(\frac{a_{\rm p}}{\au} \right)^{-3} 
\\
\times \frac{\left(1 - e_{\rm p}^2 \right)^{3/2}}{e_{\rm p}} \left(\frac{M_{\rm p, stir}}{\mJup} \right)^{-1} \left(\frac{M_*}{\mSun} \right)^{1/2},
\label{eq: pltStirringTimescale}
\end{multline}

\noindent where $a_2$ is the semimajor axis of the outermost disc particle.

If the planet stirs the disc then the timescale from Eq. \ref{eq: pltStirringTimescale} must be shorter than the stellar age. Also, since we assume that the planet is interior to the inner edge of the disc, we have the additional constraint that ${a_{\rm p}(1+e_{\rm p}) < Q_{\rm i}}$, where $Q_{\rm i}$ is the apocentre of the disc inner edge (for axisymmetric discs this is just the inner edge location). Combining these two constraints with Eq. \ref{eq: pltStirringTimescale} yields the the minimum mass of an eccentric planet required to stir an external disc as

\begin{multline}
M_{\rm p, stir} \gtrsim 5.07 \times 10^{-5} \mJup \left(\frac{q_{\rm o}}{\au} \right)^{9/2} \left(\frac{Q_{\rm i}}{\au} \right)^{-3} 
\\
\times  \left(\frac{t_*}{\myr} \right)^{-1} \left(\frac{M_*}{\mSun} \right)^{1/2} \frac{(1+e_{\rm p})^3 (1-e_{\rm p}^2)^{3/2}}{e_{\rm p}}.
\label{eq: pltMassToStir}
\end{multline}

\noindent To make Eq. \ref{eq: pltMassToStir} general for both axisymmetric and eccentric discs we have set ${a_2 = q_{\rm o}}$, the pericentre of the disc outer edge, because if disc eccentricity is driven by an eccentric planet then this is the semimajor axis of the outermost debris particle (equation 7 in \citealt{Pearce2014}). Note that the minimum-mass stirring planet would have apocentre at the disc inner edge; planets located further inwards would require larger masses to stir the disc.

Eq. \ref{eq: pltMassToStir} depends on planet eccentricity; the higher the eccentricity, the lower the planet mass required for stirring. However, this eccentricity is unknown. If the planet also sculpts the disc inner edge then a lower bound on planet eccentricity is given by Eq. \ref{eq: lowerLimitOnPlanetEccentricity}, but this is only a lower limit because a planet with higher eccentricity located further inwards could also sculpt the disc to the same degree (Eq. \ref{eq: planetSemimajorAxisEccentricityFromDiscEdges}). Also, since the great majority of our discs are either unresolved or axisymmetric, we cannot bound the eccentricity of a stirring planet in most systems. In order to estimate the minimum planet mass required to stir the disc, we must therefore assume a maximum planet eccentricity. Denoting the eccentricity term in Eq. \ref{eq: pltMassToStir} as ${f(e_{\rm p}) \equiv (1+e_{\rm p})^3 (1-e_{\rm p}^2)^{3/2}/e_{\rm p}}$, we see that for small planet eccentricities ${f(e_{\rm p}) \propto e_{\rm p}^{-1}}$, whilst for larger eccentricities the scaling is roughly ${f(e_{\rm p}) \propto (1-e_{\rm p})}$. The transition between these two regimes occurs around ${e_{\rm p} \approx 0.3}$. For the planet-stirring analysis we therefore assume that the eccentricity of the stirring planet is less than $0.3$, because for smaller eccentricities the minimum planet mass is very sensitive to the exact choice of eccentricity, whilst for larger eccentricities the dependence is less severe. A maximum eccentricity of 0.3 is also consistent with the planet eccentricities required to sculpt the inner edges of the discs (Sect. \ref{subsec: singlePlanet}). We therefore assume a maximum planet eccentricity of 0.3 for this analysis; for comparison, if we had assumed a maximum eccentricity of 0.1 then the minimum planet masses required for stirring would be doubled. Note that we expect the great majority of planets to actually have eccentricities smaller than 0.3, so assuming 0.3 as an upper limit will set a lower-bound on the stirring masses required. 

Assuming that the stirring planets have eccentricities smaller than 0.3, we use Eq. \ref{eq: pltMassToStir} with ${e_{\rm p} = 0.3}$ to find the minimum planet masses required to stir the discs. For resolved discs we take $Q_{\rm i}$ and $q_{\rm o}$ from the observed disc edges, and for discs with only SED data we calculate conservative lower limits on the stirring planet masses by assuming that those discs have zero width, are located at the corrected blackbody radii, and are axisymmetric (i.e. ${Q_{\rm i} = q_{\rm o} = r_{\rm SED}}$; non-zero widths would increase the required stirring planet masses). Using this method we plot the planet-stirring constraints for ${49 \; \rm Cet}$ (${\HD9672}$) on Fig. \ref{fig: hd9672PltCnstrnts} as an example (line 7); if a single planet sculpts this disc then that planet could also stir it, but if multiple planets instead sculpt the disc, then those planets may be too small to perform the stirring themselves.

%---------------------------------------------------------
\subsubsection{Stirring the debris discs with planets: results}
\label{subsec: planetStirringResults} 

\begin{figure}
  \centering
   \includegraphics[width=8cm]{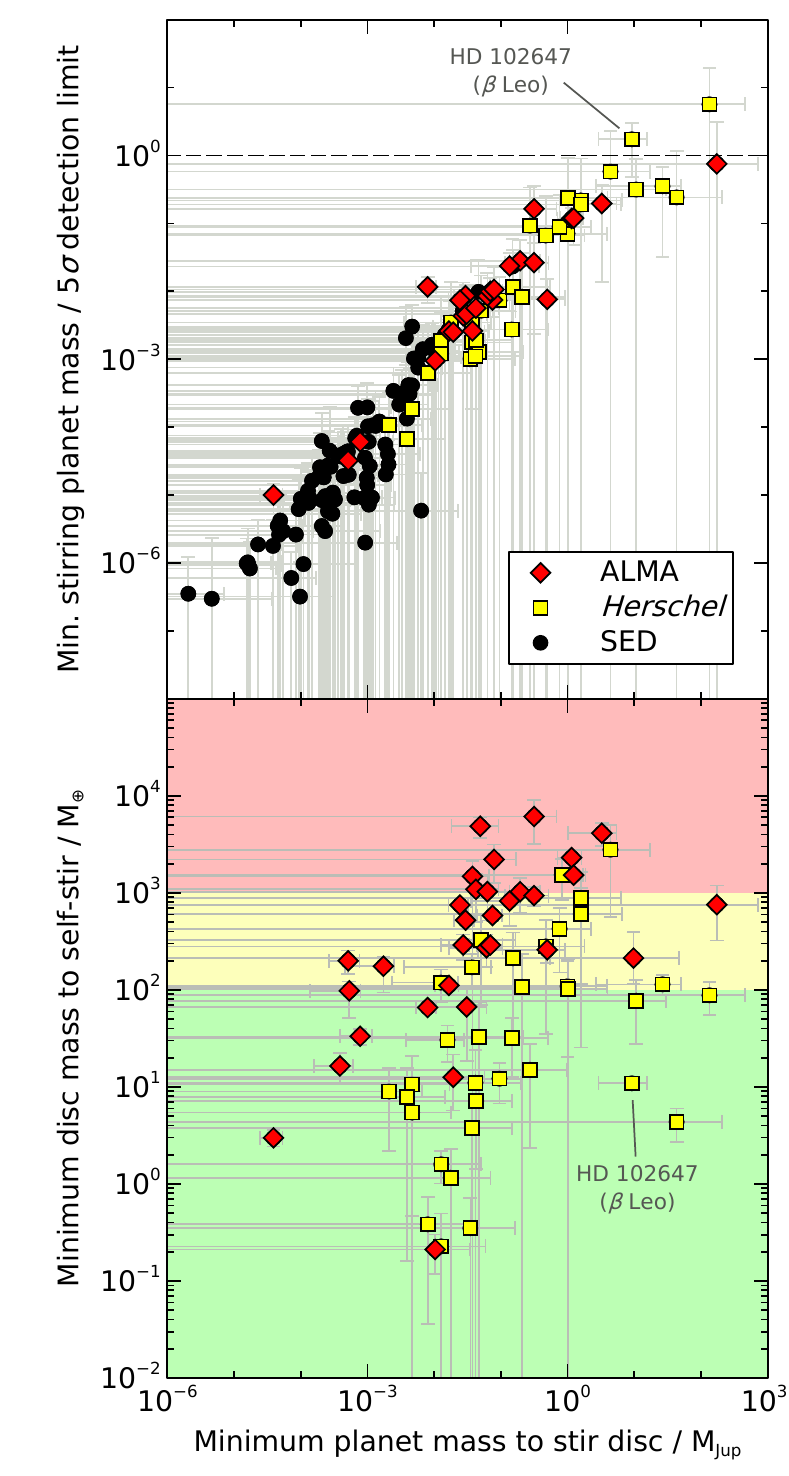}
   \caption{Horizontal axes: minimum planet masses required to stir the discs, if those planets have eccentricities smaller than 0.3 as described in Sect. \ref{subsec: planetStirring} (higher eccentricities would require lower masses). Top plot: vertical axis shows these minimum stirring-planet masses, divided by our L$'$-band planet detection limits at the apocentres of those planets. The minimum stirring-planet masses are all consistent (to ${1\sigma}$) with lying below the detection limits (i.e. below the dashed line showing unity), so all of the discs can potentially be stirred by unseen planets. The stirring planet predictions for systems where discs are unresolved (black circles) are conservative lower limits, since these discs are modelled as having zero width. Bottom plot: vertical axis shows the minimum masses of the resolved discs required for those discs to be self-stirred, from Fig. \ref{fig: selfStirringAndTruncatingPlanets}. Many systems would require disc masses greater than ${100-1000 \mEarth}$ to self-stir, which may be unfeasibly high \citep{Krivov2021}; however, the top plot shows that these systems could still be stirred by unseen planets. This suggests that self-stirring is not the dominant stirring mechanism in at least some debris discs. No unresolved discs are shown on the bottom plot, because these were omitted from the self-stirring calculation owing to a lack of width information. Also note that a few systems on the bottom plot are missing from the top plot because their predicted planet locations are not covered by our observations. The uncertainties on stirring planet masses are large, owing to the very strong dependence of Eq. \ref{eq: pltMassToStir} on disc edge locations.}
   \label{fig: planetStirringAndSelfStirring}
\end{figure}

We now apply the above method to all of our systems, and list the resulting minimum planet masses for stirring in Table \ref{tab: systemAnalysisData}. On Fig. \ref{fig: planetStirringAndSelfStirring} we compare these minimum stirring masses to our L$'$-band observational upper limits on planet masses, and also to the disc masses required for self-stirring from Sect. \ref{subsec: selfStirring}.

The top plot of Fig. \ref{fig: planetStirringAndSelfStirring} shows the minimum planet masses required to stir the discs, divided by our L$'$-band planet detection limits at the semimajor axes of those planets. The plot shows that \textit{all} of our discs could be stirred by unseen planets (i.e. the minimum planet masses required are all consistent with being below the detection limits, to ${1\sigma}$). Almost all of the required planet masses are orders of magnitude too small to be detectable with current instrumentation, with only one system (${\beta \; \rm Leo}$: ${\HD102647}$) needing a planet mass greater than ${10\percent}$ of our L$'$-band detection limit (to ${1\sigma}$) to stir the disc (this system is discussed below). For a further 17 systems, the stirrers could have masses above ${10\percent}$ of our detection limits too, but their large uncertainties mean they could also be below ${10\percent}$ of the detection limits (to ${1\sigma}$). This leaves 160 of our 178 systems where the minimum planet masses required for stirring are well below current detection limits; therefore, even order-of-magnitude improvements in mass-detection limits may not be sufficient to confirm or exclude planetary stirrers, unless those planets also sculpt the discs (see below). 

The bottom plot of Fig. \ref{fig: planetStirringAndSelfStirring} compares the minimum planet masses required for planet-stirring to the minimum disc masses required for self-stirring. Generally, discs requiring larger masses to self-stir also need larger planets to planet-stir, since both mass requirements increase with disc size and decrease with stellar age. However, the discs that require unfeasibly high masses to self-stir could alternatively be stirred by moderate-mass planets; for example, the outer disc of ${49 \; \rm Cet}$ (${\HD9672}$) requires a disc mass of at least ${2300\pm200\mEarth}$ (${7.3 \pm 0.7\mJup}$) to self-stir, but could alternatively be stirred by a ${1.2 \pm 0.5 \mJup}$ planet with eccentricity ${\leq0.3}$. Table \ref{tab: largeSelfStirringDiscMasses} shows that all systems potentially requiring disc masses above ${1000 \mEarth}$ to self-stir could instead be stirred by planets with masses of ${1 \mJup}$ or less (to $1\sigma$). Since all minimum-mass stirring planets are consistent with being below the detection limits (Fig. \ref{fig: planetStirringAndSelfStirring}, top plot), it appears that whilst not all debris discs can be self-stirred (unless they have very high masses, argued against by \citealt{Krivov2021}), all discs (in our sample at least) could be planet- or companion-stirred. This could point towards general conclusions about the relative prevalence of self-stirring and planet/companion-stirring across all debris discs. However, we urge caution on this point; many of our discs could still be self-stirred with masses below ${100-1000 \mEarth}$.

The bottom plot of Fig. \ref{fig: planetMassesHistogram} shows the distribution of minimum planet masses required to stir our discs. Whilst these masses appear considerably smaller than those required to sculpt the discs, the stirring planet mass distribution is biased towards unrealistically small values because SED discs were modelled as having zero width. It is still useful to provide these estimates as conservative lower bounds for the stirring planet masses in specific systems without resolved disc data, but more realistic estimates of the `typical' planet masses required for stirring arise by considering only discs resolved by ALMA or \textit{Herschel} (although note that these also may not be fully representative, since ALMA and \textit{Herschel} observations are biased towards the largest, brightest discs). For these systems, the median planet mass required for stirring is ${0.05\mJup}$ (${20\mEarth}$), with first and third quartiles of ${0.02\mJup}$ (${5\mEarth}$) and ${0.6\mJup}$, respectively.

\begin{figure*}
  \centering
   \includegraphics[width=17cm]{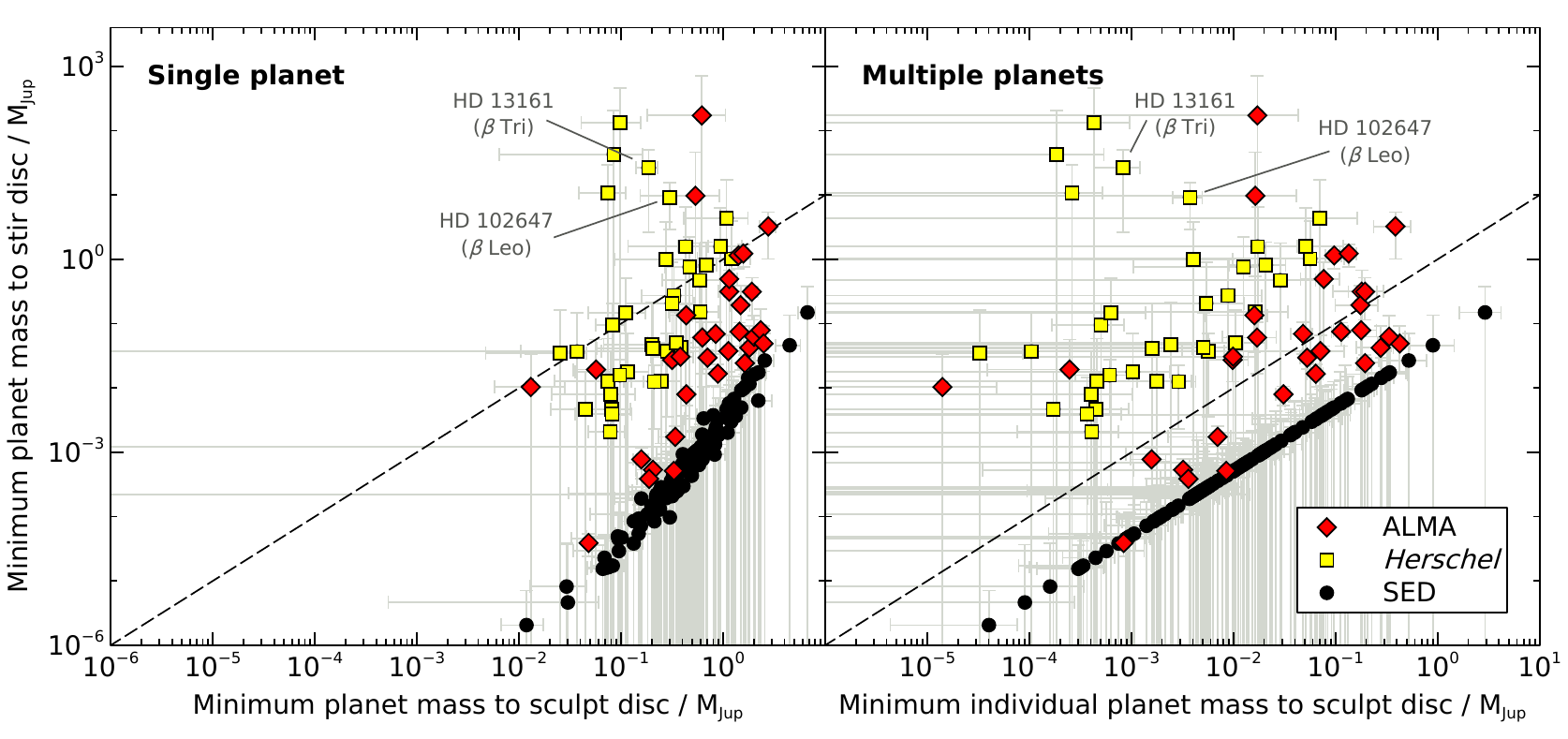}
   \caption{Comparison of the planet masses required to stir and sculpt debris discs. Vertical axis: minimum planet masses required to stir each disc (Sect. \ref{subsec: planetStirring}). Horizontal axes: minimum planet masses required to sculpt the inner edges of those discs (Sect. \ref{sec: discTruncation}), if sculpting is performed by one planet (left plot) or by multiple, equal-mass planets (right plot). Symbols were defined on previous figures, and the dashed lines denote 1:1 relations. If one planet sculpts each disc inner edge (left plot), then for almost all systems that planet could also stir the disc (either the sculpting planet mass is greater than that required for stirring, or the two are consistent to ${1\sigma}$). The only exception is ${\beta \; \rm Tri}$ (${\HD13161}$), which is a binary system where the binary could instead stir the disc (for the other labelled system, ${\beta \; \rm Leo}$ (${\HD102647}$), we likely underestimate the disc inner edge and therefore overestimate the required stirring planet mass). Conversely, if multiple planets sculpt each disc (right plot), then for several systems the minimum-mass sculpting planets would be too small to stir those discs (to ${1\sigma}$); some, but not all, of those discs could be feasibly self-stirred, so either the planets in those systems are more massive than required by the multi-planet sculpting model, or disc sculpting and stirring are performed by different entities. Note that the stirring planet masses derived from SEDs of unresolved discs (black circles) are conservative lower limits, since these discs were assumed to have zero width for this analysis; in this case, the planet masses required for sculpting and stirring have similar or identical dependencies on system parameters, resulting in the tight relationships between the two. As on previous plots, the large uncertainties are due to the strong dependencies of Eqs. \ref{eq: shannonPltMassK16} and \ref{eq: pltMassToStir} on disc edge locations.}
   \label{fig: planetStirringAndTruncation}
\end{figure*}

Fig. \ref{fig: planetStirringAndTruncation} compares the planet masses required to stir the discs to those required to sculpt the discs, for the two scenarios where either one or multiple planets perform the sculpting. The left plot shows that, for all but two of our systems, if a single planet sculpts the disc then that planet could also stir the disc (i.e. the minimum mass required by a single planet to sculpt the disc is greater than that required to stir the disc, or the two are consistent to $1\sigma$). The two exceptions (both \textit{Herschel} discs) are ${\beta \; \rm Tri}$ (${\HD13161}$) and ${\beta \; \rm Leo}$ (${\HD102647}$). However, ${\beta \; \rm Tri}$ is a close binary with a circumbinary disc, and \cite{Kennedy2012} show that the binary would be capable of stirring the disc. For ${\beta \; \rm Leo}$, \cite{Churcher2011HD102647} argue that multiple belts exist, with one or two inner belts plus an outer disc spanning ${15-70\au}$ (compared to the ${7-79 \au}$ that we use in our single-disc model); it is therefore likely that we underestimate the outer disc inner edge, and taking this to be 15 rather than ${7\au}$ reduces the required stirring planet mass by an order of magnitude, below that needed to sculpt the disc (the general effect of underestimating the disc inner edges in our \textit{Herschel} models is discussed in Sect. \ref{subsec: herschelDiscussion}). In conclusion, if the inner edges of our discs are each sculpted by a single planet, then these planets could also fully stir the discs (except for ${\beta \; \rm Tri}$, where stirring could instead be performed by the binary).

Conversely, the right plot of Fig. \ref{fig: planetStirringAndTruncation} shows that, if multiple planets sculpt each disc, then for eight systems the minimum-mass sculpting planets would be too small to also stir the discs (to ${1\sigma}$). These systems are ${49 \rm \; Cet}$ (${\HD9672}$), ${\beta \rm \; Tri}$ (${\HD13161}$), ${\psi^{5} \; \rm Aur}$ (${\HD48682}$), ${\HD92945}$, ${\beta \rm \; Leo}$ (${\HD102647}$), ${\HD107146}$, ${\HD131835}$, and ${\HD138813}$. If the inner edges of these systems are carved by multiple planets, then those planets must either be larger than the minimum masses derived in Sect. \ref{subsec: multiplePlanets} (in which case they could also stir the discs), or the stirring is performed by some other process(es). Some of these discs could be self-stirred without requiring disc masses greater than ${1000\mEarth}$, so for these systems a mixture of self-stirring and multi-planet sculpting could be occurring. Alternatively, the planet(s) responsible for stirring the discs could be different to those sculpting them; larger planets located further in could stir the discs, whilst smaller ones at the disc edges could sculpt them (discussed further in Sect. \ref{subsec: comparisonToSpecificPlanets}). Another possibility is that, since some of these discs contain gaps that could host planets, planets located in the gaps could be responsible for stirring, whilst other planets sculpt the inner edges. However, two of the above discs (${49 \rm \; Cet}$ and ${\HD138813}$) have no observed gaps, and would require masses of more than ${1000\mEarth}$ to be self-stirred, which are possibly unfeasibly high; for these systems it seems likely that the sculpting planets are larger than the minimum masses derived assuming multi-planet sculpting, and that these planets also stir the discs, or that other interior planets stir the discs. Note that both of these discs also have gas detected \citep{Hughes2017, LiemanSifry2016}, so dust-gas interactions could also be occurring in them (e.g. \citealt{Pearce2020}). In summary, if one planet sculpts each of our discs, then the discs could also all be stirred by that planet or a known companion; conversely, if the discs are sculpted by multiple planets, then for some systems the minimum planet masses required for sculpting are smaller than those needed for stirring.

%%%%%%%%%%%%%%%%%%%%%%%%%%%%%%%%%%%%%%%%%%%%%%%%%%%%%%%%%%%%%%%%%%%%%%%%%%%%%%%%%%%%%%%%%%%%%%%%%%%
\section{Discussion}
\label{sec: discussion}

\noindent We used sculpting and stirring arguments to predict planet parameters in 178 debris-disc systems. In this section we compare our predictions to known planets and those inferred from protoplanetary discs (Sect. \ref{subsec: comparisonToOtherPltPopulations}), consider the systems best-suited for future planet searches (Sect. \ref{subsec: futureObs}), and discuss several \textit{caveats} that must be considered when interpreting our results.

%--------------------------------------------------------------------------------------------------
\subsection{Comparisons to known planet populations}
\label{subsec: comparisonToOtherPltPopulations}

\noindent Here we compare our planet predictions to both the detected planet population, and to planets inferred to be forming in protoplanetary discs.

%--------------------------------------------------------------------------------------------------
\subsubsection{Predicted planets vs. detected and forming planet populations}
\label{subsec: generalPopulationComparison}

\begin{figure*}
  \centering
   \includegraphics[width=17cm]{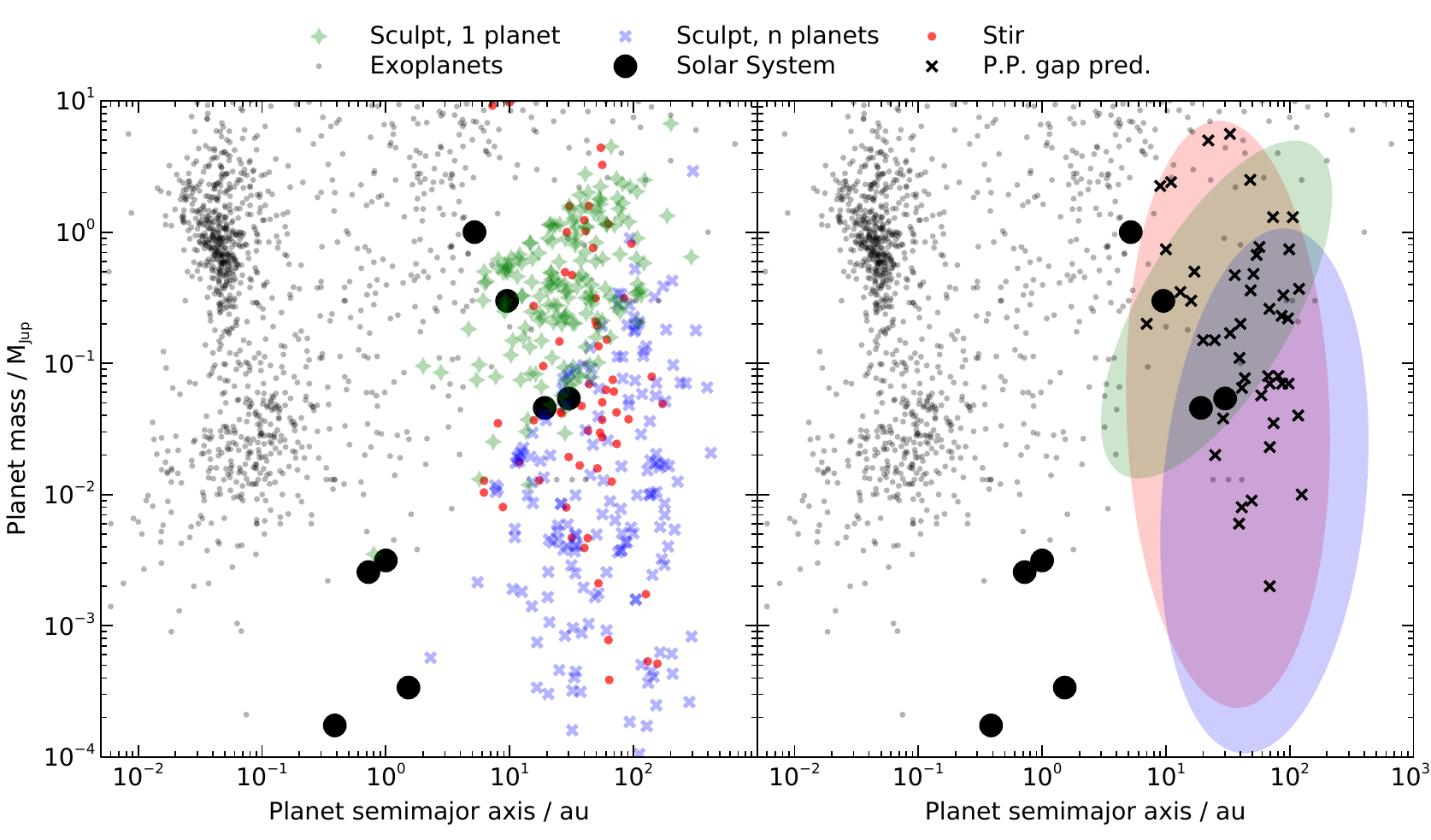}
   \caption{Planets predicted from our debris analyses, compared to known planets and those inferred to be forming in protoplanetary discs (Sect. \ref{subsec: generalPopulationComparison}). Small black circles are known exoplanets from Exoplanet.eu, and large circles are Solar System planets. The left plot shows the planetary populations inferred from our debris analyses; green diamonds and blue crosses are the planets required to sculpt the discs, if sculpting is performed by one or multiple planets respectively (Sects. \ref{subsec: singlePlanet} and \ref{subsec: multiplePlanets}). Red circles are planets required to stir the discs, (only systems with resolved discs are shown; Sect. \ref{subsec: planetStirring}). The plot shows that debris disc morphologies imply planetary populations that are poorly explored with current instruments, so the hypotheses that planets sculpt and stir debris discs are consistent with the lack of planet detections in these systems. Right plot: comparison of our planets inferred from debris discs to those inferred to be forming in protoplanetary discs (black crosses; \citealt{Lodato2019}). To make the plot clearer, we have replaced our individual planet predictions with rough areas representing the three populations. Our planetary population inferred from debris discs well-matches the population inferred to be forming in protoplanetary discs; if the forming planets remain \textit{in situ}, then they could later join the population responsible for sculpting and stirring known debris discs. Alternatively, it is possible that the forming planets rapidly sculpt debris discs before migrating inwards. Error bars are omitted for clarity, but can be seen on our other figures.}
   \label{fig: planetsInWholePlanetPop}
\end{figure*}

\noindent Our debris disc analysis predicts that planets of at least super-Earth- to Jovian-masses commonly reside between ${10-100\au}$. We show our planet predictions from sculpting and stirring arguments on Fig. \ref{fig: planetsInWholePlanetPop}, compared to the known planetary population. Our predicted planets occupy a different parameter space to known exoplanets, in a region that is difficult to probe with current instrumentation and consequently hosts few known objects. However, our analysis predicts that this parameter space should harbour a rich family of planets, some of which should be detectable with next-generation instruments (Sect. \ref{subsec: futureObs}). Whilst dissimilar to known exoplanets, our predicted planets are more similar to those in the outer Solar System; Uranus and Neptune in their current positions would be capable of sculpting and stirring about half of our discs, and many of these systems may well contain Uranus or Neptune analogues. For the remaining systems, we typically require either ice giants at 2-4 times the distances of Uranus and Neptune, or Jovian planets at 5-10 times the distance of Jupiter; such bodies have no known Solar System analogues.

Whilst dissimilar to the known exoplanets, our predicted planets are similar to those inferred to be forming in protoplanetary discs; the crosses on the right plot of Fig. \ref{fig: planetsInWholePlanetPop} show the young planets inferred to be carving gaps in ALMA-resolved protoplanetary discs \citep{Lodato2019}. Both planetary and non-planetary origins have been suggested for such gaps (e.g. \citealt{Okuzumi2016, Dullemond2018, Richert2018, Zhang2018, Pinte2020}), but if these gaps are created by forming planets then we may be witnessing planets being born that will later join the population responsible for sculpting and stirring known debris discs. Of course, this picture could be complicated by planetary migration both in and beyond the gas disc phase, as predicted from population synthesis models (e.g. \citealt{Alibert2013, Pfyffer2015, Mordasini2018}). However, it is interesting that the planets potentially forming in protoplanetary disc gaps, if they remained \textit{in situ}, overlap very strongly with those that we infer to exist from debris disc observations. If these inferred forming planets are destined to migrate to inwards of ${10\au}$ as suggested (e.g. \citealt{Lodato2019, Wang2021}), then either a different population of planets is required to later interact with debris discs, or these forming planets must quickly sculpt and stir the debris discs before migrating inwards (see Sect. \ref{subsec: comparisonToSpecificPlanets}), or debris discs are neither sculpted nor stirred by planets. Alternatively, since the forming planets already have the masses and locations of planets inferred from debris discs, this could be evidence that the forming planets do not migrate as far as previously thought.

%--------------------------------------------------------------------------------------------------
\subsubsection{Predictions for systems with known planets}
\label{subsec: comparisonToSpecificPlanets}

Planets have already been detected in some of our systems. Here we compare these to our inferred planets, to assess whether the known planets could be those interacting with our discs.

Of our 178 debris disc systems, 15 host at least one planet residing interior to the disc inner edge. These planets were detected using a variety of techniques, and specific details for individual systems are given in Appendix \ref{app: specificSystems}. On Fig. \ref{fig: detectedAndPredictedPlanets} we compare the masses and semimajor axes of these detected planets to those required to sculpt the associated discs; we consider single-planet sculpting here (Sect. \ref{subsec: singlePlanet}), because it predicts the required planet mass as a function of its semimajor axis (rather than assuming that the outermost planet resides exactly at the disc inner edge, as in the multi-planet model). If several planets have been detected in a system then we only consider the outermost planet, and we estimate the masses of radial velocity (RV) planets by assuming that they share the same inclination as the associated debris disc. 

\begin{figure}
  \centering
   \includegraphics[width=8cm]{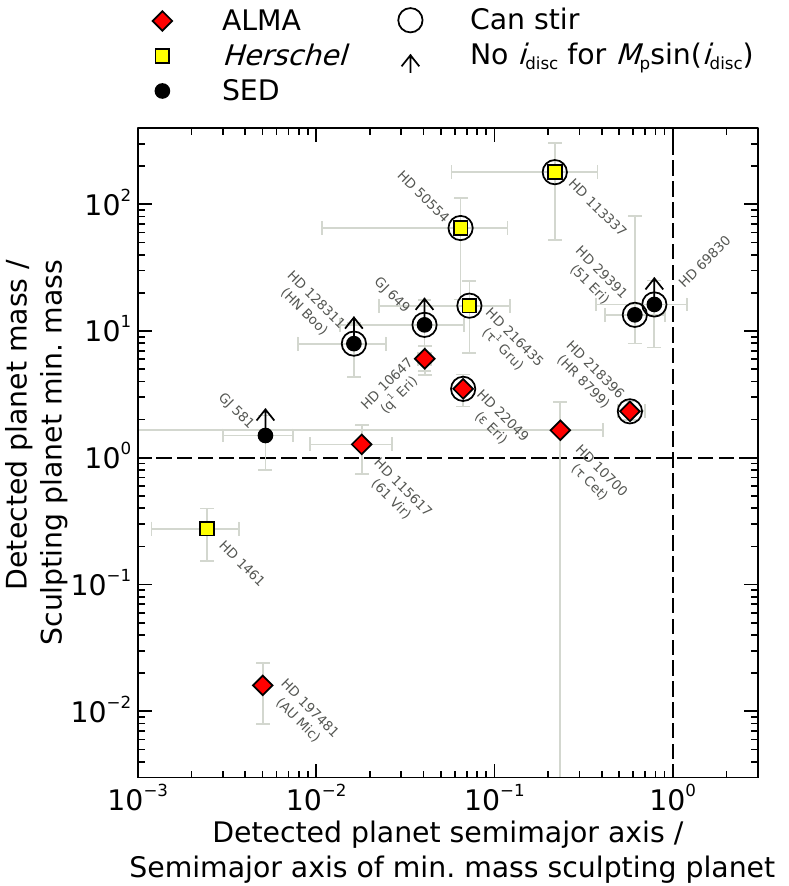}
   \caption{Comparison of detected and predicted planets in our debris disc systems, as described in Sect. \ref{subsec: comparisonToSpecificPlanets}. For systems with a detected planet, the vertical axis shows the mass of one detected planet divided by the mass required by a single planet to sculpt the disc inner edge (Sect. \ref{subsec: singlePlanet}). We calculate the masses of RV planets by assuming that they are coplanar with the discs. The horizontal axis shows the semimajor axis of the detected planet, divided by the semimajor axis that the minimum-mass sculpting planet would have. For systems with multiple known planets, only the outermost detected planet is shown. Dashed lines denote unity. With the possible exceptions of ${\rm 51 \; Eri}$ (${\HD29391}$) and ${\HD69830}$, the detected planets are unlikely to be massive enough to sculpt the disc inner edges \textit{in situ}; whilst some are larger than the minimum masses required for sculpting, these detected planets lie far interior to the discs, so would need significantly greater masses to sculpt the discs \textit{in situ} (line 4 on Fig. \ref{fig: hd9672PltCnstrnts}). This means that either additional undetected planets also reside in these systems, or the known planets formed in the outer regions and rapidly sculpted the discs before migrating inwards. Conversely, some detected planets are massive enough to have stirred the discs \textit{in situ}, despite their smaller semimajor axes; systems where detected planets could stir the discs are circled. Arrows show lower limits on detected planet masses for RV planets with no disc inclination data, and the remaining symbols were defined on previous figures.}
   \label{fig: detectedAndPredictedPlanets}
\end{figure}

Fig. \ref{fig: detectedAndPredictedPlanets} shows that, whilst many known planets would have masses sufficient to sculpt their associated discs were the planets located further out in those systems, these detected planets are mostly located too far inwards to be sculpting the discs \textit{in situ}; the mass required by a sculpting planet sharply increases as its semimajor axis decreases (line 4 on Fig. \ref{fig: hd9672PltCnstrnts}). The possible exceptions are ${\rm 51 \; Eri}$ (${\HD29391}$) and ${\HD69830}$, where the outermost planets may be capable of sculpting the disc inner edges \textit{in situ}; for the remaining systems, if disc inner edges are sculpted by planets then these planets probably reside exterior to the known ones. However, for systems above the horizontal line on Fig. \ref{fig: detectedAndPredictedPlanets}, it is also possible that the detected planets formed in the outer regions, sculpted and stirred the discs, and have since migrated inwards to their current locations. The latter picture is consistent with the expectation that young planets inferred to reside in gaps in ALMA-resolved protoplanetary discs will later migrate inwards of ${10\au}$ (see Sect. \ref{subsec: generalPopulationComparison}), although in this case some planets would have to be considerably more massive than our predictions to rapidly sculpt/stir debris discs before migrating.

Whilst most of the known planets are unlikely to be sculpting discs \textit{in situ}, a larger fraction of them could be stirring the discs. The circles on Fig. \ref{fig: detectedAndPredictedPlanets} mark systems where the known planet would, in their present location, have sufficient mass to stir the disc within the system age according to Eq. \ref{eq: pltStirringTimescale} (assuming the planet eccentricities are smaller than 0.3, as in Sect. \ref{subsec: planetStirring}). This is possible because the planet masses required for stirring typically have a much shallower dependence on semimajor axis than the masses required for sculpting (compare lines 4 and 7 on Fig. \ref{fig: hd9672PltCnstrnts}), since it is often easier to drive debris orbit crossings at the disc outer edge than it is to eject all material from the disc inner edge. Some detected planets could therefore still be interacting with the discs by stirring them, even if they are located too far inwards to sculpt them; this emphasises the general point that planets responsible for stirring discs need not be located at the disc inner edges, but could reside much closer to the star.

%--------------------------------------------------------------------------------------------------
\subsection{Potential for future observations}
\label{subsec: futureObs}

\noindent Many of our predicted planets lie just below current detection limits, so could be observable in the near future. We now discuss the detectability of our predicted planets, and identify some of the more promising targets for future direct-imaging observations.

To detect a planet via direct imaging, it must be bright enough to lie above the contrast curve at its projected separation from the star. This naturally favours massive planets, and there is also a bias towards younger planets because these are more luminous \citep{Baraffe2003, Phillips2020}. For our predicted planets these two considerations are linked, because younger systems require higher-mass planets to sculpt and stir the discs within the system lifetime; our systems requiring the most-massive planets are therefore also the youngest, so these should be strongly favoured for future observations. For a planet to be detectable it must also have a wide angular separation from its host star at the moment of observation; this not only favours planets on wide orbits, but also those with low sky-plane inclinations (because these would be well-separated from their stars in projection for greater fractions of their orbits). Since we assume that our predicted planets are coplanar with the debris discs, we therefore also favour systems with low disc-to-sky inclinations as targets for future observations.

\begin{figure*}
  \centering
   \includegraphics[width=17cm]{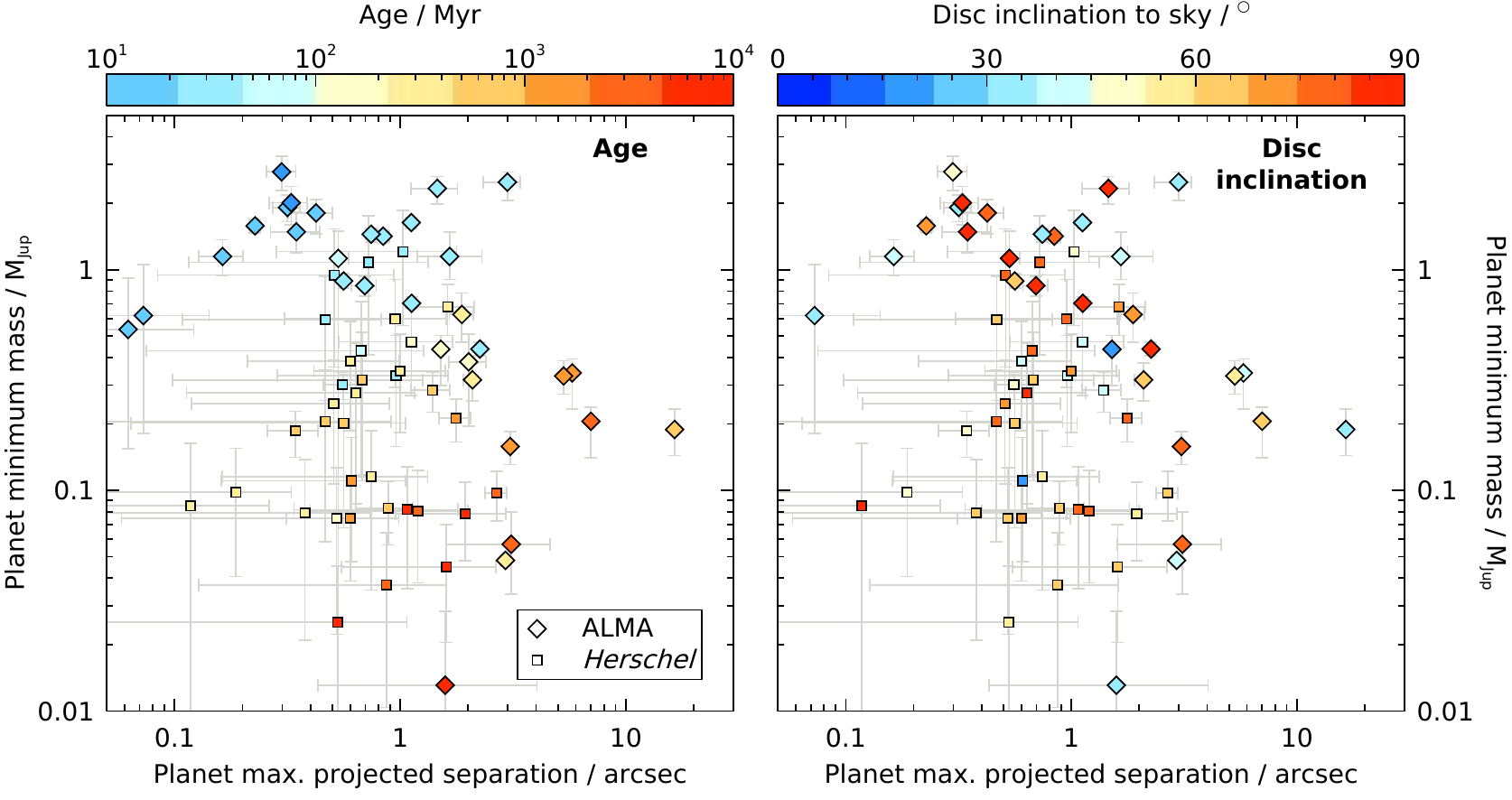}
   \caption{Identification of our predicted planets with the best chances of future detection (Sect. \ref{subsec: futureObs}). Points show the minimum masses and maximum projected separations of the planets predicted to be sculpting the inner edges of our discs, if one planet is responsible in each system (Sect. \ref{subsec: singlePlanet}). Only systems where the disc is resolved by ALMA (diamonds) or \textit{Herschel} (squares) are shown. Left plot: systems coloured by age. Right plot: systems coloured by disc inclination to the sky. The best targets have predicted planets with high masses, which are well-separated from their stars, are young (so the planets are brightest), and have low-inclination discs (so a coplanar planet would be separated from its star in projection over a large fraction of its orbit). \textit{JWST} will be capable of imaging ${0.1-1 \mJup}$ planets at ${\sim \rm arcsec}$ separations \citep{Beichman2010, Carter2021}, and this should be complemented at small angular separations by upcoming instruments such as VLT/ERIS \citep{Davies2018}, so many of these predicted planets could be observable in the near future. The plots show the single planets required to sculpt each disc, but if multiple planets were instead responsible (Sect. \ref{subsec: multiplePlanets}), then those planets could be at least an order of magnitude less massive than those shown here. The planets required to stir each disc are typically smaller than those shown (Sect. \ref{subsec: planetStirring}). Note that some systems are missing from the right plot because they lack inclination information.}
   \label{fig: futureObsSystems}
\end{figure*}

Fig. \ref{fig: futureObsSystems} shows the minimum masses, maximum projected separations, ages and inclinations of the planets that we predict to be sculpting our discs (assuming that one planet is responsible in each case; Sect. \ref{subsec: singlePlanet}). We only plot systems where the disc has been resolved by ALMA or \textit{Herschel}, since these provide the most-reliable planet constraints. The figure shows that we predict planets of up to several Jupiter masses at arcsecond separations, that our most-massive predicted planets are also young, and that a reasonable number are expected on low-inclination orbits which maximise the chances of observations. These compare favourably with the observational capabilities of the upcoming \textit{JWST}; ${0.1-1 \mJup}$ planets at arcsecond separations around young stars such as those in the ${\beta \rm \; Pic}$ and TW Hya moving groups would have detection thresholds of ${50\percent}$ when observed for 1-hour with MIRI \citep{Beichman2010, Carter2021}. Other upcoming instruments should also be able probe our planet parameter space; these include the Enhanced Resolution Imager and Spectrograph (ERIS, \citealt{Davies2018}) at the VLT, due to start operations in 2022. Considered the successor of NaCo, it will provide enhanced adaptive optics and improved sensitivity in the J to M bands. In addition, the grating vector Apodizing Phase Plate (gvAPP, \citealt{Boehle2018}) should enable high-contrast imaging observations, and should complement \textit{JWST} by probing small angular separations, where some of our most-massive planets are predicted to reside. Further ahead, the Mid-infrared E-ELT Imager and Spectrograph (METIS, \citealt{Brandl2018}) should detect planets down to several Earth masses at small angular separations \citep{Quanz2015}, thanks to its ${39 \m}$ mirror that increases angular resolution by a factor of ${\sim4-5}$ compared to current telescopes. The expected background-limited magnitude for a one-hour observation in L$'$ with METIS is ${21.2 \magnitude}$ \citep{Bowens2021}, which is several orders-of-magnitude deeper than current NaCo limits (${\sim16-17 \magnitude}$, \citealt{Launhardt2020}). Taken together, many of our predicted planets should be observable by upcoming instrumentation.

Table \ref{tab: bestObsCandidates} shows all ALMA and \textit{Herschel} systems where, if the discs are each sculpted by one planet, then that planet would be more massive than ${0.5\mJup}$ (to ${1\sigma}$). Many of these systems would be good targets for future planet searches. In particular, we can be reasonably confident that large planets are interacting with the discs of ${49 \rm \; Cet}$ (${\HD9672}$), ${\HD121617}$, ${\HD131835}$ and ${\HD138813}$; these all require large sculpting planets (${\gtrsim 1\mJup}$ if one planet, ${\gtrsim0.1\mJup}$ if multiple planets), are unlikely to have disc masses sufficient for self-stirring, and would need ${\gtrsim 0.1 \mJup}$ planets for planet-stirring. These systems are also all young, and the discs of ${\HD121617}$ and ${\HD138813}$ have low-to-moderate inclinations, making these good candidates for future observations. Note, however, that these systems all have gas detected, so the grain dynamics could also be more complicated than assumed in this paper (e.g. \citealt{Pearce2020}). Another promising target is ${\HR8799}$ (${\HD218396}$), which is young, has low inclination and requires a well-separated, ${\gtrsim0.1\mJup}$ planet to sculpt its disc (in addition to the four known planets; see Sect. \ref{systemNotesSec: HR8799}).

\begin{table*}
\caption{Systems where, if the debris disc inner edge is sculpted by a single planet, then that planet must be more massive than ${0.5\mJup}$ (to ${1\sigma}$). These systems would make good targets for future planet searches. The best targets require high planet masses to sculpt and stir the discs ($M_{\rm p}$, $M_{\rm p,n}$, $M_{\rm p,stir}$), require planet orbits that are well-separated from the star at their maximum elongation ($Q_{\rm p}$), would require potentially unfeasible disc masses to self stir ($M_{\rm disc,stir}$), are young ($t_*$), and whose discs have low sky-plane inclinations ($i_{\rm disc}$). The systems are ordered by the mass of the single planet required to sculpt the disc inner edge, $M_{\rm p}$. Only systems with resolved discs are shown; all are resolved by ALMA except ${\HD188228}$ and ${\HD192425}$, which are resolved by \textit{Herschel}. Uncertainties are ${1\sigma}$.}
\begin{tabular}{r r r r r r r r r}
\hline
\multicolumn{1}{c}{System} & \multicolumn{1}{c}{$M_{\rm p}$ / $\rm M_{Jup}$} & \multicolumn{1}{c}{$Q_{\rm p}$ / $\rm arcsec$} & \multicolumn{1}{c}{$M_{\rm p, n}$ / $\rm M_{Jup}$} & \multicolumn{1}{c}{$M_{\rm p, stir}$ / $\rm M_{Jup}$} & \multicolumn{1}{c}{$M_{\rm disc, stir}$ / $\rm M_\oplus$} & \multicolumn{1}{c}{$M_*$ / $\rm M_\odot$} & \multicolumn{1}{c}{$t_*$ / Myr} & \multicolumn{1}{c}{$i_{\rm disc}$ / $^\circ$} \\
\hline
HD 138813 & $2.8\pm 0.5$ & $0.30^{+0.05}_{-0.04}$ & $0.4\pm 0.2$ & $3\pm 2$ & $4000\pm 1000$ & $2.15^{+0.06}_{-0.07}$ & $10\pm 3$ & $45^{+5}_{-6}$ \\
HD 218396 & $2.5^{+0.3}_{-0.4}$ & $3.0^{+0.4}_{-0.6}$ & $0.4^{+0.1}_{-0.2}$ & $0.05^{+0.04}_{-0.03}$ & $5000\pm 1000$ & $1.59\pm 0.02$ & $42^{+6}_{-4}$ & $31\pm 3$ \\
HD 38206 & $2.3\pm 0.3$ & $1.5\pm 0.3$ & $0.18^{+0.07}_{-0.08}$ & $0.08\pm 0.09$ & $2200^{+900}_{-1000}$ & $2.36\pm 0.02$ & $42^{+6}_{-4}$ & $83\pm 1$ \\
HD 146897 & $2.0\pm 0.4$ & $0.33^{+0.06}_{-0.07}$ & $0.3^{+0.1}_{-0.2}$ & $0.06\pm 0.07$ & $1000\pm 500$ & $1.32\pm 0.03$ & $10\pm 3$ & $89\pm 5$ \\
HD 121617 & $1.9\pm 0.3$ & $0.32\pm 0.04$ & $0.19\pm 0.06$ & $0.3\pm 0.2$ & $900\pm 200$ & $1.90^{+0.06}_{-0.05}$ & $16\pm 3$ & $37^{+7}_{-10}$ \\
HD 146181 & $1.8^{+0.3}_{-0.4}$ & $0.42^{+0.08}_{-0.1}$ & $0.28^{+0.09}_{-0.1}$ & $0.04\pm 0.05$ & $1100^{+500}_{-600}$ & $1.28^{+0.09}_{-0.08}$ & $16\pm 2$ & $79\pm 7$ \\
HD 181327 & $1.6\pm 0.1$ & $1.120\pm 0.009$ & $0.20\pm 0.03$ & $0.024\pm 0.008$ & $750\pm 70$ & $1.41\pm 0.02$ & $24\pm 3$ & $30.0\pm 0.5$ \\
HD 131835 & $1.6\pm 0.1$ & $0.23\pm 0.01$ & $0.13\pm 0.02$ & $1.2\pm 0.5$ & $1500\pm 100$ & $1.81^{+0.05}_{-0.04}$ & $16\pm 2$ & $73.7\pm 0.4$ \\
HD 111520 & $1.5\pm 0.3$ & $0.35^{+0.09}_{-0.08}$ & $0.18^{+0.09}_{-0.08}$ & $0.2\pm 0.2$ & $1000\pm 400$ & $1.32^{+0.1}_{-0.07}$ & $15\pm 3$ & $84^{+4}_{-5}$ \\
HD 21997 & $1.45^{+0.09}_{-0.1}$ & $0.74\pm 0.04$ & $0.11\pm 0.02$ & $0.07\pm 0.03$ & $580^{+60}_{-80}$ & $1.83\pm 0.02$ & $42^{+6}_{-4}$ & $34^{+3}_{-4}$ \\
HD 9672 & $1.4\pm 0.1$ & $0.84^{+0.05}_{-0.06}$ & $0.10\pm 0.02$ & $1.2^{+0.5}_{-0.4}$ & $2300\pm 200$ & $1.98\pm 0.01$ & $45\pm 5$ & $79.1\pm 0.4$ \\
HD 188228 & $1.2\pm 0.6$ & $1.0\pm 0.7$ & $0.06\pm 0.07$ & $1\pm 3$ & $100\pm 100$ & $2.34\pm 0.02$ & $45\pm 5$ & $50\pm 20$ \\
HD 121191 & $1.1\pm 0.2$ & $0.16\pm 0.04$ & $0.08\pm 0.03$ & $0.5\pm 0.4$ & $260^{+60}_{-70}$ & $1.63^{+0.02}_{-0.04}$ & $16\pm 3$ & $40^{+8}_{-10}$ \\
CPD-72 2713 & $1.1^{+0.3}_{-0.2}$ & $1.7^{+0.6}_{-0.5}$ & $0.18^{+0.1}_{-0.09}$ & $0.3^{+0.4}_{-0.5}$ & $6000\pm 3000$ & $0.80\pm 0.02$ & $24\pm 3$ & $40^{+10}_{-20}$ \\
HD 191089 & $0.89^{+0.07}_{-0.06}$ & $0.56\pm 0.03$ & $0.06\pm 0.01$ & $0.017\pm 0.006$ & $110\pm 10$ & $1.20^{+0.06}_{-0.03}$ & $24\pm 3$ & $60\pm 1$ \\
HD 15115 & $0.85\pm 0.09$ & $0.70\pm 0.09$ & $0.05\pm 0.01$ & $0.07\pm 0.04$ & $290\pm 30$ & $1.40\pm 0.02$ & $45\pm 4$ & $86.2\pm 0.5$ \\
HD 61005 & $0.70^{+0.05}_{-0.04}$ & $1.12\pm 0.02$ & $0.053\pm 0.008$ & $0.03\pm 0.01$ & $520^{+40}_{-50}$ & $0.92^{+0.05}_{-0.02}$ & $45\pm 5$ & $85.7\pm 0.2$ \\
HD 192425 & $0.7\pm 0.2$ & $1.6\pm 0.5$ & $0.02\pm 0.01$ & $0.8\pm 1.0$ & $1500\pm 700$ & $1.94\pm 0.04$ & $400\pm 100$ & $71\pm 5$ \\
\hline
\end{tabular}
\label{tab: bestObsCandidates}
\end{table*}

Fig. \ref{fig: futureObsSystems} shows our predicted minimum-mass planets if the discs are each sculpted by a single planet. If we instead consider our more conservative, multi-planet model (Sect. \ref{subsec: multiplePlanets}) then the sculpting planets could be less massive by an order-of-magnitude or more, whilst their projected maximum separations would be similar to those plotted. Some of these would still be in the parameter space probed by upcoming instruments, but some would not be. Non-detections of the above planets would therefore not necessarily rule out planet sculpting in some cases, but could potentially differentiate between the scenario where one planet dominates the truncation from that where sculpting is performed by multiple planets in tightly packed arrangements (or scenarios where sculpting was performed by a massive planet which has since migrated inwards). Nonetheless, the smaller planets predicted by the multi-planet sculpting model are conservative lower bounds on the required planet masses for disc sculpting, and in many cases would be incapable of stirring the discs (Sect. \ref{subsec: planetStirring}); in practice, we would therefore expect upcoming instruments such as \textit{JWST} to detect planets in many of our systems.

%--------------------------------------------------------------------------------------------------
\subsection{\textit{Caveats}}
\label{subsec: caveats}

Finally, we discuss several \textit{caveats} that must be considered when interpreting our results: the potential effects of non-zero disc mass (Sect. \ref{subsec: discMassDiscussion}), \textit{Herschel}'s spatial resolution (Sect. \ref{subsec: herschelDiscussion}) and close binaries (Sect. \ref{subsec: binaryDiscussion}) on our planet predictions.

%--------------------------------------------------------------------------------------------------
\subsubsection{Omission of disc mass in truncation and planet-stirring models}
\label{subsec: discMassDiscussion}

Whilst disc mass is explicitly included in our self-stirring model (Sect. \ref{subsec: selfStirring}), it is omitted from our truncation and planet-stirring models (Sects. \ref{sec: discTruncation} and \ref{subsec: planetStirring}). These models therefore implicitly assume that the mass of the planet(s) is significantly greater than that of the disc. We assume this because debris disc masses are fundamentally unknown (e.g. \citealt{Krivov2021}), and because the lower limits on perturbing-planet mass arise if the disc has zero mass (i.e. the planet need not be massive enough to overcome disc gravity). In reality, planets interacting with a massive debris disc would migrate as they scattered particles \citep{Ida2000, Kirsh2009, Pearce2014, Pearce2015HD107146}, and disc self-gravity could allow debris to resist deformation and stirring (L\"{o}hne et al., in prep.). The result would be that truncating planets could clear a significantly wider region (and hence be located much further inwards) than our massless-disc models imply \citep{Friebe2022}, or that the planets required to sculpt and stir discs could be more massive than our predictions.

%--------------------------------------------------------------------------------------------------
\subsubsection{Limitations of \textit{Herschel} vs. ALMA data}
\label{subsec: herschelDiscussion}

The disc widths derived from \textit{Herschel} PACS are systematically larger than those from ALMA, as expected from the lower resolution of the former. This, combined with our neglecting any additional inner emission in the  \textit{Herschel} fitting (Sect. \ref{subsec: herschelSystems}), implies that the inner edges of \textit{Herschel} discs are likely underestimated. This would lead to the masses and locations of sculpting planets also being underestimated, as appears to be the case from Fig. \ref{fig: truncatingPlanetsMassVsSemimajorAxis}, where the planet predictions for \textit{Herschel} systems are offset from those for ALMA systems. Similarly, underestimating the disc inner edges would both increase the predicted planet masses needed for planet-stirring, and (for wide discs) decrease the disc masses required for self-stirring. Both of these effects appear to occur, as evidenced by the distributions of ALMA and \textit{Herschel} points on the bottom plot of Fig. \ref{fig: planetStirringAndSelfStirring}. Whilst observational biases are also present (the largest, brightest discs are likely to be those selected for ALMA observations, and these would be expected to need large masses if they were to self-stir), the reader should be aware of the above \textit{caveats} when considering the dynamical predictions for \textit{Herschel} systems.

%--------------------------------------------------------------------------------------------------
\subsubsection{Effects of close binaries on predictions}
\label{subsec: binaryDiscussion}

Several of our systems are possible close binaries, which could impact our dynamical predictions in several ways. Our measured disc parameters should be unaffected by binarity, because dust emission is determined by temperature, which is set by the stellar luminosity; since our luminosities correspond to the \textit{combined} star luminosities in binary systems, the dust masses and locations that we use should not depend on binarity. However, the effective central mass is affected by close binarity, which in turn affects our dynamical calculations; for such systems, both the discs and our predicted planets are assumed to be circumbinary, and so the star masses $M_*$ used in our equations should be the combined binary mass. Close binarity could also affect our stellar mass estimates, as well as our ages for field stars, which would in turn affect our derived planet masses. Finally, potentially the largest effect is that some binaries could sculpt and/or stir their debris discs, removing the need for planets in those systems altogether. Whilst it is unlikely that our potential binaries could sculpt their debris discs, it is possible for some to have stirred them (e.g. ${\beta \; \rm Tri}$; ${\HD13161}$).

%%%%%%%%%%%%%%%%%%%%%%%%%%%%%%%%%%%%%%%%%%%%%%%%%%%%%%%%%%%%%%%%%%%%%%%%%%%%%%%%%%%%%%%%%%%%%%%%%%%

\section{Conclusions}
\label{sec: conclusions}

We analysed 178 debris-disc systems in the ISPY, LEECH and LIStEN planet searches, using dynamical arguments to constrain the properties of unseen planets from disc morphologies. We show that a `typical' cold debris disc requires a Neptune- to Saturn-mass perturbing planet located at ${10-100\au}$, with many needing Jupiter-mass perturbers. All of our discs could be both sculpted and stirred by either known companions or unseen planets, which is consistent with current non-detections of planets in many of these systems. Our predicted stirring planets are often consistent with those required to sculpt the discs, suggesting that in many systems a single planet may be responsible for both sculpting and stirring debris. Conversely, we show that many discs cannot be self-stirred without requiring unreasonable disc masses. Our predicted planet populations are quite different from those currently detected, both in the Solar System and beyond, but they are very similar to those inferred to be forming in protoplanetary discs; this could imply that these are the same planet population if the forming planets do not undergo significant migration. For systems with known planets, few could be sculpting their associated discs \textit{in situ}, although many could have done so in the past and since migrated inwards. Conversely, many of the known planets could be stirring their associated discs \textit{in situ}.  Finally, we provide a catalogue of planet predictions for our 178 systems, show that \textit{JWST} should be able to find many of these planets, and identify the most promising targets for future planet searches.

%%%%%%%%%%%%%%%%%%%%%%%%%%%%%%%%%%%%%%%%%%%%%%%%%%%%%%%%%%%%%%%%%%%%%%%%%%%%%%%%%%%%%%%%%%%%%%%%%%%

\begin{acknowledgements}

We thank Luca Matr{\`a} for sharing unpublished REASONS modelling results, Anne-Lise Maire and \'{E}lodie Choquet for their $\HR8799$ and ${\rm 49 \; Cet}$ planet-detection limits respectively, Virginie Faramaz for the ${\HD202628}$ ALMA image, and Johan Olofsson for helpful comments. We also thank Andr\'{e} M\"{u}ller and Ingo Stilz, whose previous efforts made this work possible, and Nicole Pawellek and Sebasti\'{a}n Marino for useful discussions. We are grateful to the anonymous referee, whose comments and suggestions improved the paper. TDP, MB and AVK are supported by the Deutsche Forschungsgemeinschaft (DFG) grants \mbox{Kr 2164/13-2}, \mbox{Kr 2164/14-2}, and \mbox{Kr 2164/15-2}. GMK is supported by the Royal Society as a Royal Society University Research Fellow. GC thanks the Swiss National Science Foundation for financial support under grant number 200021\texttt{\_}169131. TH acknowledges support from the European Research Council under the Horizon 2020 Framework Program via the ERC Advanced Grant Origins \mbox{83 24 28}. ECM acknowledges support from the Swiss National Science Foundation (SNSF); this work has been carried out in part within the framework of the National Centre for Competence in Research PlanetS supported by SNSF. We made use of the Data \& Analysis Center for Exoplanets (DACE) at the University of Geneva (CH), a platform of the Swiss National Centre of Competence in Research (NCCR) PlanetS. The DACE platform is available at \mbox{https://dace.unige.ch}.

\end{acknowledgements}

%%%%%%%%%%%%%%%%%%%%%%%%%%%%%%%%%%%%%%%%%%%%%%%%%%%%%%%%%%%%%%%%%%%%%%%%%%%%%%%%%%%%%%%%%%%%%%%%%%%
\bibliographystyle{aa}
\bibliography{bib_ispyDebsPltCnstrnts}

%%%%%%%%%%%%%%%%%%%%%%%%%%%%%%%%%%%%%%%%%%%%%%%%%%%%%%%%%%%%%%%%%%%%%%%%%%%%%%%%%%%%%%%%%%%%%%%%%%%

\begin{appendix}

%--------------------------------------------------------------------------------------------------

\section{System data}
\label{app: systemData}

Table \ref{tab: systemObsData} lists the parameters that we use for each system, and Table \ref{tab: systemAnalysisData} gives the planet and disc constraints that we calculate.

\onecolumn
\begin{landscape}

\begin{longtable}{r r r r r r r r r r}
\caption{\label{tab: systemObsData} System parameters used in this paper, with ${1\sigma}$ uncertainties. Systems where our stellar parameter estimates may be affected by close binarity are marked $^*$ (Sect. \ref{subsec: stellarParameters}). The `Disc location and extent' column describes the location and shape of the debris disc inner and outer edges: if the disc is resolved and fitted with an asymmetric model, then the column shows the inner edge pericentre $q_{\rm i}$, inner edge apocentre $Q_{\rm i}$, outer edge pericentre $q_{\rm o}$ and outer edge apocentre $Q_{\rm o}$ as `${q_{\rm i}, Q_{\rm i} \ \rightarrow q_{\rm o}, Q_{\rm o}}$'. Alternatively, if the disc is resolved and fitted with an axisymmetric model, then the column shows the disc inner edge $a_{\rm i}$ and outer edge $a_{\rm o}$ as `${a_{\rm i} \rightarrow a_{\rm o}}$'. Finally, if the disc location is estimated from SED data, then only the corrected blackbody radius is shown. Note that if multiple discs are present, then this column refers to the \textit{outermost} disc. The `Disc data' column gives the data source used to determine the debris disc properties (plus PACS wavelength for \textit{Herschel} data). `Age refs.' show the references considered in our age assignment; targets with high probabilities of association membership have the association and age reference shown (we also use association ages for seven targets with lower membership probabilities, \mbox{marked ${^\dagger}$)}, `Iso.' means our isochrone fit (performed for all targets, but only quoted if no other valid age estimates were available), and literature references are as follows. \mbox{1: \citet{AguileraGomez2018}}. \mbox{2: \citet{Bell2015}}. \mbox{3: \citet{Brandt2015}}. \mbox{4: \citet{Casagrande2011}}. \mbox{5: \citet{Chen2014}}. \mbox{6: \citet{David2015}}. \mbox{7: \citet{Dobbie2010}}. \mbox{8: \citet{Eiroa2013}}. \mbox{9: \citet{Elliott2016}}. \mbox{10: \citet{Esposito2020}}. \mbox{11: \citet{Faramaz2019}}. \mbox{12: \citet{Gaspar2013}}. \mbox{13: \citet{Gaspar2016}}. \mbox{14: \citet{Holland2017}}. \mbox{15: \citet{Jones2015}}. \mbox{16: \citet{Kains2011}}. \mbox{17: \citet{Lestrade2012}}. \mbox{18: \citet{MacGregor2018}}. \mbox{19: \citet{Mamajek2008}}. \mbox{20: \citet{Mamajek2013}}. \mbox{21: \citet{Moor2009}}. \mbox{22: \citet{Moor2016}}. \mbox{23: \citet{MoroMartin2015}}. \mbox{24: \citet{Murphy2015}}. \mbox{25: \citet{Nielsen2013}}. \mbox{26: \citet{Nielsen2019}}. \mbox{27: \citet{Nissen2020}}. \mbox{28: \citet{Pecaut2016}}. \mbox{29: \citet{Ramirez2012}}. \mbox{30: \citet{Rieke2005}}. \mbox{31: \citet{Sierchio2014}}. \mbox{32: \citet{StanfordMoore2020}}. \mbox{33: \citet{Stone2018}}. \mbox{34: \citet{Veyette2018}}. \mbox{35: \citet{Vican2012}}. \mbox{36: \citet{Weise2010}}. \mbox{37: \citet{Zuckerman2006}}. \mbox{38: \citet{Zuckerman2011}}. \mbox{39: \citet{Zuckerman2019}}.}\\
\hline\hline
\multicolumn{1}{c}{System} & \multicolumn{1}{c}{Name} & \multicolumn{1}{c}{$d$ / $\rm pc$} & \multicolumn{1}{c}{$\log_{10}(L_*$ / $\rm L_\odot)$} & \multicolumn{1}{c}{$T_{\rm eff}$ / ${\rm K}$} & \multicolumn{1}{c}{$M_*$ / $\rm M_\odot$} & \multicolumn{1}{c}{$t_*$ / Myr} & \multicolumn{1}{c}{Disc location and extent / au} & \multicolumn{1}{c}{Disc data} & \multicolumn{1}{c}{Age refs.} \\
\hline
\endfirsthead
\caption{continued.}\\
\hline\hline
\multicolumn{1}{c}{System} & \multicolumn{1}{c}{Name} & \multicolumn{1}{c}{$d$ / $\rm pc$} & \multicolumn{1}{c}{$\log_{10}(L_*$ / $\rm L_\odot)$} & \multicolumn{1}{c}{$T_{\rm eff}$ / ${\rm K}$} & \multicolumn{1}{c}{$M_*$ / $\rm M_\odot$} & \multicolumn{1}{c}{$t_*$ / Myr} & \multicolumn{1}{c}{Disc location and extent / au} & \multicolumn{1}{c}{Disc data} & \multicolumn{1}{c}{Age refs.} \\
\hline
\endhead
\hline
\endfoot

CD-52 2472{$^*$} & \multicolumn{1}{c}{-} & $152.6\pm 0.6$ & $0.196\pm 0.006$ & $5740\pm 70$ & $1.06^{+0.07}_{-0.1}$ & $50^{+50}_{-10}$ & $11\pm 3$ & \multicolumn{1}{l}{\phantom{$^*$}SED} & \multicolumn{1}{c}{5} \\
CPD-72 2713\phantom{$^*$} & \multicolumn{1}{c}{-} & $36.62\pm 0.03$ & --$0.730\pm 0.005$ & $3910\pm 70$ & $0.80\pm 0.02$ & $24\pm 3$ & $80^{+30}_{-20}$ $\rightarrow$ $180\pm 30$ & \multicolumn{1}{l}{\phantom{$^*$}ALMA} & \multicolumn{1}{c}{BPMG, 2} \\
GJ 581{$^*$} & \multicolumn{1}{c}{-} & $6.30\pm 0.01$ & --$1.919\pm 0.009$ & $3220\pm 70$ & $0.25^{+0.06}_{-0.03}$ & $5000\pm 3000$ & $16\pm 7$ & \multicolumn{1}{l}{\phantom{$^*$}SED} & \multicolumn{1}{c}{17} \\
GJ 649\phantom{$^*$} & \multicolumn{1}{c}{-} & $10.380\pm 0.003$ & --$1.345\pm 0.003$ & $3600\pm 70$ & $0.40\pm 0.06$ & $4000\pm 2000$ & $30\pm 20$ & \multicolumn{1}{l}{\phantom{$^*$}SED} & \multicolumn{1}{c}{34} \\
HD 166\phantom{$^*$} & \multicolumn{1}{c}{-} & $13.77\pm 0.01$ & --$0.198\pm 0.003$ & $5480\pm 70$ & $0.99^{+0.01}_{-0.02}$ & $200^{+200}_{-100}$ & $10\pm 10$ $\rightarrow$ $40\pm 20$ & \multicolumn{1}{l}{\phantom{$^*$}\textit{Herschel} $70 \um$} & \multicolumn{1}{c}{12, 19, 29, 32, 33, 35} \\
HD 203\phantom{$^*$} & \multicolumn{1}{c}{-} & $39.9\pm 0.1$ & $0.630\pm 0.003$ & $6840\pm 70$ & $1.43\pm 0.01$ & $24\pm 3$ & $29\pm 6$ & \multicolumn{1}{l}{\phantom{$^*$}SED} & \multicolumn{1}{c}{BPMG, 2} \\
HD 377\phantom{$^*$} & \multicolumn{1}{c}{-} & $38.48\pm 0.09$ & $0.068\pm 0.004$ & $5780\pm 70$ & $1.07\pm 0.05$ & $220\pm 50$ & $60\pm 10$ & \multicolumn{1}{l}{\phantom{$^*$}SED} & \multicolumn{1}{c}{10, 31, 32} \\
HD 870\phantom{$^*$} & \multicolumn{1}{c}{-} & $20.62\pm 0.01$ & --$0.296\pm 0.003$ & $5360\pm 70$ & $0.87^{+0.02}_{-0.04}$ & $3000\pm 1000$ & $60\pm 10$ & \multicolumn{1}{l}{\phantom{$^*$}SED} & \multicolumn{1}{c}{5, 31, 32} \\
\end{longtable}

\begin{longtable}{r r r r r r r r r r}
\caption{\label{tab: systemAnalysisData} Planet and debris disc constraints calculated in this paper, with columns as follows. $M_{\rm p}$, $a_{\rm p}$ and \mbox{$e_{\rm p}$: minimum} mass, maximum semimajor axis and minimum eccentricity respectively that a single planet requires to sculpt the disc inner edge, as derived in Sect. \ref{subsec: singlePlanet} (a dash for $e_{\rm p}$ means the planet is not required to be eccentric because the disc model is axisymmetric). \mbox{$Q_{\rm p}$: maximum} apocentre distance (in arcsec) that a single sculpting planet can have. \mbox{$M_{\rm p, n}$: minimum} mass that each individual planet requires if instead multiple planets sculpt the disc inner edge (Sect. \ref{subsec: multiplePlanets}). \mbox{$r_{\rm p, n}$: maximum} projected separation (in arcsec) of the outermost planet in the multi-planet scenario. \mbox{$M_{\rm p, stir}$: minimum} mass that a planet requires to stir the disc, assuming the planet eccentricity \mbox{is ${\leq0.3}$} (Sect. \ref{subsec: planetStirring}). \mbox{$M_{\rm disc, stir}$: minimum} mass that the disc requires to self-stir, as derived in Sect. \ref{subsec: selfStirring} (a dash means this value could not be calculated because no disc width information was available). Uncertainties are ${1\sigma}$.}\\
\hline\hline
\multicolumn{1}{c}{System} & \multicolumn{1}{c}{Name} & \multicolumn{1}{c}{$M_{\rm p}$ / $\rm M_{Jup}$} & \multicolumn{1}{c}{$a_{\rm p}$ / $\rm au$} & \multicolumn{1}{c}{$e_{\rm p}$} & \multicolumn{1}{c}{$Q_{\rm p}$ / $\rm arcsec$} & \multicolumn{1}{c}{$M_{\rm p, n}$ / $\rm M_{Jup}$} & \multicolumn{1}{c}{$r_{\rm p, n}$ / $\rm arcsec$} & \multicolumn{1}{c}{$M_{\rm p, stir}$ / $\rm M_{Jup}$} & \multicolumn{1}{c}{$M_{\rm disc, stir}$ / $\rm M_\oplus$} \\
\hline
\endfirsthead
\caption{continued.}\\
\hline\hline
\multicolumn{1}{c}{System} & \multicolumn{1}{c}{Name} & \multicolumn{1}{c}{$M_{\rm p}$ / $\rm M_{Jup}$} & \multicolumn{1}{c}{$a_{\rm p}$ / $\rm au$} & \multicolumn{1}{c}{$e_{\rm p}$} & \multicolumn{1}{c}{$Q_{\rm p}$ / $\rm arcsec$} & \multicolumn{1}{c}{$M_{\rm p, n}$ / $\rm M_{Jup}$} & \multicolumn{1}{c}{$r_{\rm p, n}$ / $\rm arcsec$} & \multicolumn{1}{c}{$M_{\rm p, stir}$ / $\rm M_{Jup}$} & \multicolumn{1}{c}{$M_{\rm disc, stir}$ / $\rm M_\oplus$} \\
\hline
\endhead
\hline
\endfoot

CD-52 2472 & \multicolumn{1}{c}{-} & $0.25^{+0.06}_{-0.1}$ & $9\pm 3$ & \multicolumn{1}{c}{-} & $0.06\pm 0.02$ & $(5^{+2}_{-5}) \times 10^{-3}$ & $0.07\pm 0.02$ & $(2^{+4}_{-5}) \times 10^{-4}$ & \multicolumn{1}{c}{-} \\
CPD-72 2713 & \multicolumn{1}{c}{-} & $1.1^{+0.3}_{-0.2}$ & $60\pm 20$ & \multicolumn{1}{c}{-} & $1.7^{+0.6}_{-0.5}$ & $0.18^{+0.1}_{-0.09}$ & $2.3^{+1.0}_{-0.7}$ & $0.3^{+0.4}_{-0.5}$ & $6000\pm 3000$ \\
GJ 581 & \multicolumn{1}{c}{-} & $0.012^{+0.006}_{-0.005}$ & $14\pm 6$ & \multicolumn{1}{c}{-} & $2\pm 1$ & $(4\pm 4) \times 10^{-5}$ & $3\pm 1$ & $(2\pm 5) \times 10^{-6}$ & \multicolumn{1}{c}{-} \\
GJ 649 & \multicolumn{1}{c}{-} & $0.03\pm 0.02$ & $30\pm 20$ & \multicolumn{1}{c}{-} & $3\pm 2$ & $(2\pm 2) \times 10^{-4}$ & $3\pm 2$ & $(8\pm 30) \times 10^{-6}$ & \multicolumn{1}{c}{-} \\
HD 166 & \multicolumn{1}{c}{-} & $0.12^{+0.07}_{-0.08}$ & $10\pm 8$ & \multicolumn{1}{c}{-} & $0.7\pm 0.6$ & $(1^{+1}_{-2}) \times 10^{-3}$ & $0.9\pm 0.7$ & $0.02\pm 0.05$ & $1\pm 1$ \\
HD 203 & \multicolumn{1}{c}{-} & $0.8\pm 0.1$ & $22\pm 4$ & \multicolumn{1}{c}{-} & $0.6\pm 0.1$ & $0.05\pm 0.02$ & $0.7\pm 0.1$ & $(2\pm 3) \times 10^{-3}$ & \multicolumn{1}{c}{-} \\
HD 377 & \multicolumn{1}{c}{-} & $0.40\pm 0.07$ & $46\pm 8$ & \multicolumn{1}{c}{-} & $1.2\pm 0.2$ & $0.013\pm 0.005$ & $1.5\pm 0.3$ & $(7\pm 7) \times 10^{-4}$ & \multicolumn{1}{c}{-} \\
HD 870 & \multicolumn{1}{c}{-} & $0.10\pm 0.02$ & $50\pm 10$ & \multicolumn{1}{c}{-} & $2.5\pm 0.5$ & $(9\pm 4) \times 10^{-4}$ & $2.9\pm 0.6$ & $(5\pm 6) \times 10^{-5}$ & \multicolumn{1}{c}{-} \\
\end{longtable}

\end{landscape}

\twocolumn

%%%%%%%%%%%%%%%%%%%%%%%%%%%%%%%%%%%%%%%%%%%%%%%%%%%%%%%%%%%%%%%%%%%%%%%%%%%%%%%%%%%%%%%%%%%%%%%%%%%
\section{Notes on specific systems}
\label{app: specificSystems}

This section gives additional notes for some systems. We mainly discuss dynamical considerations that may affect our system-specific planet predictions, such as the presence of known planets or companions.

%---------------------------------------------------------
\subsection{${\rm CPD \shorthyphen 72 \; 2713}$}
ALMA data from \cite{Moor2020}, disc model from REASONS (Matr\`{a} et al., in prep.). \cite{Moor2020} argue that the disc can can be self-stirred, using \textit{Herschel} data and the \cite{Krivov2018} method with ${S_{\rm max}=200\km}$ and ${v_{\rm frag} = 30 \mPerS}$. However, our analysis with ALMA data and self-consistent $S_{\rm max}$ and $v_{\rm frag}$ values shows that it cannot be, unless the disc mass is unfeasibly high.

%---------------------------------------------------------
\subsection{${\rm GJ \; 581}$}
Neptunes and super-Earths very close to the star (${<0.1\au}$: \citealt{Bonfils2005, Udry2007, Mayor2009, Robertson2014}). Outermost planet at ${0.074\pm 0.001 \au}$ with ${\mSinI = 5.7\pm0.4 \mEarth}$; Fig. \ref{fig: detectedAndPredictedPlanets} shows that this planet cannot be sculpting or stirring the disc \textit{in situ}.

%---------------------------------------------------------
\subsection{${\rm GJ \; 649}$}
Eccentric planet with ${\mSinI=0.33\pm0.03 \mJup}$ detected at ${1.14\pm0.04 \au}$ \citep{Johnson2010}, although its eccentricity is uncertain \citep{Wittenmyer2013}. Figure \ref{fig: detectedAndPredictedPlanets} shows that the planet could be stirring the disc, and is massive enough to have sculpted it (but not \textit{in situ}).

%---------------------------------------------------------
\subsection{${\HD1461}$}
Probably two super-Earths inside ${0.1 \au}$ \citep{Rivera2010, Diaz2016}. Outermost planet at ${0.112\pm0.004\au}$ with ${\mSinI = 5.6 \pm 0.7 \mEarth}$; Fig. \ref{fig: detectedAndPredictedPlanets} shows that this planet cannot be sculpting or stirring the disc \textit{in situ}.

%---------------------------------------------------------
\subsection{${\HD3003 \; (\beta^3 \; \rm Tuc)}$}
Multiple system with up to six components \citep{Tokovinin2018}.

%---------------------------------------------------------
\subsection{${\HD9672 \; (49 \; \rm Cet)}$}
ALMA data from \cite{Hughes2017}, disc model from REASONS (Matr\`{a} et al., in prep.). Large CO gas mass coincident with the cold disc, which could complicate grain dynamics \citep{Zuckerman1995, Moor2019}. Tentative evidence of a warp or spiral arms in the gas disc \citep{Hughes2017}. Various planet constraints are shown for this specific system on Fig. \ref{fig: hd9672PltCnstrnts}.

%---------------------------------------------------------
\subsection{${\HD10647 \; (\rm q^1 \; Eri)}$}
ALMA data from \cite{Lovell2021}, showing the disc is asymmetric with a sharp inner edge. \citet{Lovell2021} model this as a symmetric disc with a possible resonant clump, but they also show that a non-resonant, eccentric planet with ${a_{\rm p}e_{\rm p} = 2-4 \au}$ could instead drive the observed asymmetries. If the latter scenario applies, then these planet constraints agree with ours. A planet lies at ${2.0\pm0.2 \au}$ with ${\mSinI = 0.9 \pm 0.2 \mJup}$ and eccentricity ${0.15 \pm 0.08}$ \citep{Butler2006, Marmier2013}, but Fig. \ref{fig: detectedAndPredictedPlanets} shows that this inner planet cannot be sculpting or stirring the disc \textit{in situ}.

%---------------------------------------------------------
\subsection{${\HD10700 \; (\tau \; \rm Cet)}$}
ALMA data from \cite{Macgregor2016}, showing an axisymmetric disc. At least four super-Earths lie within ${1.5 \au}$ \citep{Tuomi2013, Feng2017}, the outermost at ${1.33\pm0.02 \au}$ with ${\mSinI = 4\pm1\mEarth}$ \citep{Feng2017}; Fig. \ref{fig: detectedAndPredictedPlanets} shows that this planet cannot be sculpting or stirring the disc \textit{in situ}. There are possibly more planets present \citep{Dietrich2021}, and a ${1-2\mJup}$ planet is predicted at ${3-20 \au}$ from astrometry \citep{Kervella2019}; this planet \textit{could} sculpt and stir the disc \textit{in situ}. \cite{Lawler2014} show the planetary system would be stable if an additional planet with mass below that of Neptune exists outside ${5\au}$, consistent with the planet(s) we infer.

%---------------------------------------------------------
\subsection{${\HD13161 \; (\beta \; \rm Tri)}$}
Circumbinary disc (binary separation ${0.3 \au}$, eccentricity 0.4; \citealt{Pourbaix2000}). \cite{Kennedy2012} show that the binary could stir the disc. 

%---------------------------------------------------------
\subsection{${\HD15115}$}

ALMA data from \cite{Macgregor2019} showing no significant asymmetries, although the disc has a possible eccentricity of 0.06 in the H-band \citep{Sai2015}. ALMA finds a ${14\pm6\au}$ gap at ${59\pm5\au}$, which \cite{Macgregor2019} predict is carved by an embedded ${0.2\mJup}$ planet at ${58\au}$. If a separate planet sculpts the disc inner edge, then we predict that planet has mass ${\geq 0.85 \pm0.09\mJup}$ and semimajor axis ${\leq34\pm4\au}$ (our inner planet is larger than the gap-clearing planet because we assume a higher degree of debris clearing). Alternatively, if the gap is created through a secular interaction with a precessing planet near the disc inner edge (e.g. \citealt{Pearce2015HD107146, Yelverton2018, Sefilian2021}) then that planet would need a mass of at least ${0.05\mJup}$ (${10\mEarth}$) to carve the gap (i.e. for 10 secular times to have elapsed within the system age). A large asymmetry is seen in scattered light \citep{Kalas2007, Schneider2014}, which does not necessarily imply an asymmetric parent ring \citep{Mazoyer2014}. Possible disc interaction with an M star \citep{Kalas2007}.

%---------------------------------------------------------
\subsection{${\HD17094 \; (\mu \rm \; Cet)}$}
Circumbinary disc, where the binary has separation ${3\au}$ \citep{Trilling2007}. Possibly a quadruple system \citep{Richichi2000}. The binary is unlikely to be sculpting the disc at ${\sim 100\au}$.

%---------------------------------------------------------
\subsection{${\HD17925 \; (\rm EP \; Eri)}$}
Possible close binary (two K-types at ${0.07\au}$ separation; \citealt{Rodriguez2015}), which would make the ${6-25\au}$ disc circumbinary. If so, the binary could stir the disc, but probably not sculpt it.

%---------------------------------------------------------
\subsection{${\HD21997}$}
 ALMA data and disc model from REASONS (Matr\`{a} et al., in prep.). Axisymmetric disc \citep{Kospal2013, Moor2013}. Very high CO gas mass, which could affect grain dynamics. It is questionable whether the disc can be self-stirred, as its mass would have to be ${580^{+60}_{-80} \mEarth}$ (also see \citealt{Moor2015} and \citealt{Krivov2018}).

%---------------------------------------------------------
\subsection{${\HD22049 \; (\epsilon \; \rm Eri)}$}
ALMA data from \cite{Booth2017}, using their Gaussian model. The disc is probably axisymmetric in SMA and \textit{Spitzer} images \citep{Macgregor2015, Backman2009}, although there may be low-significance clumps. A ${0.66^{+0.12}_{-0.09} \mJup}$ planet lies much closer in at ${3.52\pm0.04 \au}$, which may be significantly inclined relative to the disc \citep{LlopSayson2021}. Figure \ref{fig: detectedAndPredictedPlanets} shows this planet could be stirring the disc, and is massive enough to have sculpted it (but not \textit{in situ}). \cite{Booth2017} predict a single sculpting planet at ${>45\au}$ with mass ${\leq1.3\mJup}$, consistent with our prediction of a ${\geq0.19\pm0.04\mJup}$ planet at ${\leq52.7^{+0.9}_{-1.0}\au}$. They also show that a planet at ${48\au}$ could be in 3:2 and 2:1 resonance with the disc inner and outer edges, respectively, which is also consistent with our predicted planet.

%---------------------------------------------------------
\subsection{${\HD27290 \; (\gamma \; \rm Dor)}$}
Our \textit{Herschel} model fits a single broad disc spanning ${14 \pm 7}$ to ${280 \pm 40 \au}$, but modelling by \cite{BroekhovenFiene2013} shows that the system could be fitted by a warm ring at several au plus a cool disc spanning 55 to ${400\au}$, or two narrow rings at 70 and ${190 \au}$, or any configuration between.

%---------------------------------------------------------
\subsection{${\HD29391 \; (51 \; \rm Eri)}$}
Directly imaged object with a mass of ${2-12 \mJup}$ (depending on whether hot or cold start models used) detected with semimajor axis ${11^{+4}_{-1} \au}$, and possibly a non-zero eccentricity \citep{Macintosh2015, DeRosa2020}. Figure \ref{fig: detectedAndPredictedPlanets} shows that this object could be both sculpting and stirring the disc \textit{in situ}.

%---------------------------------------------------------
\subsection{${\HD32297}$}
ALMA data from \cite{MacGregor2018}, disc model from REASONS (Matr\`{a} et al., in prep.). Possibly some unresolved, non-axisymmetric thermal structure. A swept-back structure is visible in scattered light, similar to `The Moth' ($\HD61005$; \citealt{Schneider2014}). There is also a large CO gas mass \citep{Moor2019}, which could affect grain dynamics.

%---------------------------------------------------------
\subsection{${\HD35850 \; (\rm AF \; Lep)}$}
Close binary, with separation ${0.021\au}$ \citep{Rodriguez2012}; the binary is unlikely to be sculpting the disc at ${\sim 50\au}$.

%---------------------------------------------------------
\subsection{${\HD38206}$}
ALMA data from \cite{Booth2021}, showing a possibly asymmetric disc (eccentricity ${0.25^{+0.10}_{-0.09}}$). If a single planet both stirs and sculpts the disc, then \cite{Booth2021} predict it to have mass ${0.7^{+0.5}_{-0.3} \mJup}$, semimajor axis ${76^{+12}_{-13} \au}$ and eccentricity ${0.34^{+0.20}_{-0.13}}$; our predicted planet is at a similar location but is three times more massive, because \cite{Booth2021} use the shorter multi-planet sculpting timescale from \cite{Shannon2016} rather than the single-planet timescale from \cite{Pearce2014}.

%---------------------------------------------------------
\subsection{${\HD48682 \; (\psi^5 \; \rm Aur)}$}
Potential brightness asymmetries in \textit{Herschel} images \citep{Hengst2020}.

%---------------------------------------------------------
\subsection{${\HD50554}$}
A large, eccentric companion is present with ${\mSinI = 5 \pm 3 \mJup}$, semimajor axis ${2 \pm 1 \au}$, and eccentricity ${\sim 0.5}$ \citep{Fischer2002, Perrier2003}. Figure \ref{fig: detectedAndPredictedPlanets} shows this planet could be stirring the disc, and is massive enough to have sculpted it (but not \textit{in situ}).

%---------------------------------------------------------
\subsection{${\HD53143}$}
A potential asymmetry is visible in scattered light (\citealt{Schneider2014}, although also see \citealt{Kalas2006}). Possible clumps may be visible in micron grains, and the disc is highly asymmetric in unpublished ALMA data \citep{Stark2020}.

%---------------------------------------------------------
\subsection{${\HD61005 \; \rm ({\normalfont `}The \; Moth{\normalfont \text{'}})}$}
ALMA data from \cite{MacGregor2018}, disc model from REASONS (Matr\`{a} et al., in prep.). Very asymmetric in scattered light \citep{Schneider2014, Olofsson2016}. No significant asymmetries in ALMA, but hints at unresolved asymmetries. \cite{Esposito2016} show the structure could be caused by inclined, eccentric planet, predicting a planet with eccentricity ${\sim 0.2}$ and semimajor axis ${<50\au}$, with many possible masses. Our predicted sculpting planet with semimajor axis ${41.0^{+0.6}_{-0.5}\au}$ and mass ${0.70^{+0.05}_{-0.04}\mJup}$ is consistent with this.

%---------------------------------------------------------
\subsection{${\HD69830}$}
\label{systemNotesSec: HD69830}
The disc is close to the star. Three Neptunes exist inside ${1 \au}$, with ${\mSinI = 10-20 \mEarth}$ (${0.03-0.06 \mJup}$; \citealt{Lovis2006}). The outermost planet has semimajor axis ${0.6\pm0.3 \au}$ and ${\mSinI = 18\pm9 \mJup}$ (we assume 50 per cent uncertainties); Fig. \ref{fig: detectedAndPredictedPlanets} shows that this planet could be both sculpting and stirring the disc \textit{in situ}. However, the system dynamics may be complicated; the planets may even have migrated \textit{through} the disc \citep{Payne2009}.

%---------------------------------------------------------
\subsection{${\HD92945}$}
ALMA data from \cite{Marino2019}, disc model from REASONS (Matr\`{a} et al., in prep.). Gap at $73\au$, and small asymmetries seen (possibly due to a background galaxy). If the gap is carved by an \textit{in situ} planet, then \cite{Marino2019} argue that that planet has mass ${<0.6\mJup}$. Alternatively, if the gap is created through a secular interaction with planets near the disc inner edge (e.g. \citealt{Pearce2015HD107146, Yelverton2018, Sefilian2021}), then the outermost planet would have a semimajor axis of ${40-45\au}$ and mass of at least ${10\mEarth}$ (${0.03\mJup}$). This is consistent with our prediction of a planet sculpting the disc inner edge with mass ${\geq 0.32\pm0.06\mJup}$ and semimajor axis ${\leq 45^{+1}_{-4}\au}$.
 
%---------------------------------------------------------
\subsection{${\HD102647 \; (\beta \; \rm Leo)}$}
\label{systemNotesSec: HD102647}
Possibly separate hot, warm and cold debris components \citep{Churcher2011HD102647, Defrere2021}. \cite{Churcher2011HD102647} argue for either three discs at 2, 9 and ${30-70\au}$, or two discs at 2 and ${15-70\au}$. Our \textit{Herschel} model assumes only a single broad disc (${7-79\au}$), so we may underestimate the outer disc inner edge. A companion lies at ${440\au}$ \citep{Rodriguez2015}.

%---------------------------------------------------------
\subsection{${\HD107146}$}
\label{systemNotesSec: HD107146}
ALMA data from \cite{Marino2018}, disc model from REASONS (Matr\`{a} et al., in prep.). The disc contains a ${39\au}$ wide gap at ${76\au}$ \citep{Ricci2015, Marino2018}. \cite{Marino2018} argue that the gap could be carved by multiple ${10\mEarth}$ (${0.03\mJup}$) planets \textit{in situ}, and Eq. \ref{eq: pltMassToStir} shows that such planets could stir the disc outer edge even with small eccentricities. Alternatively, if the gap is created through a secular interaction with a precessing planet near the disc inner edge (e.g. \citealt{Pearce2015HD107146, Yelverton2018, Sefilian2021}) then that planet would need a mass of at least ${10\mEarth}$ (${0.04\mJup}$) to carve the gap (i.e. for 10 secular times to have elapsed within the system age). If there were several such planets then they could also sculpt the disc inner edge; alternatively, this could be done by a single planet with \mbox{mass ${\geq0.44^{+0.07}_{-0.1}\mJup}$} and semimajor axis ${\leq 41.6^{+1.0}_{-0.7}\au}$.

%---------------------------------------------------------
\subsection{${\HD109085 \; (\eta \; \rm Crv)}$}
ALMA data from \cite{Marino2017_EtaCorvi}, disc model from REASONS (Matr\`{a} et al., in prep.). \cite{Marino2017_EtaCorvi} predict a sculpting planet at ${75-100\au}$ with a mass of ${3-30\mEarth}$ (${0.01-0.1\mJup}$) and eccentricity ${<0.08}$. We predict a ${\geq 0.34^{+0.05}_{-0.1}\mJup}$ sculpting planet at ${\leq104^{+3}_{-2}\au}$; our planet is larger because \cite{Marino2017_EtaCorvi} use the shorter multi-planet sculpting timescale from \cite{Shannon2016} rather than the single-planet timescale from \cite{Pearce2014}.

%---------------------------------------------------------
\subsection{${\HD110058}$}
ALMA data from \cite{LiemanSifry2016}. The disc inner edge is unresolved but ${<30 \au}$ with best fit ${10 \au}$, so we use ${10 \au}$ with ${100 \percent}$ uncertainties. In scattered light, the disc beyond ${40\au}$ shows a strong warp, implying that the inner disc is misaligned with the outer disc \citep{Kasper2015}; if caused by a planet, then that planet should be aligned with either the inner or outer disc. Our predicted ${\geq0.5\pm0.4\mJup}$ sculpting planet with semimajor axis ${\leq 8\pm8\au}$ would yield a secular timescale equal to the system age at ${\sim40\au}$, so this planet, if aligned with the inner disc, could be driving the warp.

%---------------------------------------------------------
\subsection{${\HD111520}$}
ALMA data from \cite{LiemanSifry2016}, disc model from REASONS (Matr\`{a} et al., in prep.). \cite{Draper2016} find a strong brightness asymmetry in scattered light, but argue against a high disc eccentricity.

%---------------------------------------------------------
\subsection{${\HD113337}$}
One planet with ${\mSinI = 3.1\pm0.2 \mJup}$ lies at ${1.03\pm0.02 \au}$ \citep{Borgniet2014}, and a candidate planet with ${\mSinI = 7.2\pm0.5 \mJup}$ may lie at ${4.8\pm0.2 \au}$ \citep{Borgniet2019}. We take the candidate planet as the outermost planet; Fig. \ref{fig: detectedAndPredictedPlanets} shows this planet could be stirring the disc, and is massive enough to have sculpted it (but not \textit{in situ}). There is a probable misalignment between the planet(s) and the disc \citep{Xuan2020}, and an M-star companion is also present at ${4000 \au}$ \citep{Reid2007}; the misalignment could be driven by the planet(s) and stellar companion \citep{Xuan2020}.

%---------------------------------------------------------
\subsection{${\HD114082}$}
\cite{Wahhaj2016} predict a planet at ${\sim 25\au}$ with mass ${\leq 1 \mJup}$; this is similar to our predicted sculpting planet of mass ${\geq 1.1\pm0.2 \mJup}$ and semimajor axis ${22\pm4 \au}$.

%---------------------------------------------------------
\subsection{${\HD115617 \; (61 \; \rm Vir)}$}
ALMA and JCMT data from \cite{Marino2017_61Vir}. Three super-Earths/Neptunes with minimum masses ${0.02-0.07 \mJup}$ lie within ${0.5\au}$ \citep{Vogt2010}, although the outermost is not confirmed by HARPS \citep{Wyatt2012}. We consider the outermost (unconfirmed) planet at ${0.476\pm0.001\au}$, with ${\mSinI = 23\pm3 \mEarth}$ \citep{Feng2017}; Fig. \ref{fig: detectedAndPredictedPlanets} shows that this planet cannot be sculpting or stirring the disc \textit{in situ}. \cite{Marino2017_61Vir} predict a stirring planet at ${10-20\au}$ with mass ${>10 \mEarth}$ (${0.03\mJup}$), if its eccentricity is less than 0.1. This is similar to our stirring planet prediction (mass ${\geq0.02^{+0.04}_{-0.03}\mJup}$), although the planet would have to be slightly more massive (${\geq0.06 \pm 0.02\mJup}$) to also sculpt the disc inner edge.

%---------------------------------------------------------
\subsection{${\HD120326}$}
The SPHERE image shows a ring at ${\sim 50\au}$ and a possible second structure at ${130\au}$ \citep{Bonnefoy2017}; this scattered-light ring is slightly larger than the ${31\pm6\au}$ SED-derived radius that we use.

%---------------------------------------------------------
\subsection{${\HD121191}$}
ALMA data from \cite{Moor2017} and \cite{Kral2020}, disc model from REASONS (Matr\`{a} et al., in prep.).

%---------------------------------------------------------
\subsection{${\HD121617}$}
ALMA data from \cite{Moor2017}, disc model from REASONS (Matr\`{a} et al., in prep.).

%---------------------------------------------------------
\subsection{${\HD123889}$}
A companion lies at projected separation ${200\au}$ \citep{Janson2013}, whilst the disc is at ${34 \pm 8 \au}$; it is possible that the two are interacting.

%---------------------------------------------------------
\subsection{${\HD127762 \; (\gamma \; \rm Boo)}$}
Circumbinary disc, where the binary separation is ${2\au}$ \citep{Yelverton2019}. The binary unlikely to be sculpting the disc at ${130\pm40\au}$.

%---------------------------------------------------------
\subsection{${\HD128311 \; (\rm HN \; \rm Boo)}$}
At least two Jovian planets lie within ${2 \au}$, possibly in 2:1 resonance \citep{Butler2003, Vogt2005}. We consider the outermost at ${1.8 \pm0.1\au}$ with ${\mSinI = 3.2\pm0.3 \mJup}$. Figure \ref{fig: detectedAndPredictedPlanets} shows that this planet could be stirring the disc, and is massive enough to sculpt it (but not \textit{in situ}); this agrees with \cite{MoroMartin2010} who show that the detected planets are unlikely to eject planetesimals outside ${4\au}$ (compared to the disc at ${140\pm70\au}$).

%---------------------------------------------------------
\subsection{${\HD131511 \; \rm (DE \; Boo)}$}
Circumbinary disc, where the binary comprises 0.79 and ${0.45\mSun}$ stars with a semimajor axis of ${0.2\au}$ \citep{Jancart2005}. The binary is aligned with the disc \citep{Kennedy2015HD131511}, but is unlikely to be sculpting it (the disc is at ${\sim 100\au}$).

%---------------------------------------------------------
\subsection{${\HD131835}$}
ALMA data and disc model from REASONS (Matr\`{a} et al., in prep.). In scattered light, the disc is wide and possibly comprises three concentric rings, with gaps at ${46-57\au}$ and ${71-85\au}$ \citep{Feldt2017}. Our sculpting planet predicted from ALMA data would be at ${\leq30\pm 2\au}$, which could be coincident with a tentative asymmetric inner ring seen in scattered light. 

%---------------------------------------------------------
\subsection{${\HD138813}$}
ALMA data and disc model from REASONS (Matr\`{a} et al., in prep.).

%---------------------------------------------------------
\subsection{${\HD139006 \; (\alpha \; \rm CrB)}$}
Circumbinary disc, where the binary has semimajor axis ${0.2\au}$ and eccentricity 0.37 \citep{Tomkin1986}. \cite{Kennedy2012} show that the binary could stir the disc. An inner belt may also be resolved \citep{Moerchen2010}, but is not factored in to our \textit{Herschel} single-disc model; our cold disc inner edge may therefore be underestimated. 

%---------------------------------------------------------
\subsection{${\HD139664 \; \rm (g \; Lup)}$}
ALMA data and disc model from REASONS (Matr\`{a} et al., in prep.). Disc possibly asymmetric in scattered light (\citealt{Schneider2014}, although also see \citealt{Kalas2006}).

%---------------------------------------------------------
\subsection{${\HD143894 \; (\pi \; \rm Ser)}$}
JCMT data may show a very large (${560\au}$) structure, but could be confused by background \citep{Holland2017}. This large size is comparable to our SED radius of ${240\pm70\au}$.

%---------------------------------------------------------
\subsection{${\HD146181}$}
ALMA data from \cite{LiemanSifry2016}. Their median ALMA model gives the inner edge as ${73^{+14}_{-19} \au}$ with width ${<50\au}$, whilst their best-fitting model gives the inner edge as ${83 \au}$ with width ${20 \au}$. We therefore take a width of ${20 \au}$, and assume uncertainties of ${50 \percent}$.

%---------------------------------------------------------
\subsection{${\HD146897}$}
ALMA data from \cite{LiemanSifry2016}. Their median ALMA model gives the inner edge as ${60^{+11}_{-13} \au}$ with width ${<50\au}$, whilst their best-fitting model gives the inner edge as ${71 \au}$ with width ${20 \au}$. We therefore take a width of ${20 \au}$, and assume uncertainties of ${50 \percent}$.

%---------------------------------------------------------
\subsection{${\HD156623}$}
ALMA data from \cite{LiemanSifry2016}. The inner edge is unresolved but ${<20 \au}$ with best fit ${10 \au}$, so we use ${10 \au}$ with ${100 \percent}$ uncertainties.

%---------------------------------------------------------
\subsection{${\HD161868 \; (\gamma \; \rm Oph)}$}
ALMA data and disc model from REASONS (Matr\`{a} et al., in prep.).

%---------------------------------------------------------
\subsection{${\HD172555}$}
A giant impact may have recently occurred in this system \citep{Johnson2012, Schneiderman2021}, so our dynamical models assuming a low-excitation debris disc may not apply.

%---------------------------------------------------------
\subsection{${\HD181296 \; (\eta \; \rm Tel)}$}
A brown dwarf is present beyond the disc, at 200 au \citep{Lowrance2000}; this could be interacting with the disc, which is at several tens of au.

%---------------------------------------------------------
\subsection{${\HD181327}$}
ALMA data and disc model from REASONS (Matr\`{a} et al., in prep.). There is possibly some unresolved exterior emission from another ring or a broad population \citep{Marino2016}. Asymmetric in scattered light \citep{Stark2014}. \cite{Nesvold2015} predict a sculpting planet at ${55-82\au}$, similar to our predicted sculpting planet with semimajor \mbox{axis ${\leq 53.9\pm 0.4 \au}$} and \mbox{mass ${\geq 1.6\pm 0.1\mJup}$}.

%---------------------------------------------------------
\subsection{${\HD191089}$}
ALMA data and disc model from REASONS (Matr\`{a} et al., in prep.). From scattered light images, \cite{Ren2019} give an upper limit on the ring eccentricity as 0.3. \cite{Churcher2011HD191089} predict a ${\geq 2.6\mEarth}$ planet at ${28\au}$ stirs the disc, similar to our prediction of a ${\geq5\pm2\mEarth}$ (${0.017\pm0.006\mJup}$) stirring planet. If the disc is sculpted by a single planet, then \cite{Ren2019} predict a ${<5\mJup}$ planet at ${>20\au}$, consistent with our prediction of \mbox{a ${\geq 0.89^{+0.07}_{-0.06} \mJup}$} planet at ${\leq28\pm1\au}$.

%---------------------------------------------------------
\subsection{${\HD193571 \; (\kappa^1 \rm \; Sgr)}$}
Target is a binary: an A0 star with a ${0.3-0.4\mSun}$ companion at ${\sim 11\au}$ \citep{MussoBarcucci2019}. Our models show that this companion cannot sculpt the disc at ${120\pm30\au}$, but could stir it.

%---------------------------------------------------------
\subsection{${\HD196544 \; (\iota \; \rm Del)}$}
Circumbinary disc, where the binary has separation ${0.2\au}$ \citep{Rodriguez2012}. The binary is unlikely to be sculpting the disc at ${100\pm30\au}$.

%---------------------------------------------------------
\subsection{${\HD197481 \; \rm (AU \; Mic)}$}
ALMA data from \cite{Daley2019}, disc model from REASONS (Matr\`{a} et al., in prep.). Two transiting Neptunes lie in the plane of the disc, inside ${0.1 \au}$ \citep{Plavchan2020, Martioli2021}. The outermost planet has mass ${>2 \mEarth}$ and is located at ${0.110 \pm0.002 \au}$ \citep{Martioli2021}; this mass is a lower bound so we assume a 50 per cent uncertainty. Fig. \ref{fig: detectedAndPredictedPlanets} shows that this planet cannot be stirring or sculpting the disc.

%---------------------------------------------------------
\subsection{${\HD202628}$}
ALMA data from \cite{Faramaz2019}, showing the disc to be narrow and eccentric (${e=0.09}$). Since  \cite{Faramaz2019} use offset circles (rather than ellipses) to model the disc edges, for most of our calculations we approximate the edges as ellipses with eccentricity 0.09 and semimajor axes equal to the offset circle radii. However, this approximation fails when calculating the self-stirring disc mass from Eq. \ref{eq: selfStirringMinimumMass}, so for that equation we take the semimajor axes of the innermost and outermost disc particles to be the radii of the inner and outer offset circles, respectively. Our predicted sculpting planet (mass ${\geq 0.33 \pm 0.06 \mJup}$, semimajor axis ${\leq 116 \pm 2 \au}$, eccentricity ${\geq 0.09 \pm 0.01}$) agrees with various literature predictions; \cite{Faramaz2019} predict mass ${0.8\mJup}$, semimajor axis ${110\au}$ and eccentricity ${0.1}$, \cite{Thilliez2016} predict mass ${3\mJup}$, semimajor axis ${100\au}$ and eccentricity ${0.2}$, and \cite{Nesvold2015} predict mass ${< 15\mJup}$ and semimajor axis ${86-160\au}$.

%---------------------------------------------------------
\subsection{${\HD206860 \; \rm (HN \; Peg)}$}
A companion T dwarf (${0.02 \mSun}$, ${20 \mJup}$) lies at ${800 \au}$ \citep{Luhman2007}. This is unlikely to be interacting with the disc at ${32\pm8\au}$.

%---------------------------------------------------------
\subsection{${\HD207129}$}
ALMA data and disc model from REASONS (Matr\`{a} et al., in prep.). Some hints at asymmetry in \textit{Herschel} images \citep{Lohne2012}, but we use an axisymmetric ALMA model. If a single planet sculpts the disc then we predict that planet to have mass ${\geq 0.21^{+0.03}_{-0.07} \mJup}$ and semimajor axis ${\leq 109 \pm 3 \au}$, which agrees with the prediction of \cite{Thilliez2016} that the planet is at ${100\au}$ (they also expect the planet to have eccentricity 0.06 based on the possible disc asymmetry). A companion lies at ${900\au}$ \citep{Rodriguez2015}, which could potentially interact with the disc at ${130-170\au}$. 

%---------------------------------------------------------
\subsection{${\HD213398 \; (\beta \; \rm PsA)}$}
A stellar companion lies at ${1000\au}$ \citep{Rodriguez2015}, which is unlikely to be interacting with the disc at ${60\pm30\au}$.

%---------------------------------------------------------
\subsection{${\HD216435 \; (\tau^1 \; \rm Gru)}$}
A planet with ${\mSinI = 1.3 \pm 0.1 \mJup}$ lies at ${2.6 \pm0.2 \au}$ \citep{Jones2003}; Fig. \ref{fig: detectedAndPredictedPlanets} shows that this planet could be stirring the disc, and is massive enough to sculpt it (but not \textit{in situ}).

%---------------------------------------------------------
\subsection{${\HD218396 \; (\rm HR \; 8799)}$}
\label{systemNotesSec: HR8799}
ALMA data from \cite{Faramaz2021}. We use their model 3, with a smoothly rising inner edge and smoothly falling outer edge, as expected from a disc containing a high eccentricity component (e.g. a scattered disc). Emission therefore extends both inwards and especially outwards of the radii we use, so our sculpting and stirring planet masses may be underestimates. A scattered population was also inferred by \cite{Geiler2019}. Four giant planets are detected \citep{Marois2008, Marois2010}, with semimajor axes of ${15-70 \au}$ and masses ${6-7 \mJup}$ \citep{Wang2018}. We consider the outermost planet at ${70.8 \pm 0.2 \au}$ with mass ${5.8 \pm 0.5 \mJup}$; Fig. \ref{fig: detectedAndPredictedPlanets} shows that this planet could be stirring the disc, and is massive enough to sculpt it (but not \textit{in situ}). This is consistent with \cite{Booth2016}, \cite{Read2018} and \cite{Geiler2019}, who also argue that the outermost known planet cannot sculpt the disc \textit{in situ}. \cite{Read2018} predict a fifth planet, showing that a ${0.1\mJup}$ planet at ${138\au}$ could reproduce the disc morphology (although other planets are also possible). Our predicted sculpting planets are larger than this, because our models assume that planets have cleared most unstable debris; however, unless the observed planets have since migrated, the presence of a significant scattered disc implies that debris clearing is ongoing in this system, in which case the planets could be less massive than our predictions. Planet detection limits were taken from \cite{Maire2015}.

%---------------------------------------------------------
\subsection{${\HD223340}$}
Target is component C in a quadruple system involving ${\HD223352}$ (which is component A). Separated from AB by ${3100\au}$ in projection \citep{Phillips2011}.

%---------------------------------------------------------
\subsection{${\HD223352 \; (\delta \; \rm Scl)}$}
\label{systemNotesSec: HD223352}
Target is component A in a quadruple system involving ${\HD223340}$ (which is component C). AB projected separation is ${164\au}$ \citep{Phillips2011}, so B could affect the disc around A (at ${26\pm8\au}$). B is two stars \citep{deRosa2011}.

%---------------------------------------------------------
\subsection{${\rm NLTT \; 54872 \; \rm (Fomalhaut \; C)}$}
ALMA data from \cite{CroninColtsmann2021}, showing the disc to be narrow and possibly eccentric (${e=0.04^{+0.03}_{-0.02}}$). Since \cite{CroninColtsmann2021} use an offset symmetrical torus (rather than ellipses) to model the disc edges, for most of our calculations we approximate the edges as ellipses with eccentricity 0.04 and semimajor axes equal to the radii of the offset torus edges. However, this approximation fails when calculating the self-stirring disc mass from Eq. \ref{eq: selfStirringMinimumMass}, so for that equation we take the semimajor axes of the innermost and outermost disc particles to be the radii of the offset torus edges, respectively. \cite{CroninColtsmann2021} argue that, if the disc edges are bound by the 2:1 and 3:2 resonances with a planet, then that planet would lie at ${17-20\au}$. This is consistent with our prediction that a single sculpting planet  would have semimajor axis ${\leq 21.6^{+0.5}_{-0.6} \au}$ and mass ${\geq0.048 \pm 0.003\mJup}$ (${15.3\pm0.9 \mEarth}$). Part of the Fomalhaut triple system.

\end{appendix}

\end{document}

%% file: journal_defs.tex
\long\def\symbolfootnote[#1]#2{\begingroup%
\def\thefootnote{\fnsymbol{footnote}}\footnote[#1]{#2}\endgroup} 
\def\apjl{ApJ}%          % Astrophysical Journal, Letters
\def\apjs{ApJS}%          % Astrophysical Journal, Supplement
\def\aap{A\&A}%          % Astronomy and Astrophysics